\documentclass[usenatbib,useAMS]{mn2e}
\newcommand\lsim{\mathrel{\rlap{\lower4pt\hbox{\hskip1pt$\sim$}}
        \raise1pt\hbox{$<$}}}
\newcommand\gsim{\mathrel{\rlap{\lower4pt\hbox{\hskip1pt$\sim$}}
        \raise1pt\hbox{$>$}}}

\newcommand{\vr}{v_{\hat{r}}}
\newcommand{\vp}{v_{\hat{\phi}}}
\newcommand{\hr}{{\hat{r}}}
\newcommand{\hp}{{\hat{\phi}}}

\usepackage{amsmath}
\usepackage{array}

\usepackage{graphicx}
\usepackage{fixltx2e}

\title[Accretion into the Central Cavity]
  {Accretion into the Central Cavity of a Circumbinary Disc}

\author[D. J. D'Orazio, Z. Haiman, and A. MacFadyen]
  {Daniel J. D'Orazio$^1$,
   Zolt\'an~Haiman$^1$, Andrew~MacFadyen$^2$\thanks{dorazio@astro.columbia.edu; zoltan@astro.columbia.edu; macfadyen@nyu.edu}\\
     $^1$Department of Astronomy, Columbia University, 550 West 120th Street, New York, NY 10027\\
  $^2$Center for Cosmology and Particle Physics, Physics Department, New York University, New York, NY 10003}
\begin{document}

\pagerange{\pageref{firstpage}--\pageref{lastpage}} \pubyear{2012}

\maketitle

\label{firstpage}

\begin{abstract}
  A near-equal-mass binary black hole can clear a central cavity in a
  circumbinary accretion disc; however, previous works have revealed
  accretion streams entering this cavity.  Here we use 2D
  hydrodynamical simulations to study the accretion streams and their
  periodic behavior.  In particular, we perform a suite of
  simulations, covering different binary mass ratios $q=M_2/M_1$ in
  the range $0.003 \leq q \leq 1$.  In each case, we follow the system
  for several thousand binary orbits, until it relaxes to a stable
  accretion pattern.  We find the following results: (i) The
  binary is efficient in maintaining a low-density cavity. However, the time-averaged mass accretion rate into 
  the cavity, through narrow coherent accretion streams, is suppressed by 
  at most a factor of a few compared to a disc with a single BH 
  with the same mass; (ii) for $q \gsim 0.05$, the accretion rate is
  strongly modulated by the binary, and depending on the precise value
  of $q$, the power spectrum of the accretion rate shows either one,
  two, or three distinct periods; and (iii) for $q \lsim 0.05$, the
  accretion rate becomes steady, with no time-variations.  Most
  binaries produced in galactic mergers are expected to have $q\gsim
  0.05$. If the luminosity of these binaries tracks their accretion
  rate, then a periodogram of their light-curve could help in their
  identification, and to constrain their mass ratio and disc properties.
\end{abstract}


\section{Introduction}

Massive black holes (MBHs) appear to reside in the nuclei of most
nearby galaxies (see, e.g., reviews by \citealt{kr95} and
\citealt{ff05}).  In hierarchical structure formation models, galaxies
are built up by mergers between lower--mass progenitors, which deliver
nuclear MBHs (e.g. \citealt{Springel+2005,Robertson+2006}), along with
a significant amount of gas \citep{BH1992}, to the central region of
the newly born post--merger galaxy.  Since mergers are common
(e.g. \citealt{HK2002}), it follows that massive black hole binaries
(MBHBs) should also be common in galactic nuclei.

Despite this expectation, observational evidence for MBHBs remains
scarce (see, e.g., \citealt{Komossa:Rev06,Tsalmantza:2011, Eracleous:2011}).
The dearth
of MBHBs could be attributed to several factors: it is possible that
typically only one of the two BHs is active at spatially resolvable
separations; binaries may also lose their angular momentum efficiently
due to the surrounding stars and gas and quickly move to spatially
unresolvable orbital separations.  Another possible hindrance, which
we address in this paper, is that the outward gravitational torques
from the binary can balance the inward viscous torques and pressure forces, 
clearing a central cavity in a putative circumbinary gas disc 
\citep{Artymowicz:1994}, possibly rendering the system too dim for
detection.  Overall, identifying MBHBs is difficult, and a better
understanding of their expected observational signatures, especially
those based on time-variability \citep{Haiman+2009}, is
needed. Merging MBHBs should be unambiguously identifiable by
gravitational wave (GW) detectors, such as eLISA \citep{eLISA:Amaro-Seoane:2012} or
ongoing Pulsar Timing Arrays (e.g. \citealt{PTAs}).  Identifying the
electromagnetic (EM) counterparts of these GW sources (among the many
false candidate galaxies in the GW error box) will, however, likewise
require an understanding of their observational signatures.

Recent studies have explored the gas-dynamics of circumbinary accretion
discs around near-equal-mass binaries in some detail.  Since the
system is not axisymmetric, this requires a two-- or
three--dimensional treatment.  \cite{MacFadyen:2008} (hereafter MM08)
have run two--dimensional hydrodynamical simulations for an equal-mass
binary. Three--dimensional smoothed particle hydrodynamical (SPH)
simulations have been carried out for equal-mass and 2:1 mass-ratio
binaries by \cite{Hayasaki:2007} and for a 3:1 mass-ratio binary by
\cite{Cuadra:2009} and \cite{Roedig:2012}.
\cite{ShiKrolik:2012} have followed up on the work of MM08 for an
equal-mass binary by running 3D magneto-hydrodynamical (MHD)
simulations, and \cite{Noble+2012} have further added a post-Newtonian
treatment of general relativistic (GR) effects, and followed the disc
through the late stages of orbital inspiral (from orbital separation
$r=20M$ to $8M$).  \cite{FarrisShap:2011} followed the merger of an
equal-mass binary and a surrounding disc through merger in full 3D
general relativistic MHD, starting from $10M$. Finally,
\cite{FarrisGold:2012} have added gas-cooling to GRMHD simulations of an
equal-mass binary prior to decoupling, through decoupling and to
merger starting from $10M$.  A generic result of all of these studies
is that a low--density cavity is carved out by the binary torque, but
gas leaks into the cavity through non-axisymmetric streams 
(as first discussed in the SPH simulations of \citealt{ArtyLubow:1996}).  These
streams can power significant accretion onto the binary components,
and should lead to bright EM emission.

A particularly promising feature is that the accretion rate onto the
BHs can be both high and strongly variable, modulated by the binary's
orbital motion. This could allow a detection of sub-pc binaries by
looking for periodic variations in the luminosity of AGN-like objects
\citep{HKM09} or periodic shifts and intensity variations of spectral
lines (e.g. \citealt{HKM09,SL2010,Eracleous:2011} and references
therein). If the accretion remains significant and periodic down to
$\ll$pc separations, then it could also enable the identification of
EM counterparts of gravitational wave sources: either for precursors
to eLISA sources in the $M=10^5-10^7{\rm M_\odot}$ range
\citep{Kocsis+2006,Kocsis+2008} or by detecting periodic modulations
of more massive $M=10^8-10^9{\rm M_\odot}$ binaries discovered by
pulsar timing arrays (PTAs; \citealt{Tanaka:2012,Sesana+2012}).

Although existing studies have focused on near-equal-mass MBHBs, in
reality, coalescing MBHBs should have a distribution of mass ratios
$q\equiv M_2/M_1$.\footnote{\citet{Nixon:2011:LongSim} explored a
  range of mass ratios, using 3D SPH simulations, but restricted their
  study to {\em retrograde} discs.}  Mergers occur between galaxies
over a wide range of sizes, harboring central BHs of different masses,
so that MBHBs resulting from galactic mergers should have a
correspondingly wide range of mass ratios.  Studies based on
Monte-Carlo realizations of dark matter merger trees indeed find broad
distributions between $10^{-2}\lsim q < 1$, generally peaking in the
range $q\sim 0.1-1$
(e.g. \citealt{Volonteri+2003,Sesana+2005,Sesana+2012,GergelyBiermann:2012}).
However, the predictions depend on the occupation fraction of MBHs,
the redshift-evolution of the correlation between the masses of MBHs
and their host galaxies, as well as on the limit on the mass ratio of
host galaxies whose nuclear MBHs can coalesce; $q<0.1$ mergers could
in fact be most common (e.g. \citealt{Lippai+2009}).

Here we follow up on the earlier work of MM08 and move beyond the
near-equal-mass binary case. We study the periodicity and the
time-averaged rate of accretion across the central cavity, by running
2D hydrodynamical simulations of a circumbinary disc for 10 different
binary mass ratios ranging from $q=0.003$ to $q=1$.  Clearly, one
expects that in the limit $q\rightarrow 0$, the accretion rate
approaches that of an accretion disc around a single BH, and will no
longer be time-variable.  The main goal in this paper is to answer the
following basic questions: {\em How does the mean accretion rate, and
  its fluctuations, depend on the mass ratio?  In particular, down to
  what mass ratio is the mean accretion and/or its variability
  significantly affected by the binary torques?} 
We address these questions with the caveat that, throughout this paper, accretion is defined as the mass 
crossing the inner boundary of the simulation domain and not necessarily 
that accreted by either BH.

The rest of this paper is organized as follows:
In \S \ref{Details of Numerical Simulations}, we describe the setup of
our numerical simulations, including changes we made to the
public version of the Eulerian grid code FLASH and the initial and boundary conditions we adopted.
In \S \ref{Results} we present our main results, namely that  
we find four distinct patterns for the time-variability of the
accretion rate as a function of the mass ratio $q$.  
In \S \ref{Summary and Discussion} we compare our findings with 
that of MM08 as well as investigate the dependence of our results 
on the magnitude of viscosity on the resolution. We also
discuss scaling of the simulations to
physical parameters, such as black hole mass and orbital separation,
and discuss the corresponding orbital and residence times,
as well as some caveats.
Finally, in \S \ref{Conclusions} we conclude by briefly summarizing our
main results and their implications.
The Appendix details our implementation of viscosity in polar coordinates, an important addition to FLASH.

\section{Details of Numerical Simulations}
\label{Details of Numerical Simulations}

To simulate a gas disc in the gravitational field of a binary, we use
the Eulerian grid-based hydrodynamical code FLASH (Version 3.2;
\citealt{Fryxell:2000}).\footnote{We note that MM08 used an earlier
release, Version 2, of the same code.}  FLASH solves the
volume--integrated fluid equations by solving the Riemann 
problem at each cell boundary. A piece-wise parabolic representation
of the fluid variables is used to interpolate between cells, i.e.
FLASH is a PPM code, accurate to 2nd order in both space and
time. FLASH uses a monotonicity constraint, rather than artificial
viscosity, to control oscillations near discontinuities.  This makes
it well suited for following supersonic fluid dynamics in the inner
regions of circumbinary disc. FLASH also supports polar coordinates,
which is convenient for simulating discs.

\subsection{Numerical Implementation and Assumptions}
\label{Numerical Implementation and Assumptions}

We assume a geometrically thin accretion flow with angular momentum aligned with that of the binary. This
permits a decoupling of the fluid equations in the $z$ direction,
perpendicular to the plane of the disc, so that we can define
height--integrated fluid variables and set up simulations in two
dimensions.  In follow--up studies, we plan to extend these
simulations to the full three dimensions, which we expect will be
important in determining the amount of inflow into a putative
circumbinary cavity (for recent 3D grid-based simulations, see
\citealt{ShiKrolik:2012} and \citealt{Noble+2012}). In the present
study, we choose 2D polar coordinates $(r, \phi)$ and employ FLASH to
solve the following standard set of 2D hydrodynamical equations:
\begin{align}
&{ \frac{\partial \Sigma}{\partial{t}} } + \nabla \cdot (\Sigma \vec{v}) = 0     \nonumber \\ \nonumber
&{ \frac{\partial \vec{v }}{\partial{t}} } + (\vec{v} \cdot \nabla) \vec{v} =  \\ \nonumber
& \qquad -{\frac{1}{\Sigma}} \nabla P  - \nabla \Phi_{\rm bin} + \nabla \cdot \nu \nabla \vec{v}  + \nabla \left( \frac{1}{2} \nu \nabla \cdot \vec{v} \right)\\
&{ \frac{\partial (\Sigma E)}{\partial{t}} } + \nabla \cdot \left[ (\Sigma E + P)\vec{v}\right] = \Sigma \vec{v} \cdot (-\nabla \Phi_{\rm{bin}})  
\label{fluideqns}
\end{align}
Here $\Sigma$ is the vertically integrated disc surface density,
$\vec{v} = v_r \hat{r} + v_{\phi} \hat \phi$ is the fluid velocity,
$P$ is the pressure, $\nu$ is the coefficient of kinematic viscosity,
$E$ is the total internal plus kinetic energy of the fluid, $E =
\epsilon + \frac{1}{2} |\vec{v}|^2$, and $\Phi_{\rm{bin}}$ is the
gravitational potential of the binary.  The gravitational potential is
inserted into the simulation by hand, and is given by
\begin{align}
\Phi_{\rm{bin}}(r, \phi) &= -\frac{GM(1+q)^{-1}}{ \left[r^2 + \left(\frac{a}{1+q^{-1}}\right)^2 - \frac{2ra}{1+q^{-1}} \cos{\left(\phi - \Omega_{\rm{bin}} t \right)} \right]^{1/2}  }    \nonumber \\
 &  - \frac{GM(1+q^{-1})^{-1}}{ \left[ r^2 + \left(\frac{a}{1+q}\right)^2 + \frac{2ra}{1+q} \cos{\left(\phi - \Omega_{\rm{bin}} t \right)}   \right]^{1/2} }.
 \label{bpot}
\end{align}
Here $\Omega_{\rm bin} = \left( GM/a^3\right)^{1/2}$ is the binary's
orbital frequency, $a$ is the separation of the binary, $M_p > M_s$
are the masses of the primary and the secondary, $M=M_p+M_s$ is the
total mass, and $q = M_s/M_p \leq 1$ is the mass ratio.  The origin of
the coordinate system is chosen to coincide with the binary's center of
mass. In the case of a single point mass, we use the limit of equation (\ref{bpot}) as 
$q, a \rightarrow 0$, $\Phi_{\rm{bin}} = - GM/r$. 

Note that we do not evolve the orbital parameters of the binary nor do we allow its center of mass to wander; 
these simulations are numerical experiments which are physically motivated in the limit of small disc mass 
(this assumption is justified for our physical parameter choices; see discussion in \S\ref{Physical Regime} below).

We neglect self-gravity of the disc. Given the local sound speed
$c_s$, the Toomre parameter $Q\equiv c_s \Omega /\left( \pi G \Sigma
\right)$ can be written as $Q \sim (H/r)(M/M_d)$, for a disc with mass
$M_d$ and vertical scale-height $H$ in hydrostatic equilibrium. For a
thin disc $H/r \lsim 0.1$, but in all of our simulations, we choose $M
\gg M_d$ and thus $Q \gg 1$, making the disc stable to gravitational
fragmentation. This is justified for standard Shakura-Sunyaev discs,
when the simulations are scaled to physical BH masses and sufficiently
small separations (see \S\ref{Physical Regime} below).  For
comparison, we note that in their SPH simulations, \cite{Cuadra:2009}
and \cite{Roedig:2012} studied more massive discs, with $M_d
\sim 0.2 M$, making self-gravity important.

The pressure is given, as in MM08, by a locally isothermal equation of
state,
\begin{equation}
P = c_s^2(r) \Sigma
\end{equation}
where the sound speed is assumed to scale with radius as $c_s \propto
r^{-1/2}$. For a Keplerian potential, this corresponds to a constant
disc scale--height to radius ratio, $H/r = c_s / v_{\phi} \equiv
\mathcal{M}^{-1}$, where $v_{\phi}$ is the orbital velocity in the
disc and $\mathcal{M}$ is the corresponding Mach number.  Throughout
our simulations, we choose the disc sound speed such that, for a
Keplerian azimuthal velocity, $H/r=0.1$ (or $\mathcal{M}=10$)
everywhere.\footnote{Since we add the binary's quadrupole
  potential, and the non-zero pressure makes the azimuthal velocities
  slightly sub-Keplerian, in practice $\mathcal{M}$ in the simulations
  approaches $\sim 18$ near $r=a$ and becomes a constant
  $\mathcal{M}\sim 10$ at $r>5a$.}

To incorporate viscosity, FLASH calculates the momentum flux across
cell boundaries due to viscous dissipation (the last two terms in the
second of equations~\ref{fluideqns}). To compute $\nu$ we adopt the
$\alpha$ prescription, $\nu=\alpha c_s H$ \citep{SS73}, where $\alpha$
is a dimensionless parameter indicating the scale of turbulent cells,
and the scale height is computed from $H = c_s r / v_{\phi}$ with
$v_{\phi}$ being the Keplerian value.  Following MM08, we choose a fiducial,
constant $\alpha=0.01$ (although we explore the effects of increasing $\alpha$ in \S \ref{Viscosity Study}). 
Since the FLASH viscosity implementation is not fully supported in
polar coordinates, we made adjustments to the routines that compute
the momentum flux and viscous diffusion time from $\nu$, originally in Cartesian coordinates. 
These modifications and tests of our polar viscosity implementation are detailed in the Appendix.

\subsection{Numerical Parameter Choices}
\label{Numerical Parameter Choices}

The inner edge of the computational domain
is chosen at $r_{\rm min}=a$ and the outer edge at $r_{\rm max}=100a$.
Although we are only interested in the inner few $r/a$ of the disc,
extending the computations to larger radii acts as a buffer for small
initial numerical transients, and also provides a potential reservoir of gas
from which the inner regions can be fed.  As in other Eulerian codes,
FLASH uses a boundary zone of `guard cells' which enforces boundary
conditions. We choose a diode-type inner boundary condition: the values
of fluid variables in cells bordering the guard cells are copied into
the guard cells, with the restriction that no fluid enter the domain,
$v_r(r_{\rm min}) \leq 0$.  We adopt an `outflow', outer boundary
condition: this is identical to the diode version, except that flow is
allowed both into and out of the domain. In practice, with our initial
conditions in \S \ref{Initial Conditions}, we find an outflow
at the outer boundary of our disc. However, the simulation does not
run for a significant fraction of a viscous time at the outer
radius (see below) and we expect inflow at the outer boundary to establish itself eventually.

Unless specified otherwise, the spatial resolution is fixed throughout the grid, with the default
values set at $\left[ \Delta r/a, \Delta \phi/2\pi \right] \simeq
\left[0.024, 0.0078\right]$, corresponding to a grid of $\sim 4096
\times 128$ cells.  The wavelength of a density wave due to a Lindblad
resonance is of order $2 \pi H \sim0.6r$
(e.g. \citealt{DRS2011,DM2012}); as in MM08, the radial resolution is
chosen to resolve this length scale with many cells. Note that our fiducial resolution 
results in cell aspect ratios which are approximately square at $r \sim 3a$. To test
numerical convergence (see $\S \ref{Resolution Study}$ below), we
performed several additional runs, increasing the radial and azimuthal
resolutions, by factors of $1.42$ and $1.42^2\sim 2.0$, from the lowest resolution runs.
The time resolution is set to be half of the shortest propagation time
(viscous or dynamical) across a cell.  The effects of changing the
resolution are discussed in \S \ref{Resolution Study}.

\begin{table*}
\begin{center}
\caption{\em Summary of our simulation runs. Low, medium, and high radial and
azimuthal resolutions are denoted by ``Lo$\Delta r$"; ``Mid$\Delta r$";  ``Hi$\Delta r$'' 
and ``Lo$\Delta\phi$"; ``Mid$\Delta\phi$";  ``Hi$\Delta\phi$''  and correspond to 
$\triangle r/a = 0.035, 0.24, 0.017$ and $\triangle \phi/2\pi = 0.0078, 0.0052, 0.0039$, respectively. 
The viscosity parameter $\alpha$ is set to 0.01 unless otherwise specified.}
\label{Tbl_simruns}
\begin{tabular}{   l   |    l   |   l    }
  Mass ratio $q$    &   Spatial Resolution $\left[\triangle r, \triangle \phi \right]$  &  No. Orbits ($N_{\rm orb}=t_{\rm max} \Omega_{\rm bin}/2\pi$)\\ \hline
  $1.0$    &    $\left[\rm{Lo}\Delta r, \rm{Lo}\Delta \phi \right]$,  $\left[\rm{Mid}\Delta r , \rm{Lo}\Delta \phi \right]$    ,  $\left[\rm{\rm{Mid}\Delta r}, \rm{Mid}\Delta \phi \right]$,    $\left[\rm{\rm{Hi}\Delta r}, \rm{Hi}\Delta \phi \right]$,  &  5500, 6800, 4500, 4200       \\ 
  $1.0$ ($\alpha=0.02$)   &     \hbox{  } \qquad  \qquad \qquad   $\left[\rm{Mid}\Delta r , \rm{Lo}\Delta \phi \right]$    &  4500      \\ 
   $1.0$ ($\alpha=0.04$)   &    \hbox{  } \qquad  \qquad \qquad   $\left[\rm{Mid}\Delta r , \rm{Lo}\Delta \phi \right]$    &  5200      \\ 
    $1.0$ ($\alpha=0.1$)   &    \hbox{  } \qquad  \qquad \qquad   $\left[\rm{Mid}\Delta r , \rm{Lo}\Delta \phi \right]$    &  1400      \\ 
  $0.75$  &   $\left[\rm{Lo}\Delta r, \rm{Lo}\Delta \phi \right]$,  $\left[\rm{Mid}\Delta r, \rm{Lo}\Delta \phi \right]$      &  6500, 4500    \\ 
  $0.5$    &   $\left[\rm{Lo}\Delta r, \rm{Lo}\Delta \phi \right]$,  $\left[\rm{Mid}\Delta r, \rm{Lo}\Delta \phi \right]$      &  6000, 4500   \\ 
  $0.25$  &  $\left[\rm{Lo}\Delta r, \rm{Lo}\Delta \phi \right]$,  $\left[\rm{Mid}\Delta r, \rm{Lo}\Delta \phi \right]$       &  5700, 4500   \\ 
    $0.1$    &   $\left[\rm{Lo}\Delta r, \rm{Lo}\Delta \phi \right]$,  $\left[\rm{Mid}\Delta r, \rm{Lo}\Delta \phi \right]$,  $\left[\rm{\rm{Mid}\Delta r}, \rm{Mid}\Delta \phi \right]$,    $\left[\rm{\rm{Hi}\Delta r}, \rm{Hi}\Delta \phi \right]$,  &7000 , 8000, 4500, 4100     \\ 
   $0.075$    &   $\left[\rm{Lo}\Delta r, \rm{Lo}\Delta \phi \right]$,  $\left[\rm{Mid}\Delta r, \rm{Lo}\Delta \phi \right]$      &  4500, 5200       \\ 
   $0.05$  &   $\left[\rm{Lo}\Delta r, \rm{Lo}\Delta \phi \right]$,  $\left[\rm{Mid}\Delta r, \rm{Lo}\Delta \phi \right]$     ,  $\left[\rm{\rm{Mid}\Delta r}, \rm{Mid}\Delta \phi \right]$,    $\left[\rm{\rm{Hi}\Delta r}, \rm{Hi}\Delta \phi \right]$,  &  4500, 7600, 4500, 4200     \\ 
   $0.025$   &  \hbox{  } \qquad  \qquad \qquad $\left[\rm{Mid}\Delta r, \rm{Lo}\Delta \phi \right]$     &  4500       \\ 
   $0.01$  &   $\left[\rm{Lo}\Delta r, \rm{Lo}\Delta \phi \right]$,  $\left[\rm{Mid}\Delta r, \rm{Lo}\Delta \phi \right]$      &  5200, 5500     \\
   $0.003$  &     \hbox{  } \qquad  \qquad \qquad   $\left[\rm{Mid}\Delta r, \rm{Lo}\Delta \phi \right]$      &  5500     \\
     $0$ \  (All $\alpha$)  &   $\left[\rm{Lo}\Delta r, \rm{Lo}\Delta \phi \right]$,  $\left[\rm{Mid}\Delta r, \rm{Lo}\Delta \phi \right]$,  $\left[\rm{\rm{Mid}\Delta r}, \rm{Mid}\Delta \phi \right]$,    $\left[\rm{\rm{Hi}\Delta r}, \rm{Hi}\Delta \phi \right]$,  & 6000 , 8000, 4500, 4200
   \label{runs}
  \end{tabular}
 \end{center}
\end{table*}

We run our simulations for between $4\times 10^3$ and $10^4$ binary
orbits. For reference, we note that the viscous time can be related
to the orbital time as
\begin{equation}
t_{\rm visc} = \frac{2}{3}\frac{r^2}{\nu} =\frac{\mathcal{M}^2}{3 \pi \alpha} t_{\rm orb} \simeq 1060 \left( \frac{\mathcal{M}^2}{100} \right) \left(\frac{0.01}{\alpha} \right) t_{\rm orb}.
\label{tv_torb}
\end{equation}
Thus, our typical run of 4000 binary orbits corresponds to $\sim4$ viscous
times at the innermost regions, but less than one viscous time at
$r\gsim3a$.

We have performed 34 runs altogether, for 10 different binary mass
ratios between $q=0.003$ and $q=1.0$ (including control runs for
$q=0$, i.e. a single BH). The mass ratio, resolution, and the number
of binary orbits followed in each of our simulation runs are
summarized in Table~\ref{Tbl_simruns}.

\subsection{Initial Conditions}
\label{Initial Conditions}

In our initial conditions, we insert a central cavity around a binary,
and also include a density pile-up just outside the cavity wall.  Such
a surface density profile is expected to develop during the inward
migration of the secondary. When the secondary arrives at the radius
where the local disc mass is too small to absorb the secondary's
angular momentum, its migration stalls, the inner disc drains onto the
primary, and continued accretion from larger radii causes a pile-up of
gas outside the secondary's orbit
\citep{SyerClarke95,Ivanov99,Milos:Phinney:2005,Chang+2010,Rafikov2012}.
As emphasized recently by \citet{Kocsis+2012a}, the details of this
process are still uncertain, as the coupled time-dependent migration,
cavity formation, and pile-up, has not been modeled self-consistently,
even in one-dimensional calculations.  However, using self-consistent
{\em steady-state} solutions, \citet{Kocsis+2012b} showed that in many
cases, the pile-up can cause overflow already at large binary
separations.

For simplicity, and for ease of comparison, we adopt the same initial
surface density profile as in MM08. This profile is motivated by the
earlier results of \cite{Milos:Phinney:2005}; it has very little gas
inside of $r\simeq3a$, and peaks at $\sim 8a$,
\begin{equation}
\Sigma(r, t_0) = \Sigma_0 \left( \frac{r_s}{r} \right)^{3} \mbox{exp}\left[-\left(\frac{r_s}{r}\right)^2\right].
\label{Sig0}
\end{equation}
Here $\Sigma_0$ is an arbitrary constant, and $r_s=10a$. The initial
profile is shown by the [blue] dashed curves in Figure~\ref{AzAvDens}
below.

We also follow MM08, and incorporate pressure gradients and the
quadrupole contribution of the binary's potential into the initial
azimuthal velocity,
\begin{equation}
\Omega^2 = \Omega^2_K \left[ 1 + \frac{3}{4} \left(\frac{a}{r}\right)^2  \frac{q}{(1+q)^2}\right]^2 + \frac{1}{ r \Sigma } \frac{d P}{d r}
\end{equation}
and account for viscous drift in the initial radial velocities,
\begin{equation}
v_r =  \  \frac{d }{d r} \left( r^3 \nu \Sigma \frac{d \Omega}{dr} \right)  \left[r \Sigma \frac{d}{d r} \left( r^2 \Omega\right) \right]^{-1}.
\end{equation}

We emphasize that with these initial conditions, the disc is not initially in
equilibrium, and material diffuses away from the peak of the surface
density, both inward and outward, due to pressure gradients.  However,
after running the simulations for several thousand orbits, the system
relaxes to a steady pattern of accretion, and we do not expect the
initial profile to significantly influence our conclusions.

\subsection{disc Parameters and Accretion Rate}
\label{disc Parameters and Accretion Rate}

Our primary goal is to quantify the magnitude and variability of
accretion across the central cavity.  To this end we compute the
time-dependent accretion rate at the inner edge of the simulation
domain, $r_{\rm{min}} = a$
\begin{equation}
\dot{M}(t) = \int^{2\pi}_{0}{ \Sigma(r_{\rm{min}}) v_r(r_{\rm{min}})
r_{\rm{min}} d\phi }.
\end{equation}
Since we are unable to track the fate of the gas at smaller radii, nor
do we allow the masses of the BHs to increase, this mass is
effectively lost from the simulation. In practice, the total mass that
is lost is a small fraction of the total initial disc mass (at most a
few \% by the end of each run).

Because we neglect the self-gravity of the disc,
equations~(\ref{fluideqns}) are independent of $\Sigma_0$, and there
is no unique way to assign a physical normalization to
$\dot{M}_{\rm{sim}}$ without appealing to a disc model which includes
additional physics.  Instead, MM08 compare the average accretion rate
at the edge of the integration domain ($r_{\rm{min}}=a$) to that in a
disc in a Keplerian potential with the same surface density at
$r\simeq3a$,

\begin{equation}
\frac{\dot{M}_{\rm{bin}} }{\dot{M}_{\rm{free}} } \simeq \frac{
\left<\dot{M}_{\rm{sim}}(a) \right>_t }{6 \pi \alpha}
\left(\frac{H}{r}\right)^{-2} \left( GMr\right)^{-1/2}
\Sigma^{-1}(3a).
\label{Mbin_Mfree}
\end{equation} 

MM08 find ${\dot{M}_{\rm{bin}} }/{\dot{M}_{\rm{free}} } \simeq
0.2$. To make this comparison meaningful, one must assume that
$\dot{M}_{\rm{free}}(3a) \sim \dot{M}_{\rm{free}}(a)$, \textit{i.e.}
that the reference, fictitious point--mass disc is in steady-state.
However, for a steady-state disc, specifying $\alpha$ and $H/r$ (along
with the dominant source of opacity) sets the physical value
$\dot{M}_{\rm{free}}$. For our fiducial values of $M = 10^7M_{\odot}$
and $a=10^3 r_S$ (and with $\alpha=0.01$ and $H/r=0.1$), we find the
unphysically large value of $\dot{M}_{\rm{free}} \sim 10^7
\dot{M}_{\rm{Edd}}$, a result of the above choices (where $\dot{M}_{\rm{Edd}}$ is the accretion rate
that would produce the Eddington luminosity, with a radiative
efficiency of 10\%).  If we instead require $\dot{M}_{\rm{free}} \sim
\dot{M}_{\rm{Edd}}$, then this would translate to a much thinner disc,
with $H/r \simeq 4 \times 10^{-3}$.  In such a thin/cold disc,
resolving density waves with the same number of cells as in MM08 would
require a radial grid resolution $\sim 17$ times higher than in our
highest-resolution run, and would be impractical.

Rather than attempting to compare our binary simulations to a
hypothetical steady-state point-mass disc, we choose to leave the
surface density normalization $\Sigma_0$ essentially arbitrary, and
instead perform explicit reference simulations with a single BH
(i.e. $q \rightarrow 0$), with the same initial conditions as the binary runs.
This approach has the advantage of explicitly isolating the effect of
turning on/off the binary.  Note, however, that our point--mass runs
should {\em not} be expected to produce the steady-state accretion rate
for a corresponding single--BH system. This is because our initial
conditions are far from this state, and we do not run the simulations
long enough (i.e. for a few viscous time at the outer edge) to allow
it to settle to the correct steady-state.

\subsection{Tests of Code Implementation}
\label{Tests of Code Implementation}

To ensure that non-axisymmetric perturbations are not induced
artificially in the disc, we simulate a disc around a single black hole for $10^4$ binary orbits.  To
keep axisymmetry during these runs, we re-implemented the FLASH2
routine UNBIASED-GEOMETRY (UBG)\footnote{This routine is titled {\em
    clean$\_$last$\_$bits} in the FLASH2 download.} into the FLASH3
routines.  UBG cleans up round--off errors in the cell boundary
positions, and forces the polar grid to keep cell sizes uniform in the
azimuthal direction. Without UBG, small perturbations in the azimuthal
direction grow to significant size after a few hundred orbits. We have
verified that after re-implementing UBG, axisymmetry is preserved for
$10^4$ binary orbits.   

A detailed description as well as multiple tests of our viscosity implementation are laid in out in the Appendix.

\section{Results}
\label{Results}

\subsection{Equal-Mass Binary}
\label{Equal Mass Binary}

We begin by describing the results of our equal-mass binary runs in
some detail.  Although these are very similar to those of MM08, this
will serve as a useful point of comparison for our unequal-mass runs.

\subsubsection{A Toy Model with Massless Particles}
\label{A Toy Model}

\begin{figure}
\begin{center}
\includegraphics[scale=0.42]{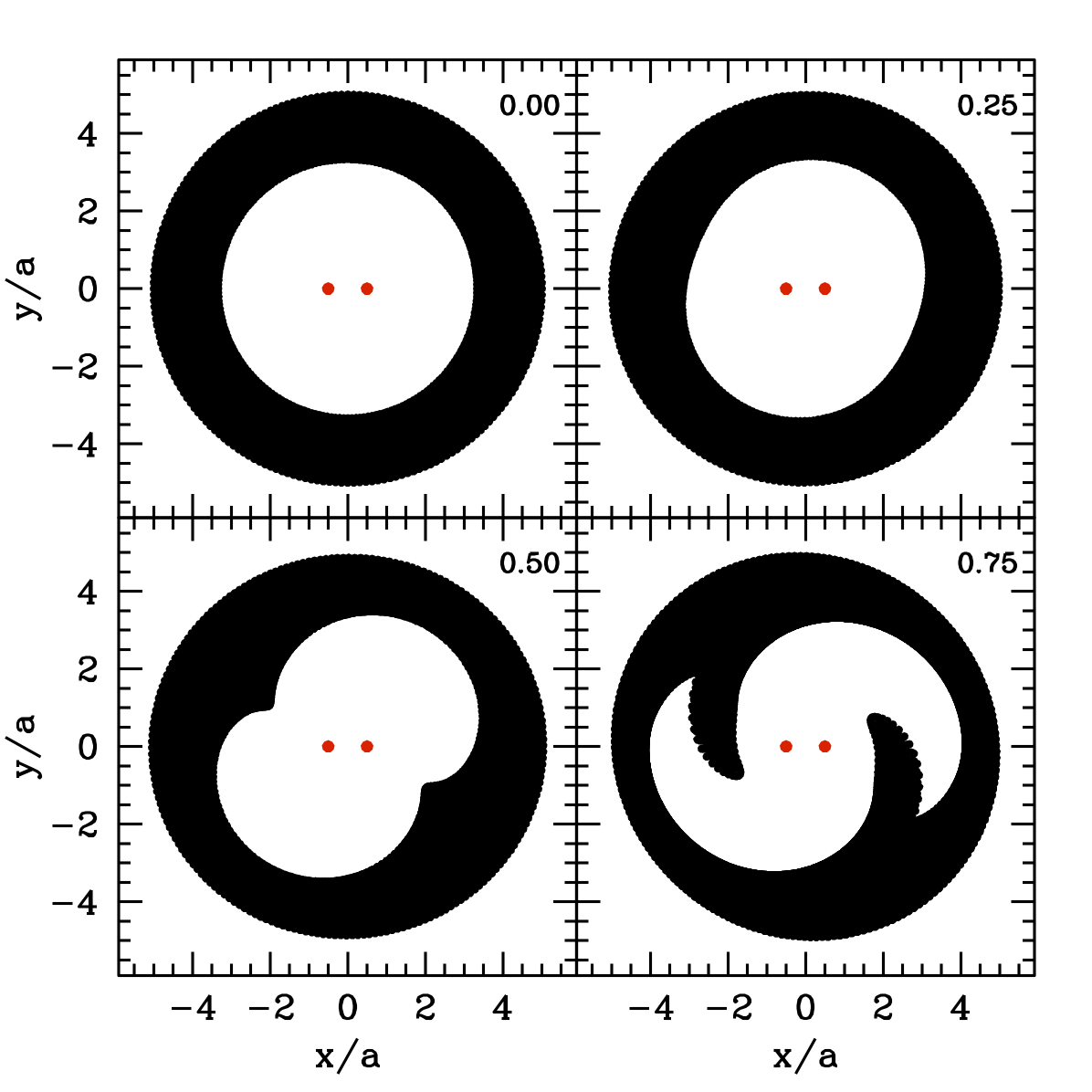} 
\caption{The distortion of a disc of massless test particles,
  initially in circular orbits around the center-of-mass of an
  equal-mass binary, with a central cavity.  The panels show snapshots
  of the locations of the disc particles after 0.25, 0.5, and 0.75
  binary orbits, as labeled.  The orbits of the particles were
  followed by solving the restricted three-body problem, and are shown
  in a frame co-rotating with the binary. The binary point masses are
  marked by the two [red] dots at $x=\pm 0.5a$.  The figure
  illustrates the tendency of the binary to create streams of
  particles entering the central cavity, due to gravity alone.}
\label{toydisc}
\end{center}
\end{figure}

Before showing the results from our simulations, we consider a simple
toy model, based on the orbits of non-interacting massless test
particles around a binary.  We populate a 2D disc with test particles,
centered on the binary's center of mass, and leave a central
cavity. We assign initial velocities equal to the Keplerian velocities
around a single point mass $M=M_p+M_s$.  We then follow the orbit of
each test particle in the rotating binary potential, by numerically
solving the restricted three-body problem for each individual particle
(for $10^5$ particles in practice, using equations 3.16 and 3.17 in
\citealt{MDbook}).

The results of this simple exercise are displayed in
Figure~\ref{toydisc}, which shows the locus of the test particles
initially, as well as after 0.25, 0.5, and 0.75 binary orbits (in a
frame co-rotating with the binary).  As this figure demonstrates,
there is a tendency for the binary to pull streams of particles into
the cavity.  This, of course, is purely a gravitational effect. As \citealt{ArtyLubow:1996} 
have pointed out, such mass flows occur near unstable 
co-rotatation equilibrium points in the binary potential. 

The toy-model can not be pushed much further in time, since after $\sim
0.75$ binary orbits, the trajectories of the test-particles cross --
this necessitates a hydrodynamical treatment.  Nevertheless, the
figure does suggest that an empty cavity can not be maintained by an
equal-mass binary, even in the absence of pressure or viscosity.
Furthermore, as we will see below, the simulations show accretion
streams with morphologies quite similar to those in the bottom right
panel of Figure~\ref{toydisc}.

\subsubsection{Hydrodynamical Evolution: Reaching Steady State}
\label{Hydrodynamical Evolution: Reaching Steady State}

We next present the results from our equal-mass binary
simulations. The disc evolves through two distinct stages.  As
explained above, the disc is set up to be out-of-equilibrium, and we
observe an initial transient state, which lasts for $\sim2500$ orbits.
We expect the details of this state to depend on the initial
conditions. The disc then settles to a quasi-steady-state, which
persists for the rest of the simulation.  In this quasi-steady-state,
the disc exhibits significant accretion, which varies periodically on
two timescales, $(1/2)t_{\rm{bin}}$ and $\sim (5 - 6)t_{\rm{bin}}$. We expect
these latter features to be robust and insensitive to initial
conditions.

{\em Transient State.}  Initially, pressure forces and viscous
stresses act to move material inside of the initial density peak at
$r\sim 8a$ inward while pressure gradients move material outside of
the density peak outward.  Once the inner disc material reaches
$r\simeq 2a$, its surface density becomes strongly perturbed and
highly non-axisymmetric.  Reminiscent of the evolution of the toy
model in \S \ref{A Toy Model}, two narrow point-symmetric streams
develop, in which the gas flow becomes nearly radial.  About 3$\%$ of
the material in these streams exits the integration domain at
$r_{\rm{min}}=a$. The rest gains angular momentum from the faster
moving black holes and is flung back toward the bulk of the disc
material at $r\simeq 2a$.  This process maintains a central cavity
within the disc, with gas streams being pulled in and pushed out on a
period of $\sim1.5 \hbox{ } \Omega_{\rm{bin}}$.

As more disc matter flows in from the initial density peak to the
inner $r\sim 2a$ region, the streams become more dense. When these
streams are flung back out and hit the opposing cavity wall, they
generate noticeable over-densities, and deform the circular shape of
the cavity to become eccentric, though still point-symmetric.  These over-densities then
rotate at the disc's orbital velocity.  They spread out and propagate
into the disc as differential rotation causes them to wind up,
creating a point-symmetric ($m=2$), rotating spiral pattern.

{\em Quasi-Steady State.}  After the initial $\sim2500$ orbits, 
the point symmetry of the transient state breaks down as stream generation
becomes preferentially stronger on one side of the cavity. This causes 
more mass, in the form of a stream, to be driven into the opposite cavity wall, 
pushing that side of the wall farther away from the binary.

This lopsided state can grow from a small initial asymmetry, through
a genuine physical instability, as follows:

Initially, in the transient point-symmetric state, the central cavity
has an elliptical (but still point-symmetric) shape, which rotates
along with the disc, its inner edge completing a rotation once every 3 binary
orbits. Streams are simultaneously pulled in from the two near sides
of this elliptical cavity. After the streams form, they are flung
across the cavity, and hit the region approximately diagonally across,
close to the azimuth where the opposite stream formed.  The above
process makes the cavity more eccentric over time, since when the
outward-going stream material hits the cavity wall, it pushes it
outward, further away from the binary's center of mass.

\begin{figure*}
\begin{center}
\includegraphics[scale=0.4]{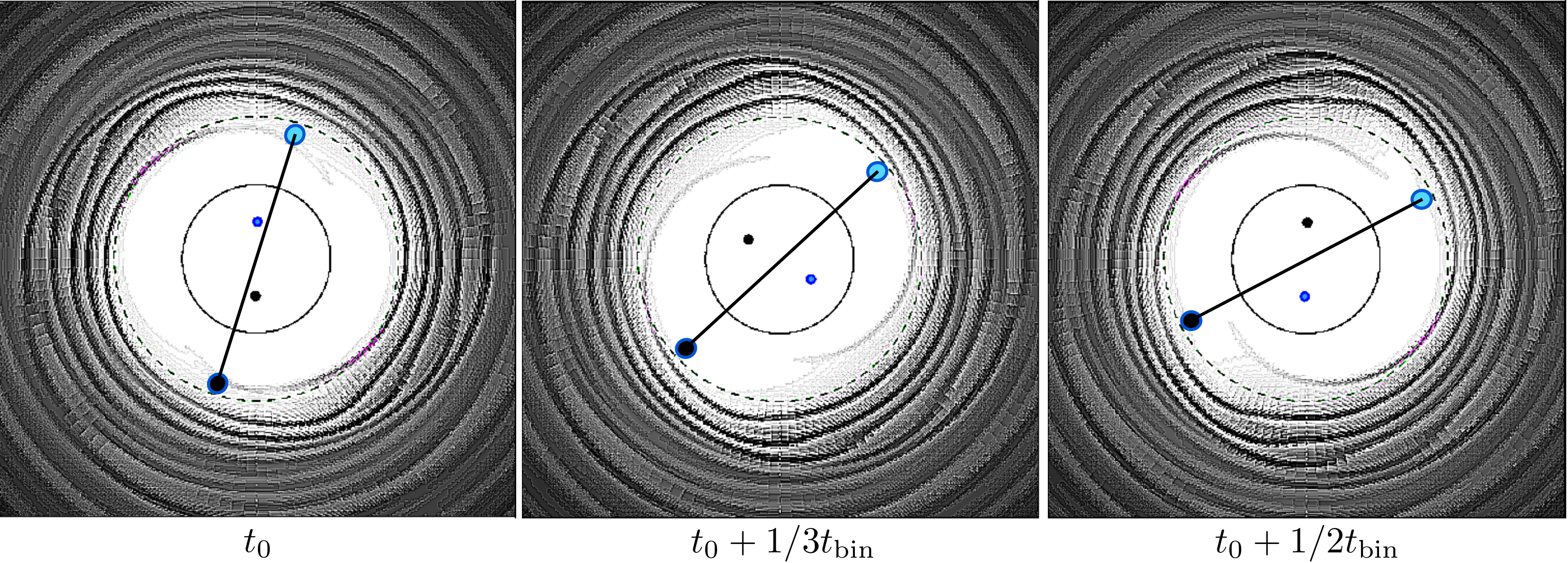} 
\caption{Snapshots of the surface density for the $q=1$ disc in the
   point-symmetric transient state after $\sim 1350$ binary orbits. The
   snapshots are sharpened (masked to include only high Fourier
   frequencies in the image) in order to see the streams more
   clearly. The connected outer circles are drawn to guide the eye. These
   circles rotate with the disc structure at a period of $\approx 3
   t_{\rm bin}$. Streams are generated on either side of the cavity,
   (shown in the left panel; the streams shown in this panel are moving
   inward, away from the cavity wall).  These streams ultimately crash
   into the cavity wall on the diagonally opposite sides of the cavity,
   near the site where the opposing stream was generated (shown in the
   right panel; the streams shown in this panel are moving outward,
   toward the cavity wall).  If a small asymmetry causes one stream to
   become stronger, then a runaway process would ensue pushing one side
   of the disc further from the binary and allowing the other side to
   come closer; this process could ultimately be responsible for the
   observed lopsided shape of the cavity.}
\label{Fig:AssymForm}
\end{center}
\end{figure*}

Figure \ref{Fig:AssymForm} shows the above, via three snapshots of the disc
in the transient, point-symmetric state, in time order from left to
right.  The first panel shows the initial formation of the two
opposing streams, demonstrating that the streams originate at
locations where the cavity wall is closest to the binary's center of
mass. The third panel shows the collision of the outward-going streams
with the opposing cavity wall, demonstrating that the streams collide
with the cavity wall approximately diagonally across their initiation
point.  The second panel shows the morphology of the streams half-way
between these time steps, for the sake of completeness.

Next, imagine that due to numerical noise, one stream (say, from side
"A" of the cavity) carries slightly more momentum than the spatially
opposite stream emanating from side "B".  This can happen due to a
small initial lopsidedness in the shape of the cavity, with side "A"
being closer to the binary than side "B" (or due to an asymmetry in
the density or velocity field).  In our simulations, this can only be
due to small numerical noise, but in reality, discs will obviously not
be perfectly symmetric, either.  The stronger stream will hit side "B"
of the cavity as before, but will push the cavity wall farther away
from the binary than the comparatively smaller counterpart stream
hitting side "A".  It is easy to see that this can lead to a runaway
behavior: side "A" of the cavity will have absorbed less momentum, and
will now be even closer to the binary's center of mass, relative to
side "B" - causing a larger asymmetry in the next pair of streams,
which further increases the lopsidedness of the cavity, and so on


This reenforcement-feedback process continues for a period of $\gsim$ 200-300 orbits, after
which the weaker stream, and its effect on the cavity wall structure
disappears entirely. The central cavity takes on a lopsided
shape, with a near-side where streams are pulled from the cavity wall
by each passage of the holes, and a far side where non-accreted
material from the streams is flung back and crashes into the cavity
wall.  At the azimuthal locations of these crashes, the far-side of
the wall develops very strong shocks, with Mach numbers up to
$\mathcal{M}\sim15$. (However, since our disc is locally
isothermal, this is likely an upper limit for the shock strength). 
The lopsided cavity precesses in the frame at rest with respect 
to the binary center of mass, completing a rotation once every 
$\sim400$ binary orbits. Excitation of a similar lopsided cavity is observed in the 3D MHD, as well as 2D hydrodynamical, simulations of \citealt{ShiKrolik:2012}, \citealt{Noble+2012}, and MM08. \cite{ShiKrolik:2012} explore the possible generation mechanisms of this mode. They find its growth to be consistent with being caused by asymmetric stream impacts described above. 

\begin{figure*}
\begin{center}$
\begin{array}{cc}
\includegraphics[scale=0.44]{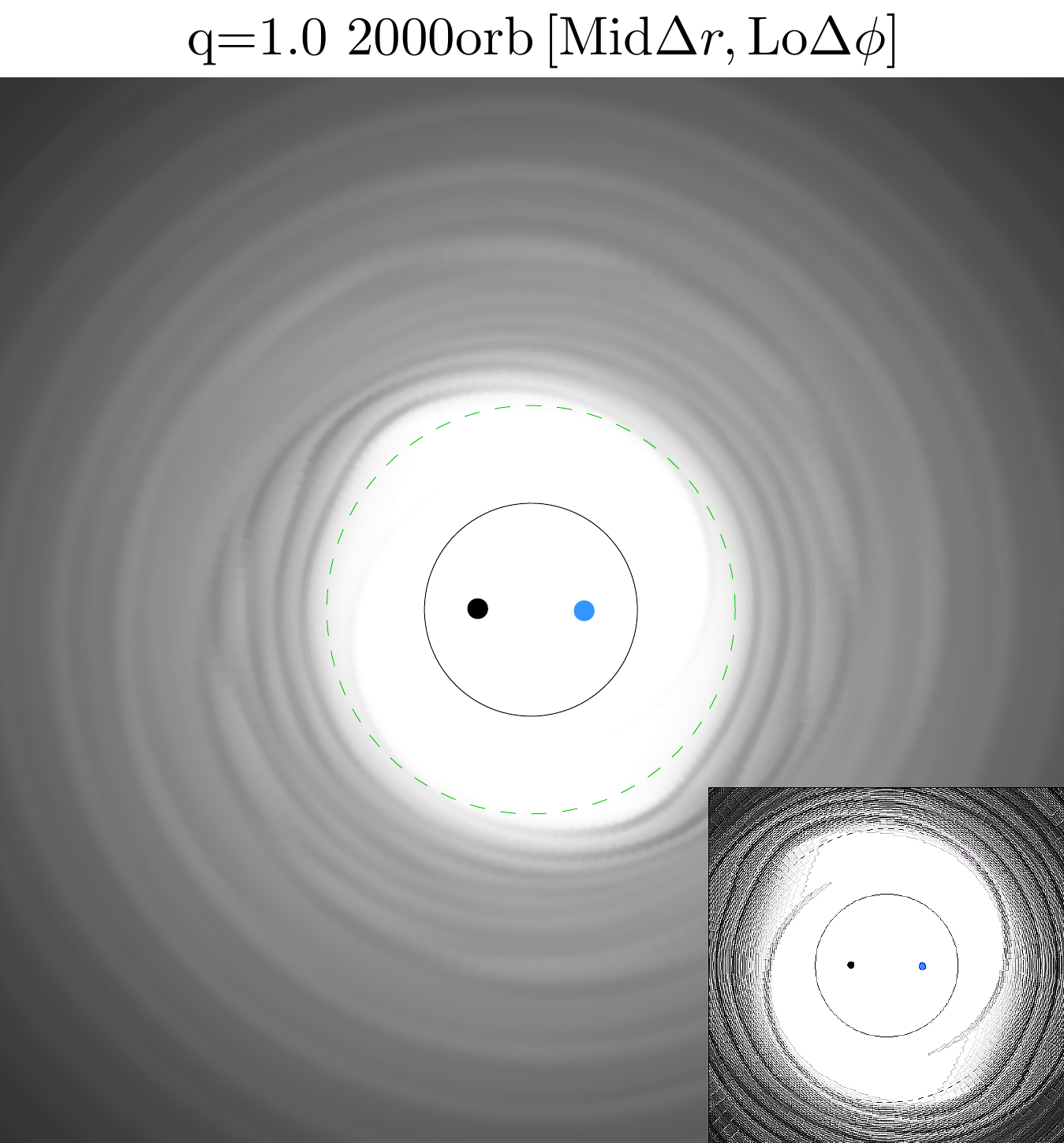}&
\includegraphics[scale=0.44]{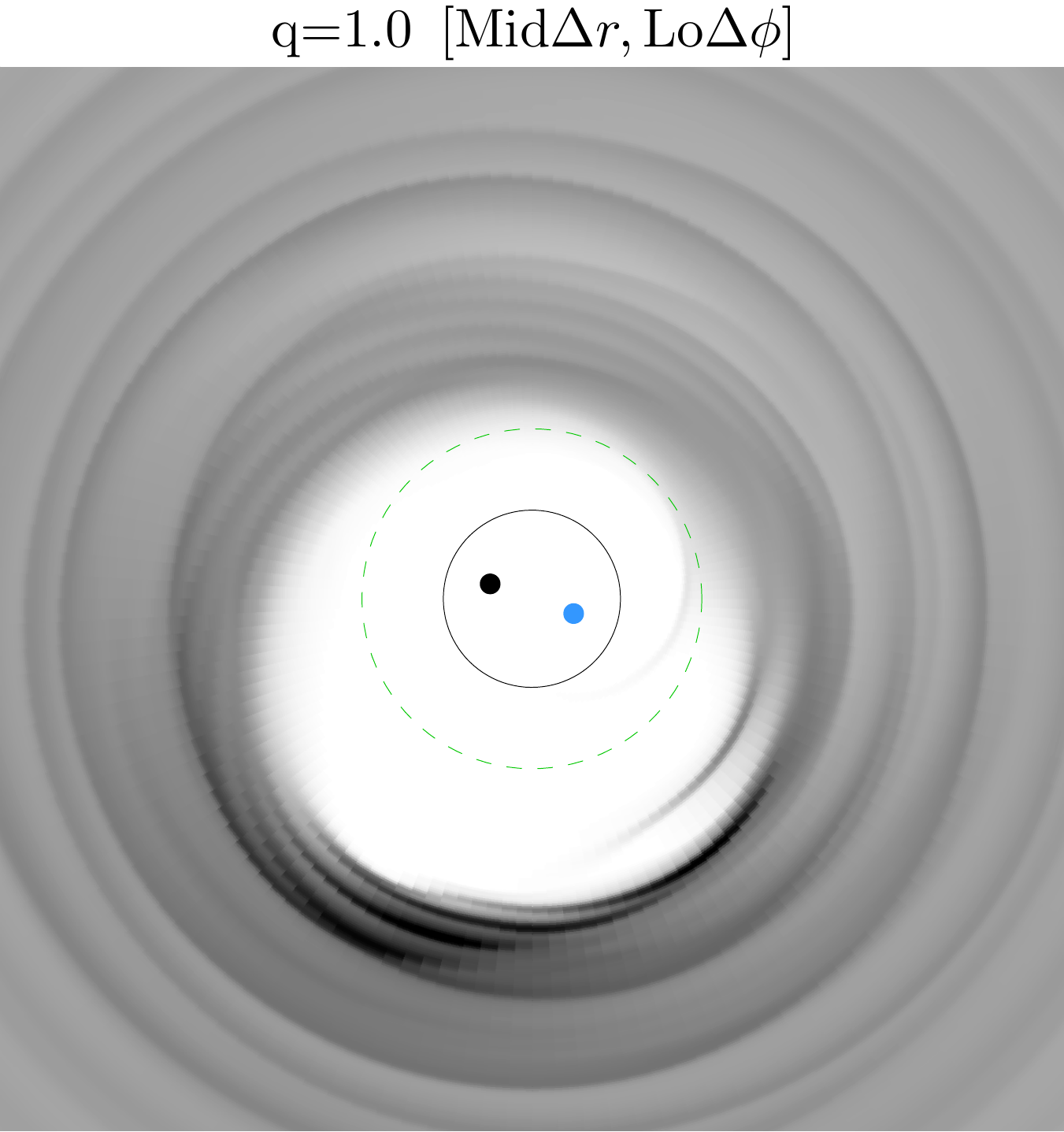}  \\ \\
\hline  \hline\\ 
\includegraphics[scale=0.44]{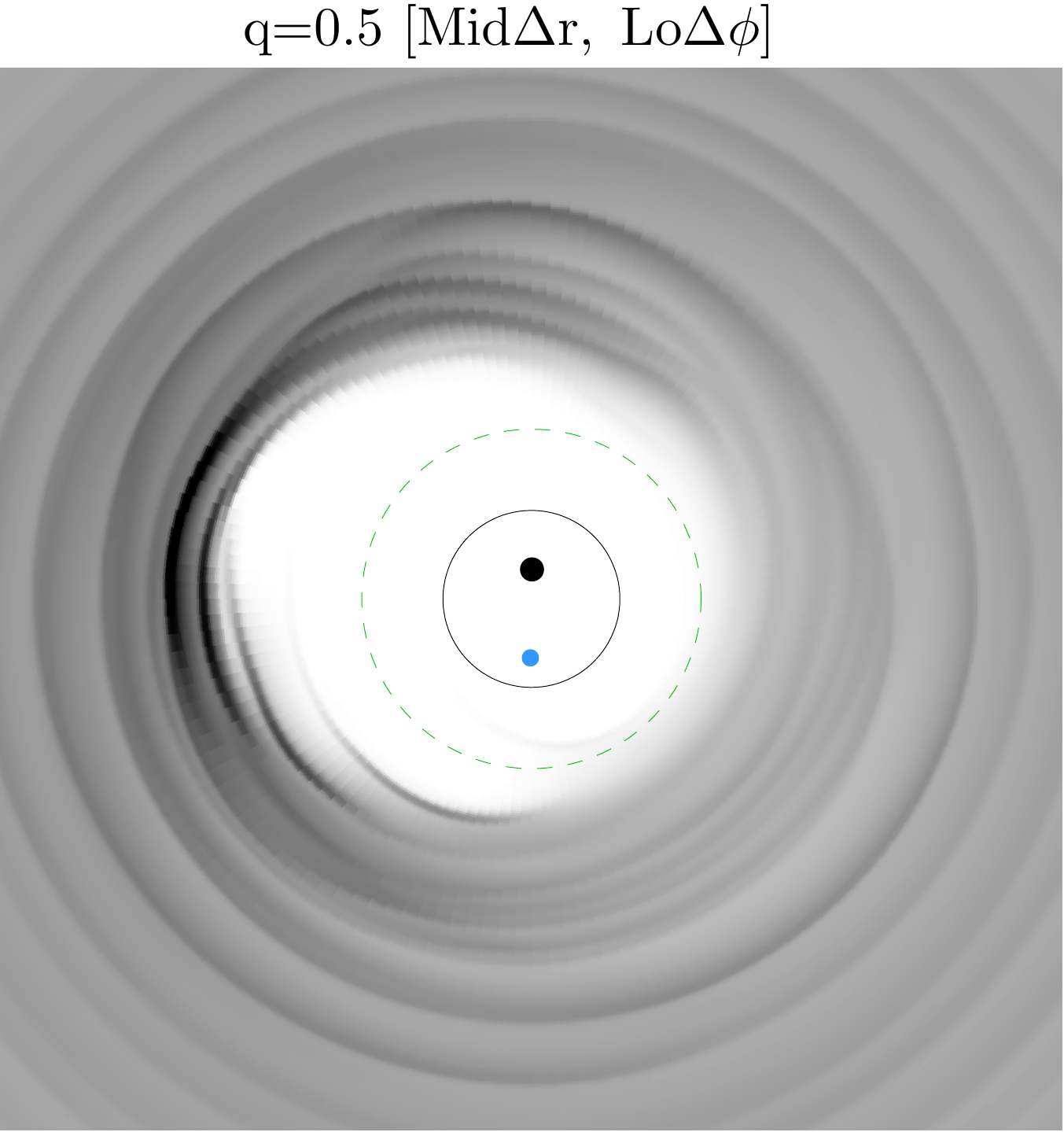} &
\includegraphics[scale=0.44]{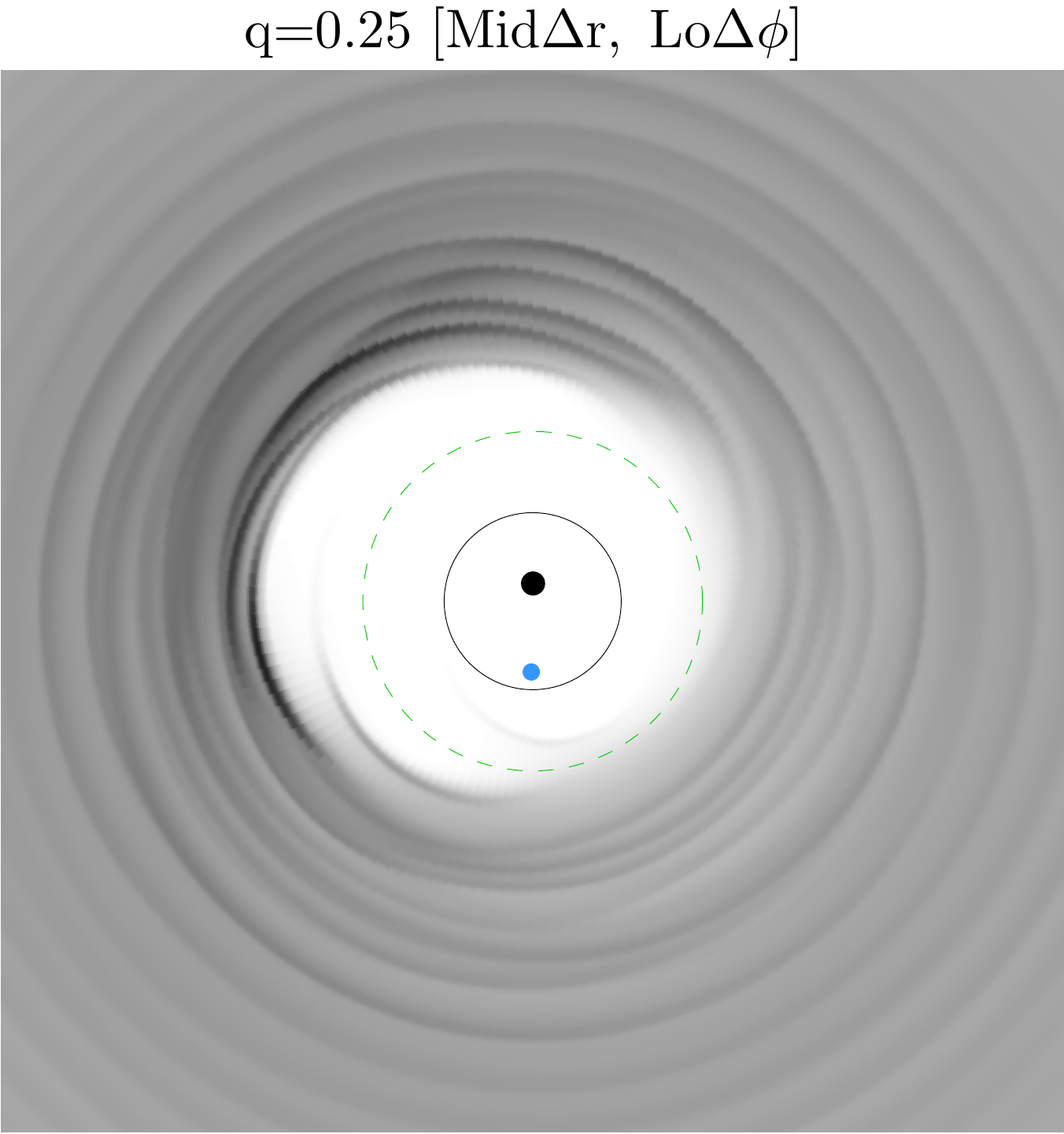}\\
\includegraphics[scale=0.44]{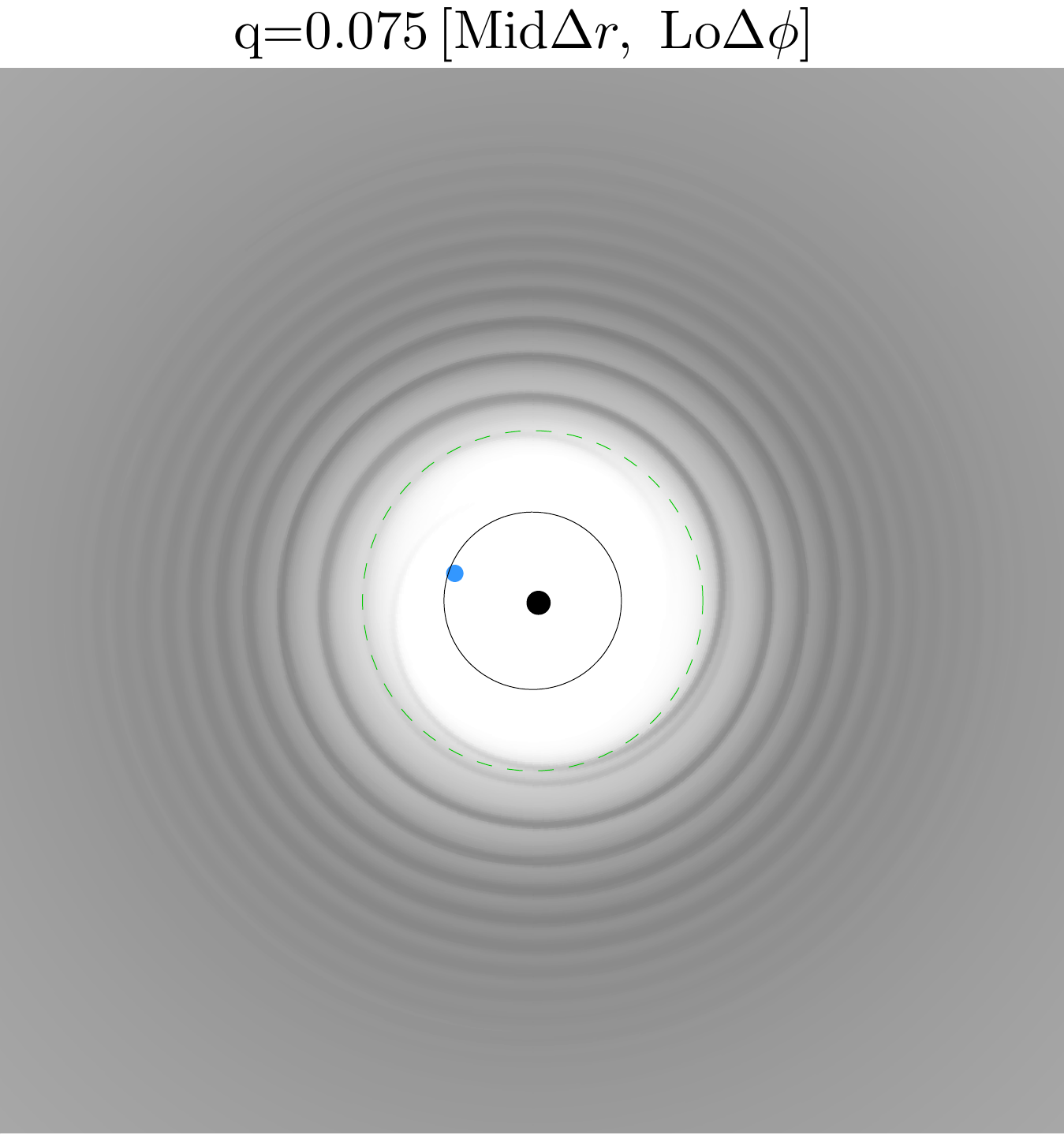} &
\includegraphics[scale=0.44]{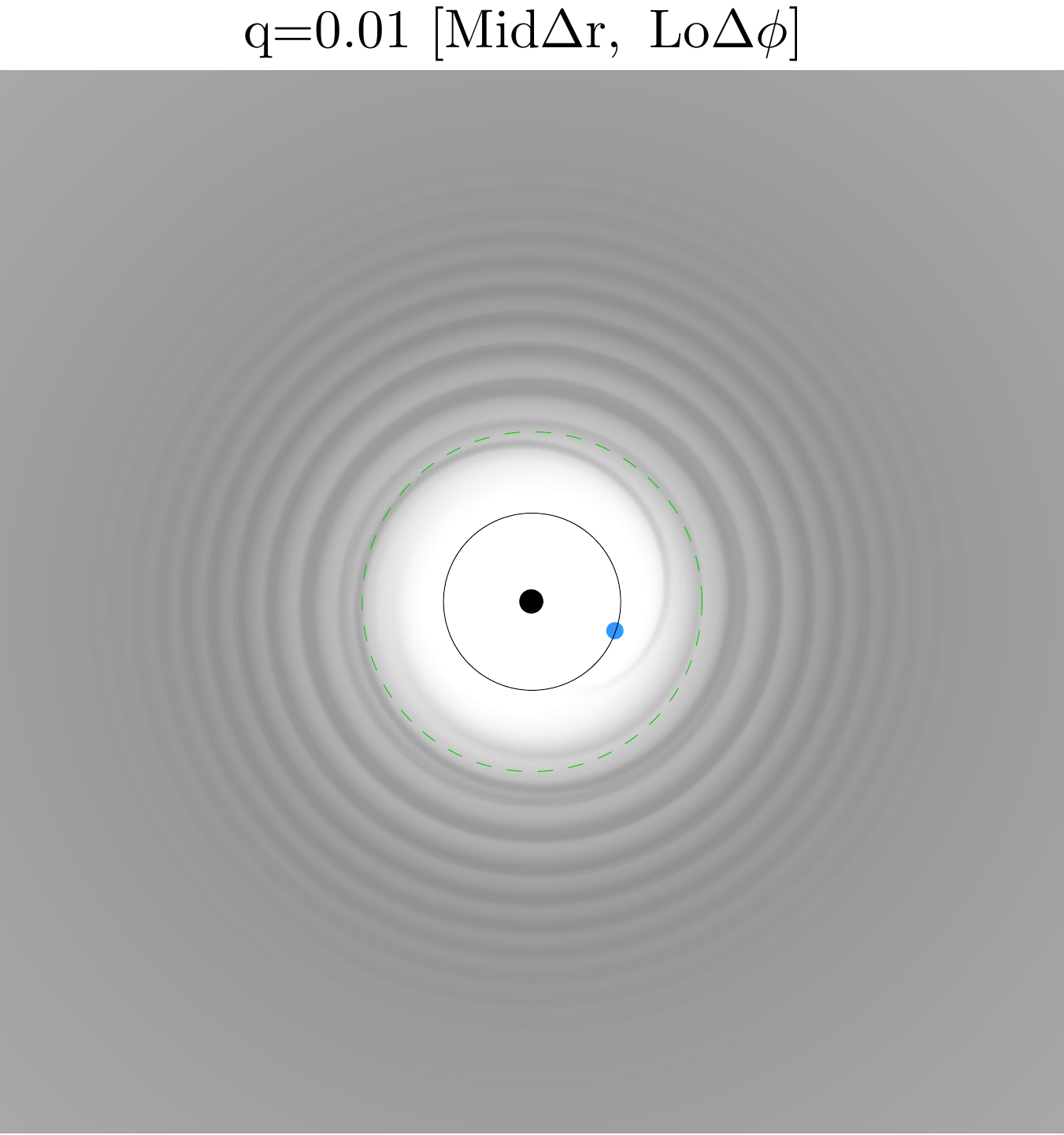} 
\end{array}$
\end{center}
\caption{\textit{Top row:} Surface density distributions for the
  equal-mass ratio ($q=1.0$) binary during a transient,
  point-symmetric state after $\sim1000$ binary orbits (left) and
  during the quasi-steady asymmetric state after $\sim$ 4000 binary
  orbits (right). The inset in the top left panel zooms in to the
  inner $\pm 2.5 r/a$ of the disc in order
  to show the stream morphology.  \textit{Bottom two rows:} snapshots
  at $\sim$ 4000 binary orbits, during the quasi-steady-state phase,
  for mass ratios $q=0.5, 0.1, 0.075$, and $0.01$, as labeled.  Each
  panel shows the inner $\sim6\%$ of the simulated disc, extending
  $\pm 6 r/a$ in both directions.  The solid circles mark the inner
  boundary of the simulation at $r=r_{\rm{min}}=a$. The larger dotted
  circle at $r\simeq2.08a$ is the position of the $(m,l)=(2,1)$ outer
  Lindblad resonance (shown only for reference).  Surface densities
  are plotted with the same linear grayscale in each panel, with the
  darkest regions corresponding to a maximum density of $0.8 \Sigma_0$
  ($0.4 \Sigma_0$ for the top left panel). Orbital motion is in the clock-wise direction.}
\label{2DDensProf}
\end{figure*}

We show examples of the two-dimensional surface density distributions
in Figure \ref{2DDensProf}.  The top row of this figure shows
snapshots at $\sim1000$ binary orbits (left) and at $\sim$4000 binary
orbits (right), of the inner 6$\%$ of the simulated disc (i.e. $\pm 6 r/a$ are
shown in both directions).  The solid circle marks the inner boundary
of the simulation domain at $r=r_{\rm{min}}=a$.  The larger dotted
circle at $r\simeq2.08a$ is the position of the $(m,l)=(2,1)$ outer
Lindblad resonance (not present, but shown for reference).  The left of these two
panels illustrates the point-symmetric, transient state. The
zoomed-in inset of this panel shows two weak point-symmetric streams reminiscent of the streams seen in the toy
model of Figure \ref{toydisc}. The right panel shows the disc after it
has settled to its quasi-steady, lopsided state, with a single stream.

\begin{figure*}
\begin{center}$
\begin{array}{cc}
\includegraphics[scale=0.39]{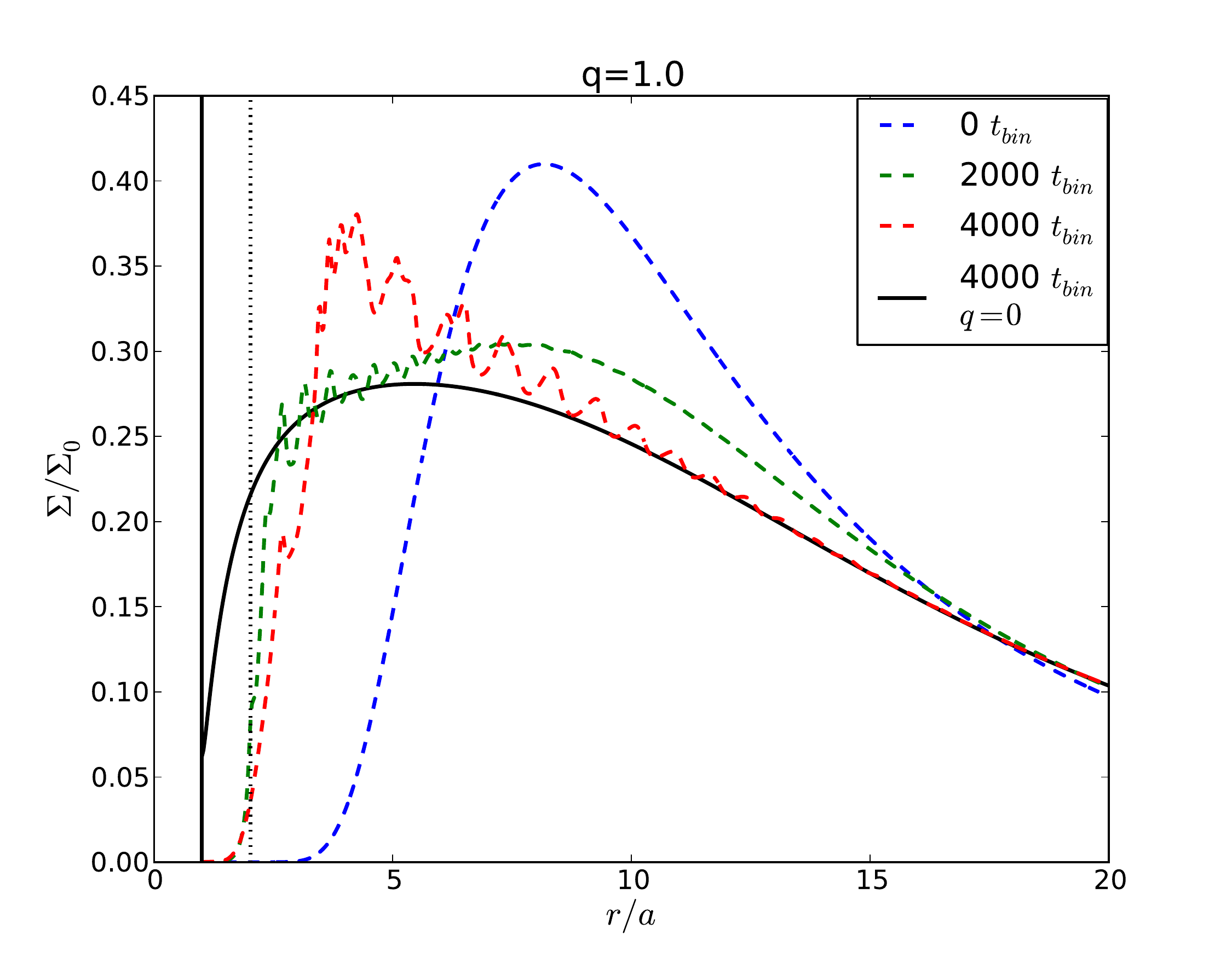} &
\includegraphics[scale=0.39]{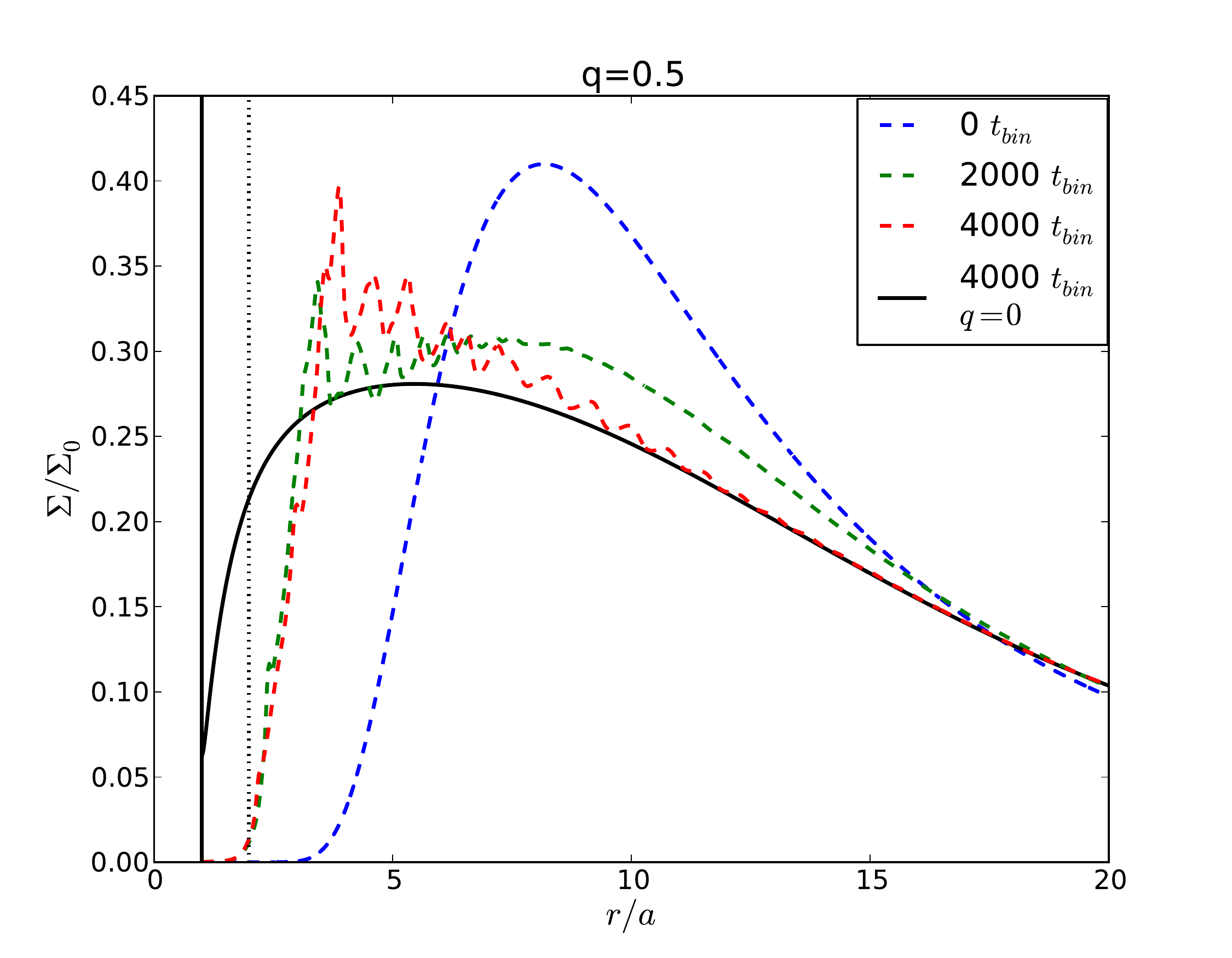} \\
\includegraphics[scale=0.39]{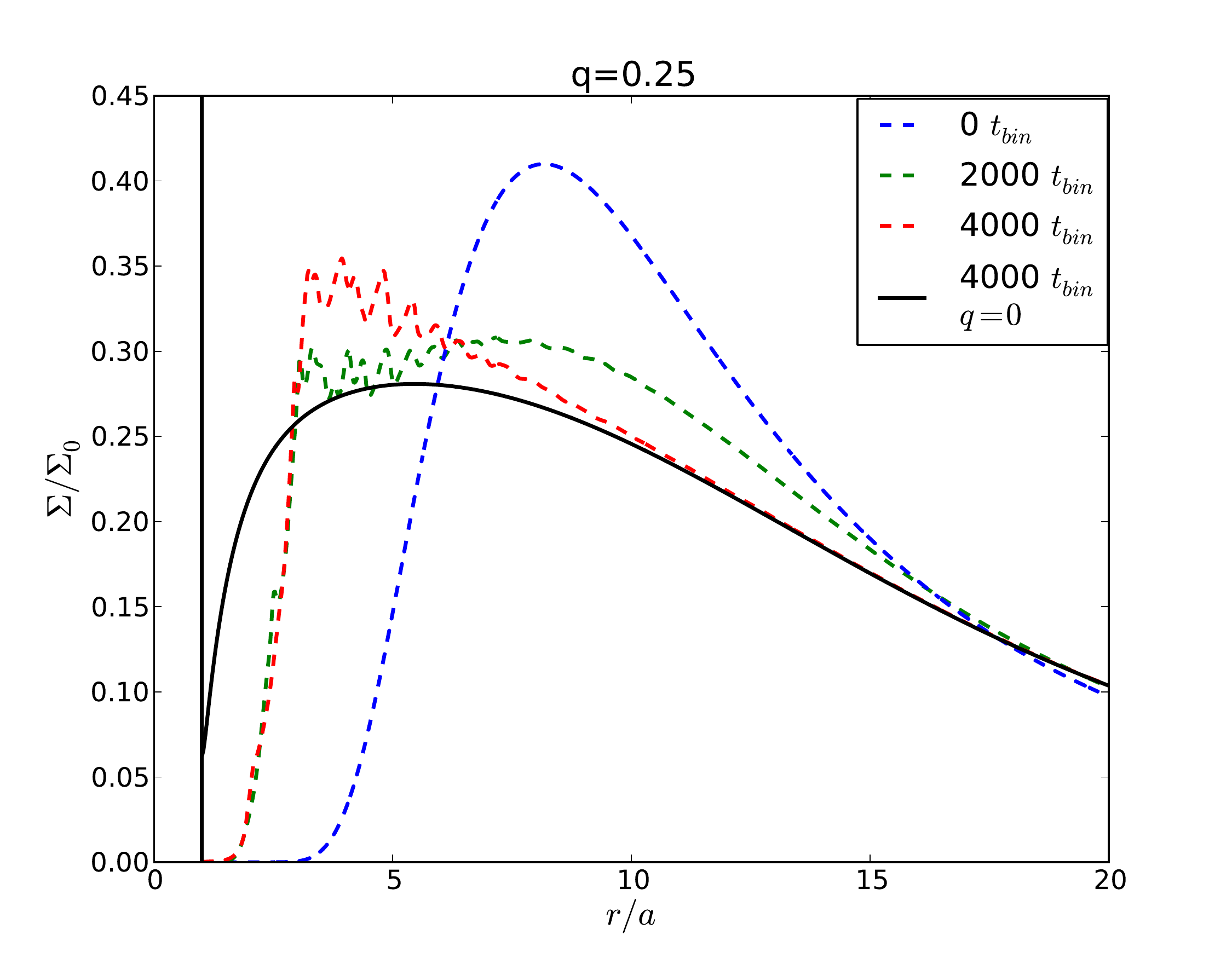} &
\includegraphics[scale=0.39]{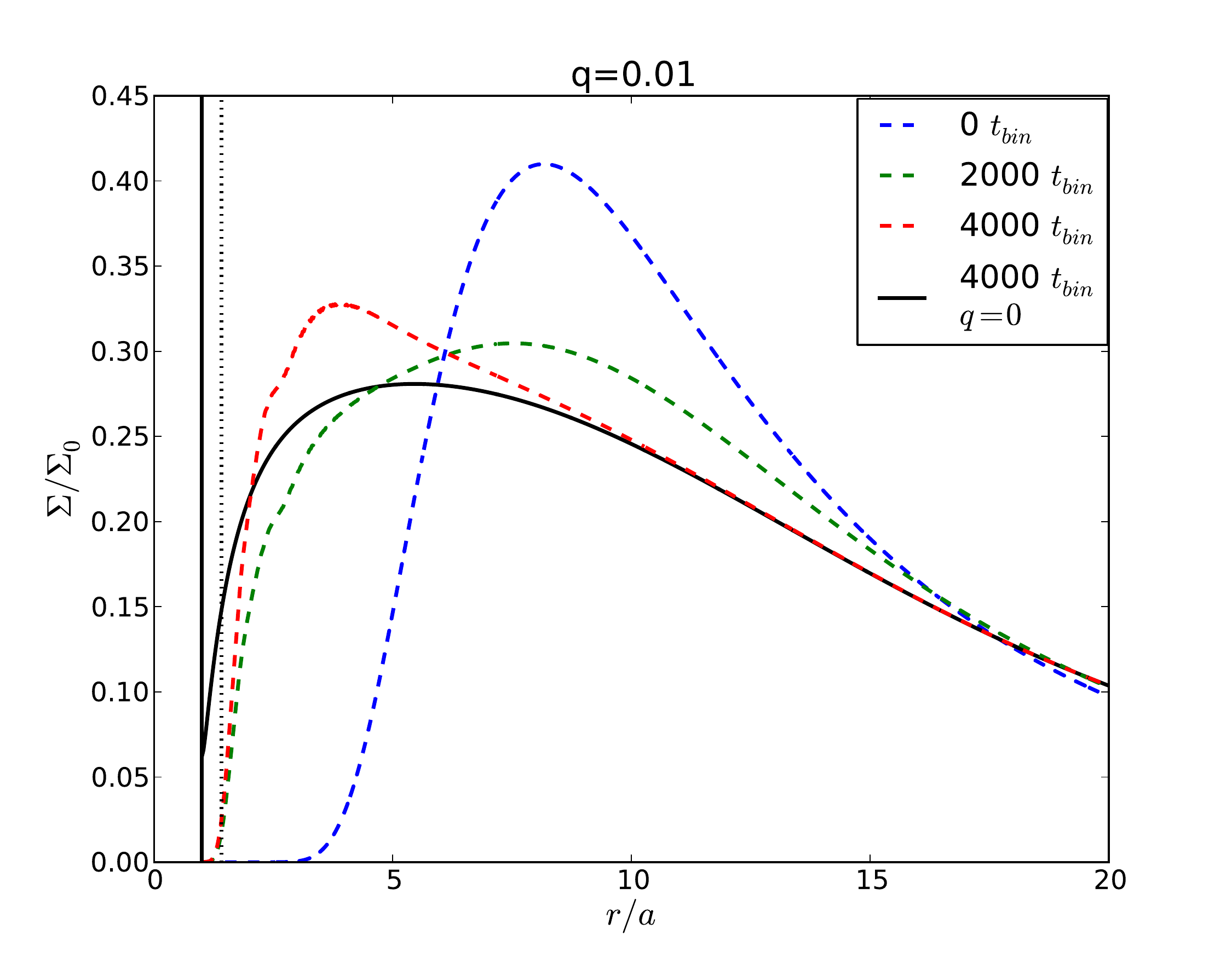}   
\end{array}$
\end{center}
\caption{Snapshots of the azimuthally averaged disc surface density at
  different times (shown by different curves in each panel, from 0
  to 4000 orbits, as labeled), and for different mass ratios (shown in
  different panels, from $q=1.0$ to $0.01$, as labeled).  In each
  panel, the solid [black] curve shows, for reference, the density
  profile in the point-mass ($q=0$) case after 4000 orbits. The
  vertical dotted lines mark the radius where binary and viscous
  torques balance (from Figure \ref{TrqDq}); these lie close to the
  observed cavity edges.  The vertical solid lines mark the inner edge
  of the integration domain ($r=r_{\rm{min}}=a$).  In each case, the
  inner circumbinary disc spreads inward with time, but the density
  profile remains sharply truncated, with a low-density central cavity
  inside $r \lsim 2a$.  }
\label{AzAvDens}
\end{figure*}

The top left panel of Figure~\ref{AzAvDens} shows snapshots of the
azimuthally averaged surface density profile of the equal-mass binary
disc at three different times, after 0, 2000, and 4000 orbits. For
comparison, the density profile is also shown for the single
point--mass ($q=0$) case, after 4000 orbits.  As the figure shows, the
inner circumbinary disc spreads inward with time. By 4000 orbits, the
disc structure at $r/a\gsim 5$, where the effect of the binary is
relatively small, closely follows the $q=0$ case.  However, the
density profile remains sharply truncated inside $r \lsim 2a$
(i.e. below the point--mass case, even at 4000 orbits).  As a result of the initial density profile, the peak
density first decreases as matter drains inward towards the holes and also
outward towards the outer boundary. With time, despite the leakage of 
streams to the binary, inward viscous diffusion causes a gas 
pileup behind the cavity wall.   The figure also shows that the
position of the cavity wall moves slightly outward between the t=2000
and t=4000 snap-shots. This is because as the disc settles to its
lopsided quasi-steady-state, the cavity size grows in the azimuthally
averaged sense.

\subsubsection{Torque Balance and the Size of the Central Cavity}
\label{Torque Balance and the Size of the Central Cavity}

As (the top left panel of) Figure \ref{AzAvDens} shows, the central
cavity around the equal-mass binary extends to $r \sim 2a$.  Here
we take the cavity edge $r_{\rm{ce}}$ to be the radius where the negative 
viscous torque density matches the binary torque density (in an azimuthally averaged sense) ,
\begin{equation}
\left[ \left(\frac{dT}{dr}\right)_{\rm{bin}}  + \left(\frac{dT}{dr}\right)_{\rm{visc}} \right]_{r_{\rm ce}} = 0.
\label{TrqBalRCE}
\end{equation}
To gain insight into the transport of angular momentum and the
clearing of the central cavity, we therefore compute the time-- and
azimuthally--averaged torque densities from the binary potential,
\begin{equation}
\left(\frac{dT}{dr}\right)_{\rm{bin}} = \left< \frac{1}{2 \pi}  \int^{2 \pi}_{0}{ \Sigma(r, \phi) \frac{d\Phi}{d\phi}(r,\phi) r d\phi }\right>_{t},
\label{DTb_num}
\end{equation}
and from viscous stresses,
\begin{equation}
\left(\frac{dT}{dr}\right)_{\rm{visc}} = 2 \pi \left<  \frac{d}{dr}\left[ r^3 \nu \left< \Sigma \frac{\partial \Omega}{\partial r} \right>_{\phi}\right] \right>_{t}.
\label{DTv_num}
\end{equation}
The outer derivative in equation (\ref{DTv_num}) is taken numerically,
and all of the above values are measured directly from the simulation
outputs, except for the binary potential derivative in equation
(\ref{DTb_num}) which is given analytically. The time averages are 
taken over $25$ binary orbits at a sample rate of 20 per orbit.

\begin{figure*}
\begin{center}$
\begin{array}{cc}
\includegraphics[scale=0.445]{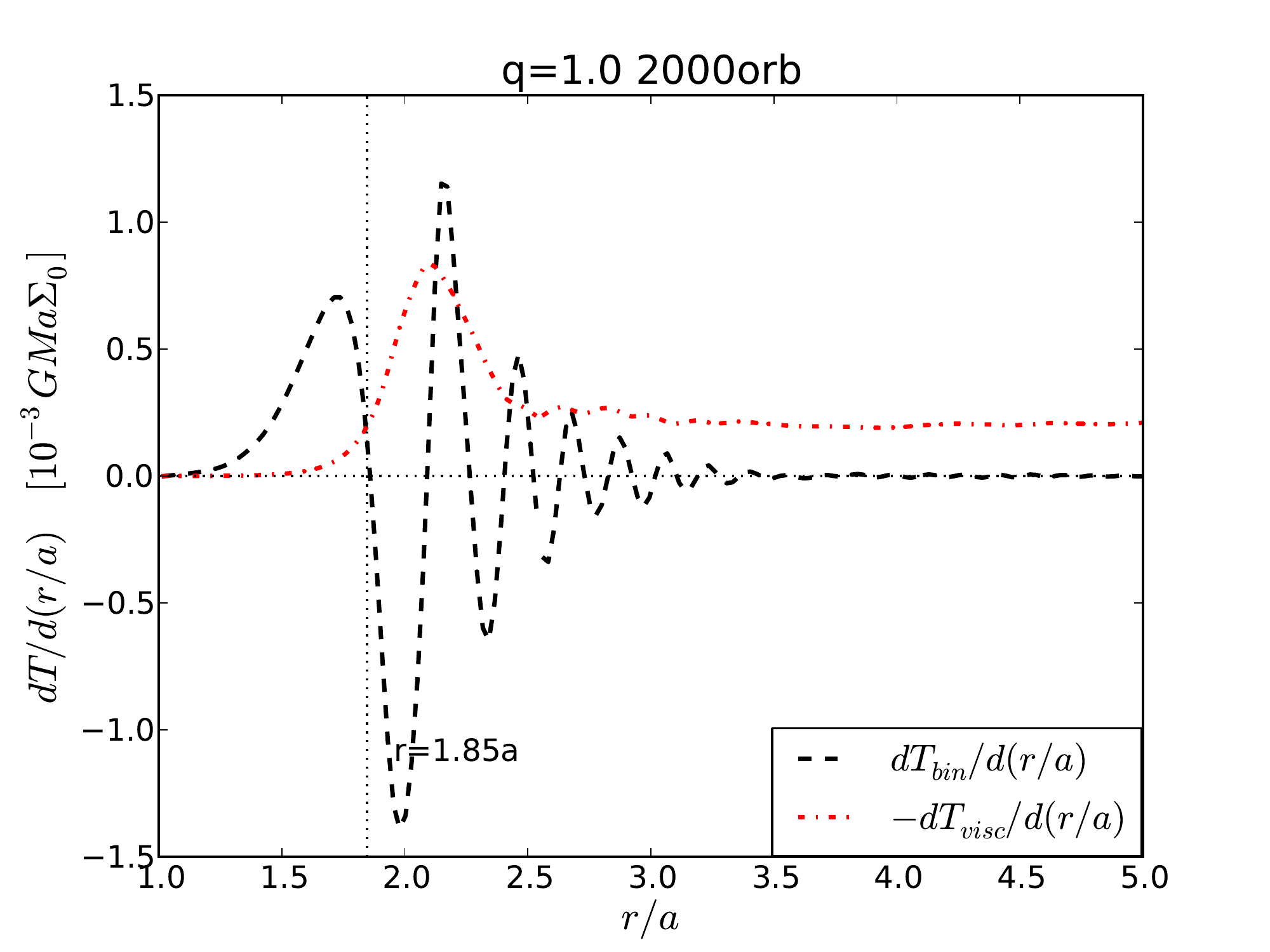} &
\includegraphics[scale=0.445]{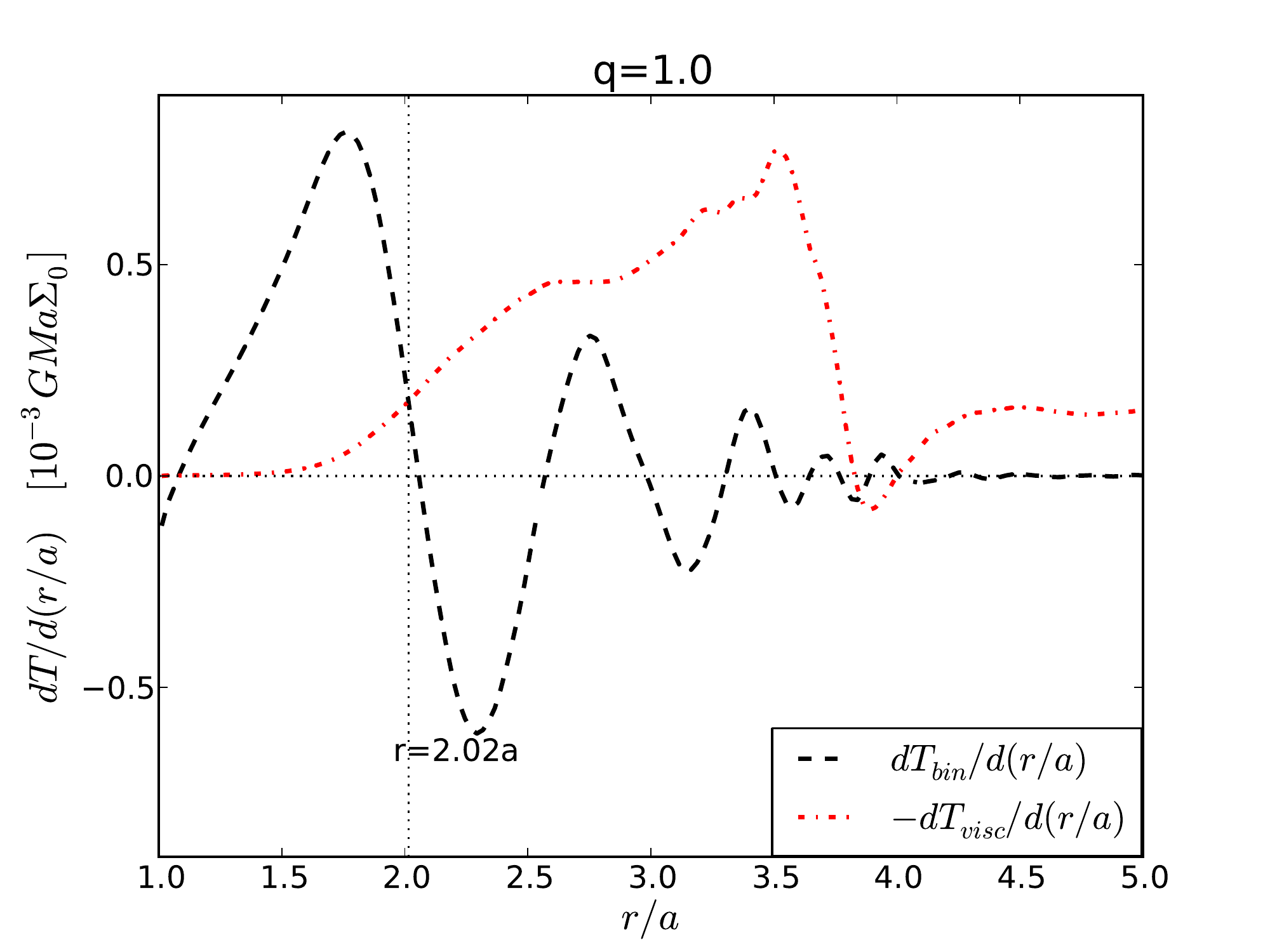} \\
\includegraphics[scale=0.445]{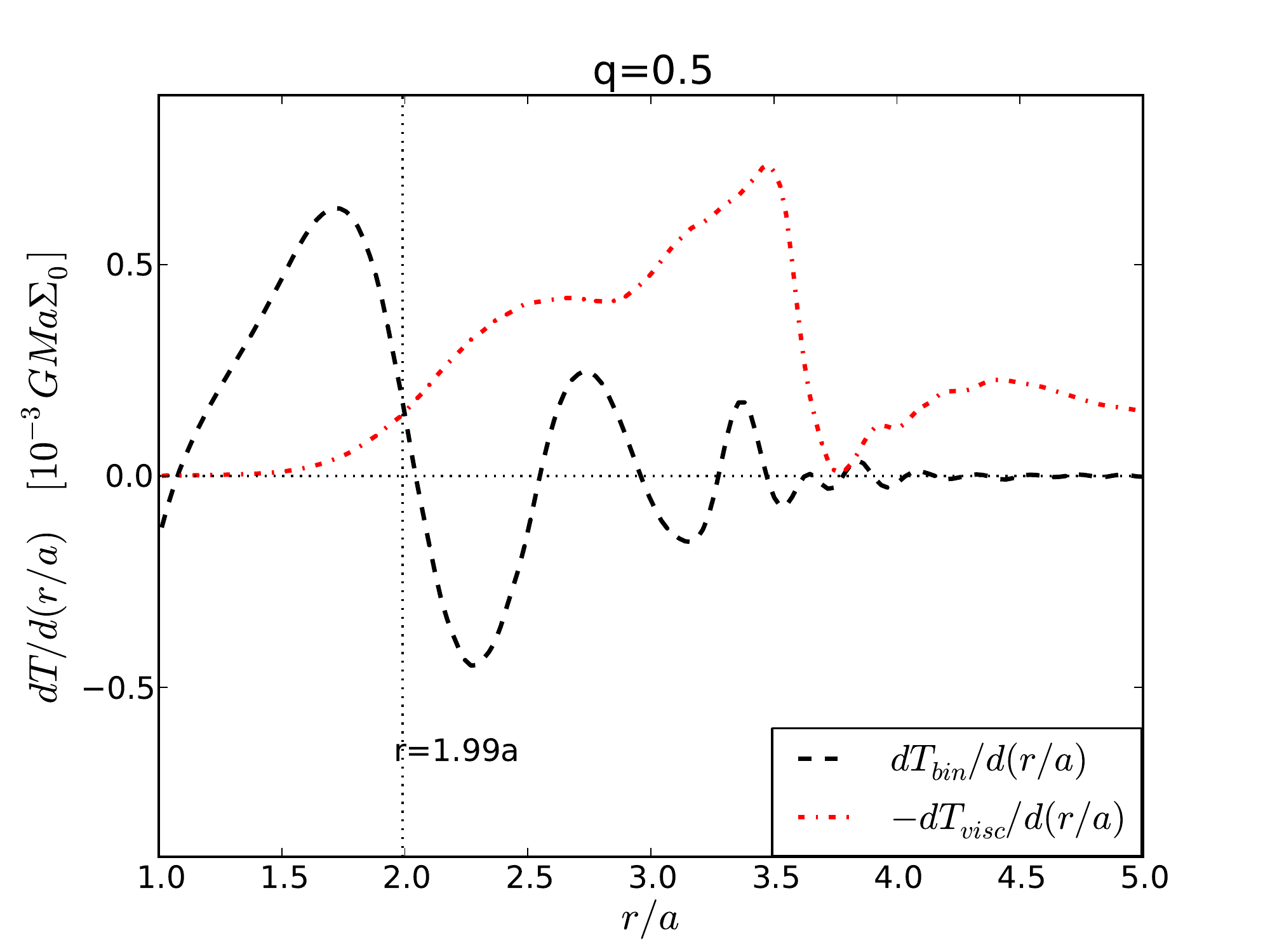} &
\includegraphics[scale=0.445]{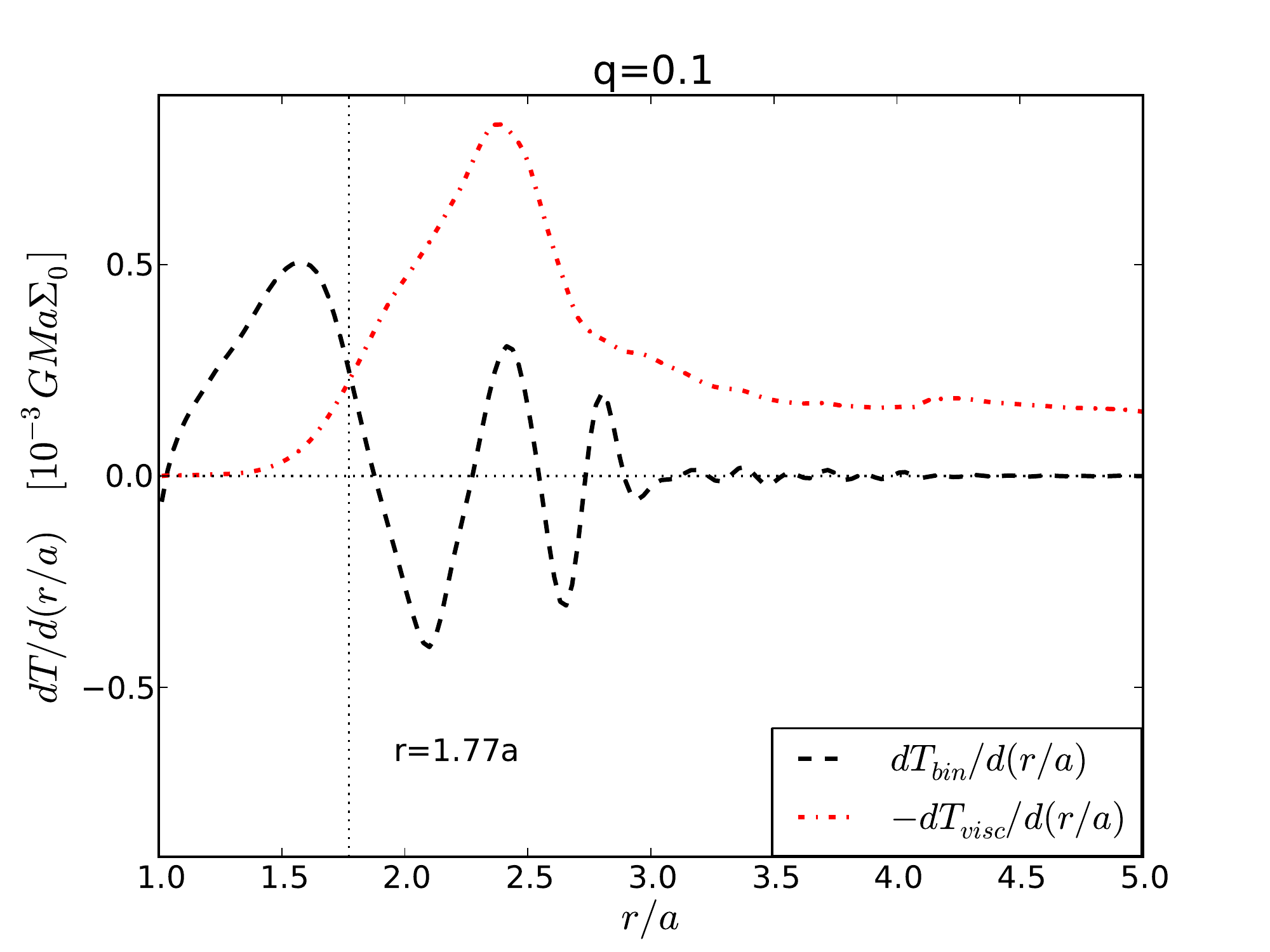} \\
\includegraphics[scale=0.445]{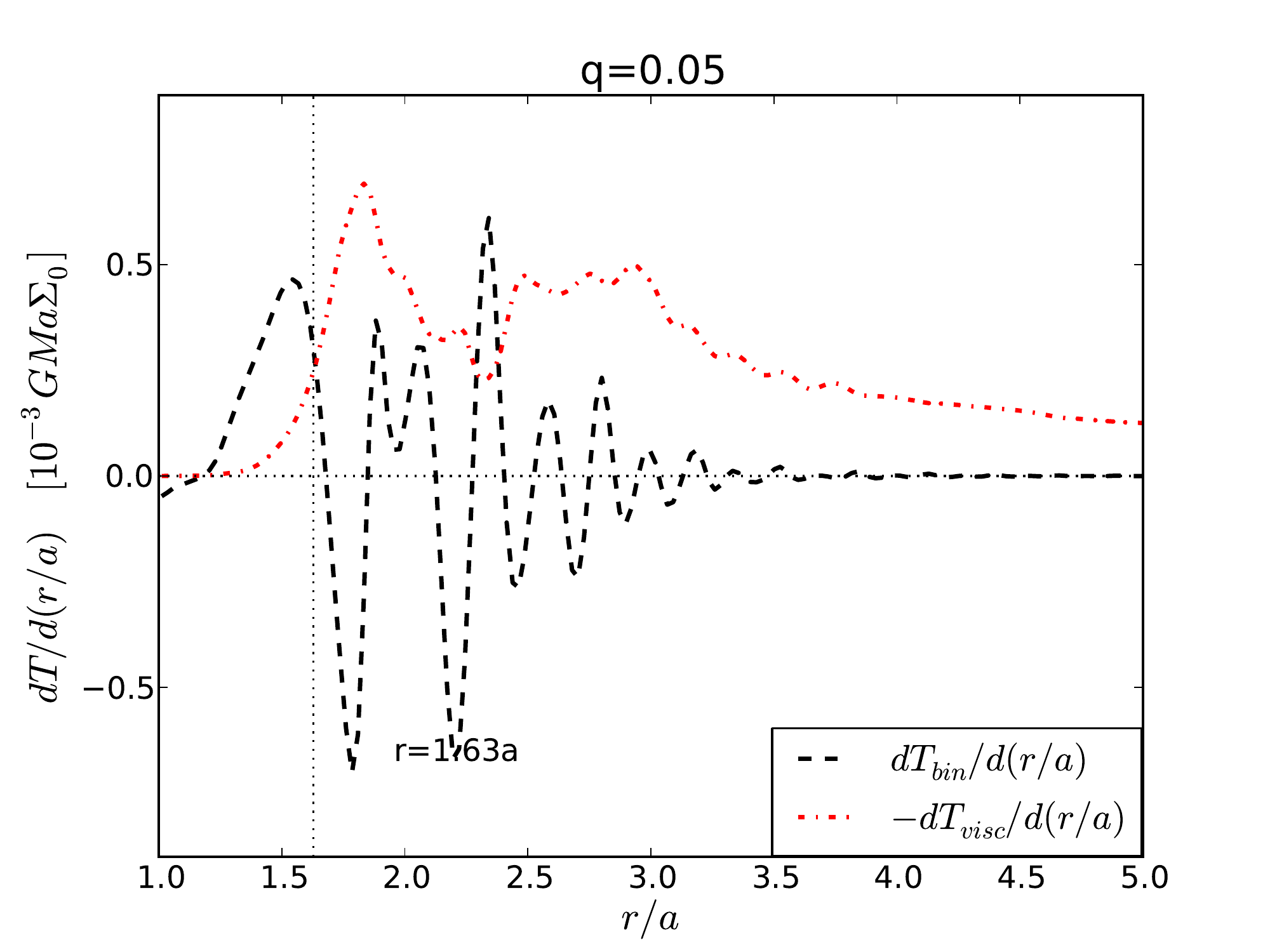} &
\includegraphics[scale=0.445]{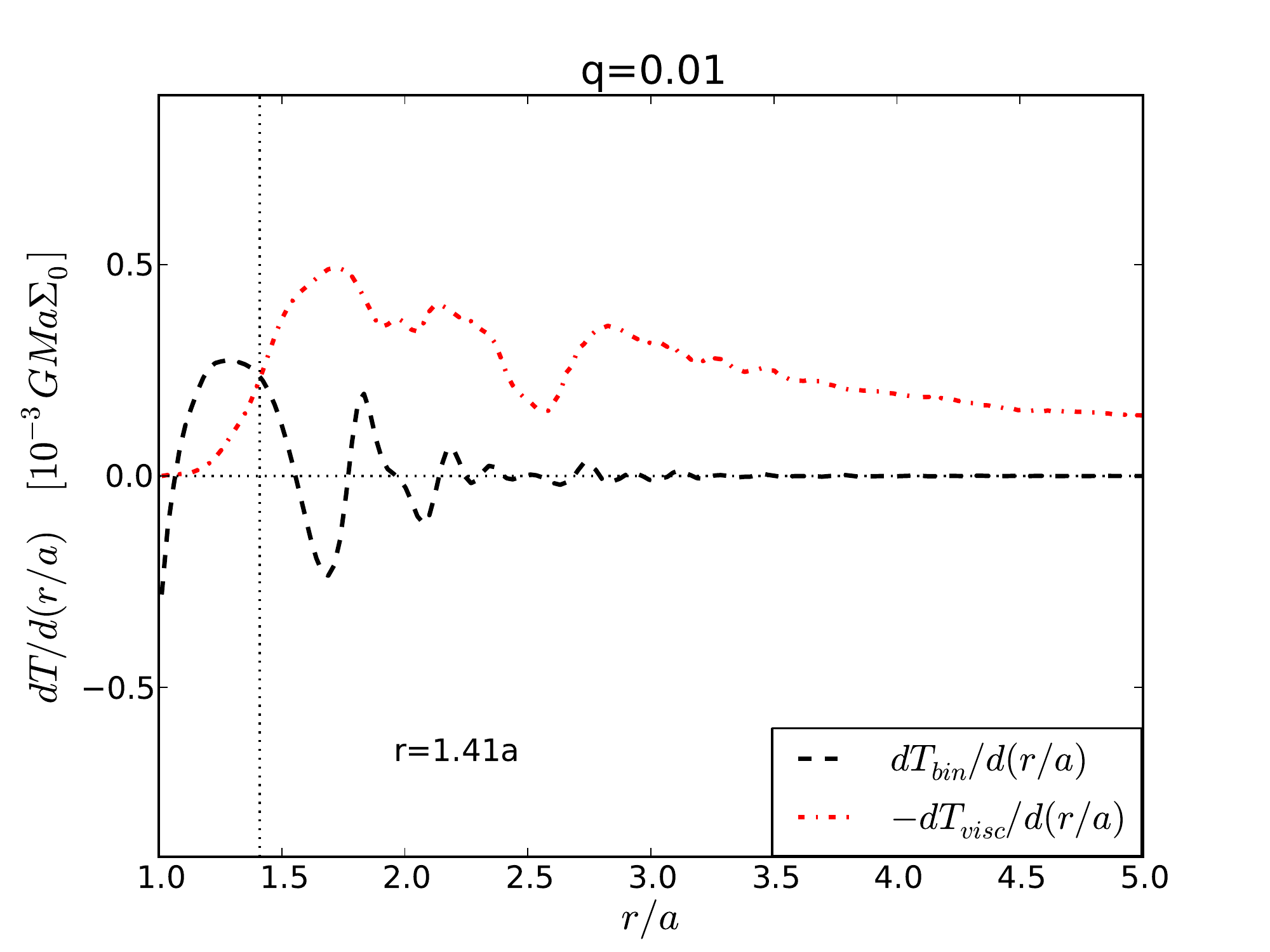} 
\end{array}$
\caption{Azimuthally- and time-averaged torque density profiles in the
  inner disc for the equal--mass binary (top two panels) and for
  unequal-mass binaries (other panels, with different mass ratios $q$
  as labeled). The top left panel corresponds to the point-symmetric
  transient stage (after $\sim$ 2000 orbits) and the top right panel to
  the asymmetric quasi-steady state (after $\sim$ 4000 orbits).  Only
  the quasi-steady state is shown for the $q<1$ cases. In each panel,
  the dashed [black] curves show the gravitational torques from the
  binary, and the red [dot-dashed] curves show the \textit{negative}
  viscous torques.  The vertical dotted line marks the radius where
  the viscous and gravitational torques balance (equation
  \ref{TrqBalRCE}); these are close to where the azimuthally-averaged
  surface density profiles are found to be truncated (see
  Fig. \ref{AzAvDens}). See Figure \ref{RCE} for a plot of this cavity edge radius vs. $q$. 
  Time averages are taken over 25 orbits at a sample rate of 20 per orbit. }
\label{TrqDq}
\end{center}
\end{figure*}

The top row of Figure \ref{TrqDq} shows the binary and viscous torque
densities for $q=1$ over the inner $5 r/a$ of the disc, during both
the initial transient state (left panel) and the subsequent
quasi-steady-state (right panel).  There is indeed a well-defined
central region, where the binary torques exceed the viscous torques
and can be expected to clear a cavity. The transition (computed via
eq. \ref{TrqBalRCE}) is located at $r_{\rm ce}\simeq1.85$ and
$r=2.02a$ in the transient and quasi-steady-state, respectively, and
is marked in both panels by a vertical dotted line.  These vertical
lines are also shown in Figure \ref{AzAvDens} and indeed lie very
close to radii where the disc surface densities remain truncated.

The small outward drift of the average position of the cavity wall
from the transient to the quasi-steady-state is also visible in Figure
\ref{AzAvDens}. As the disc transitions to the quasi-steady-state, the
binary torque-density wavelength increases in $r/a$, while keeping
approximately the same amplitude.  These effects can be attributed to
the increasingly elongated and lopsided shape of the inner cavity;
when azimuthally averaged, this results in a larger cavity size.

\subsubsection{Accretion Rates}

\begin{figure*}
\begin{center}$
\begin{array}{cc}
\includegraphics[scale=0.42]{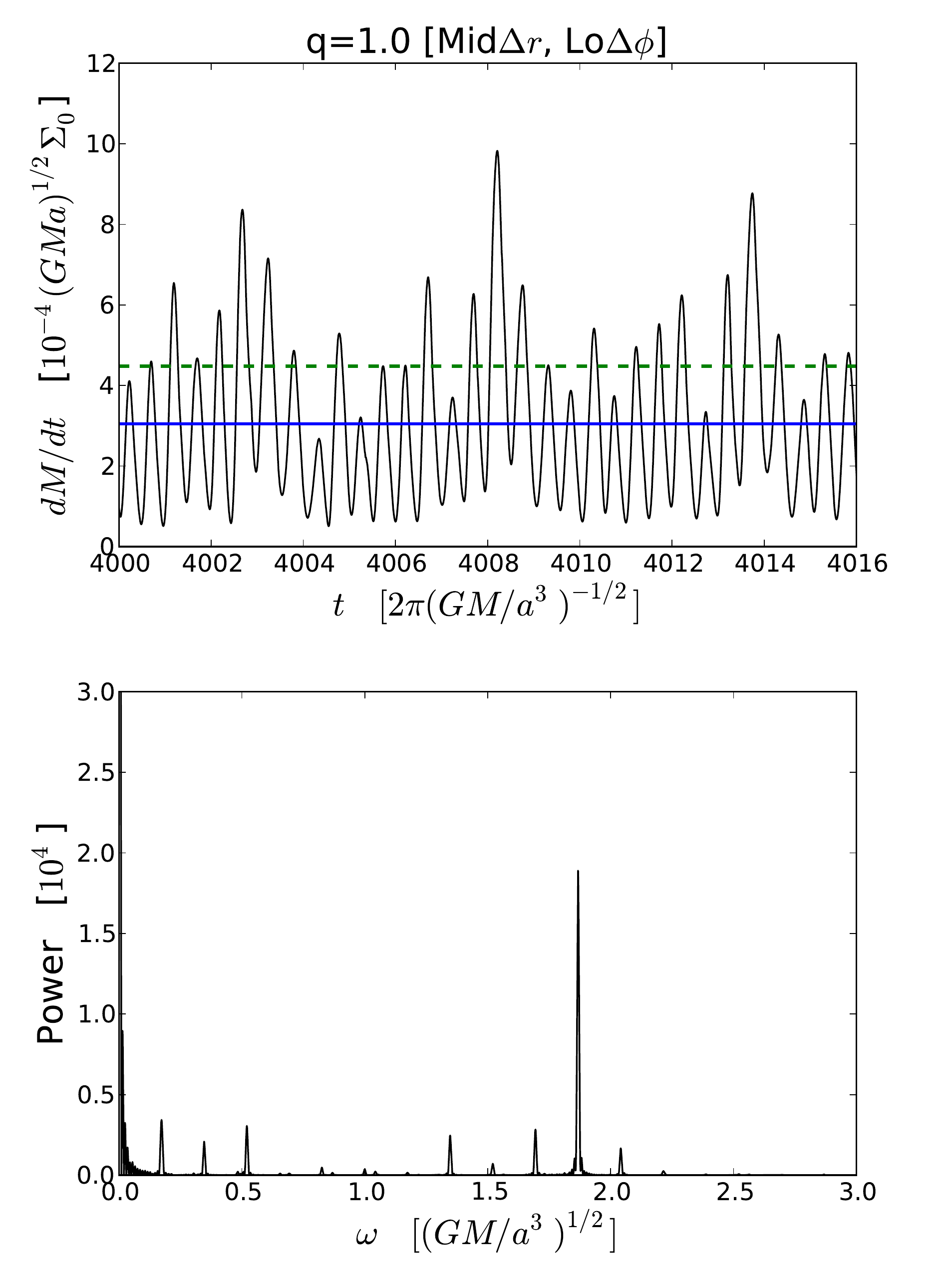} &
\includegraphics[scale=0.42]{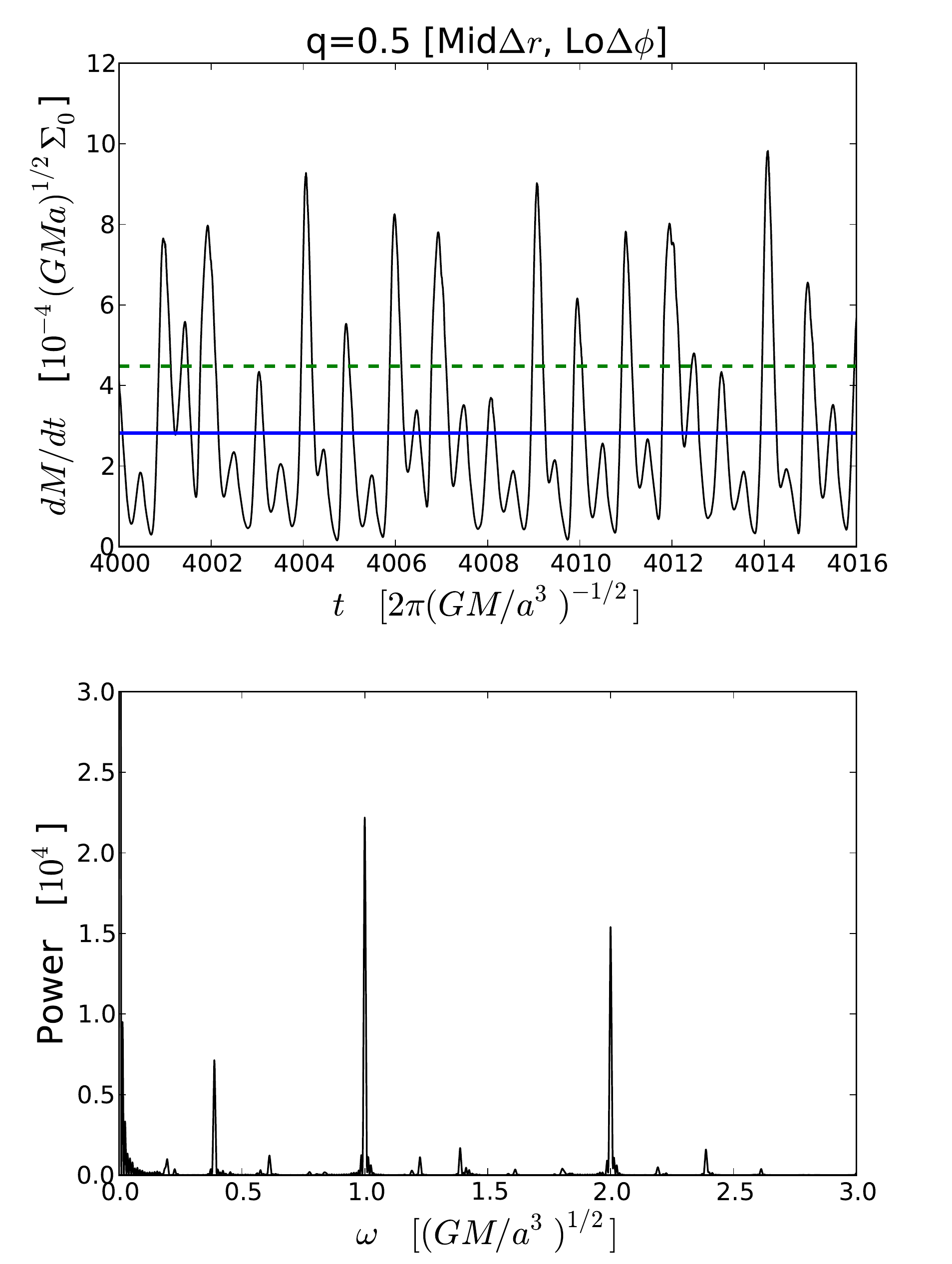} \\
\includegraphics[scale=0.42]{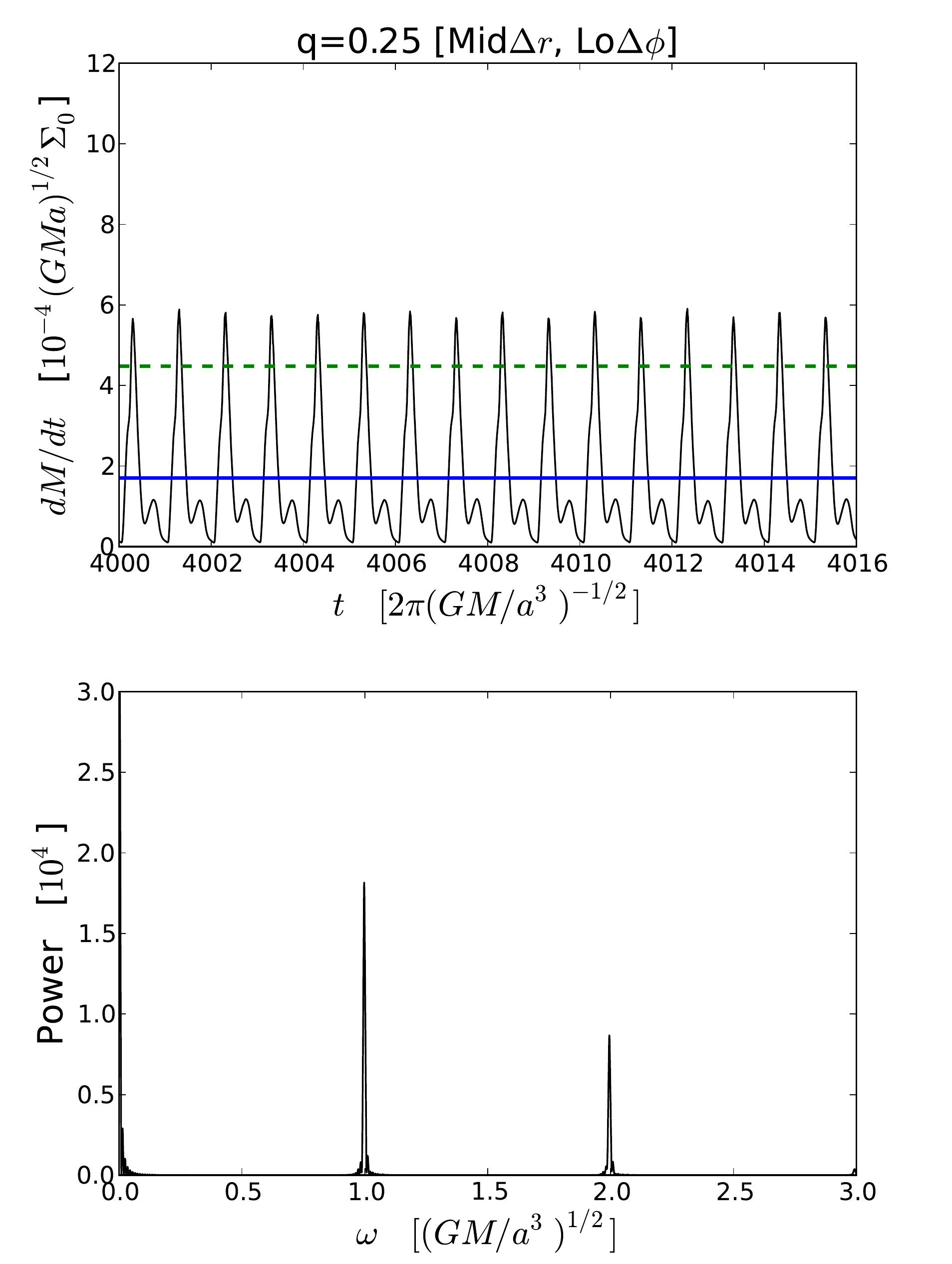}  &
\includegraphics[scale=0.42]{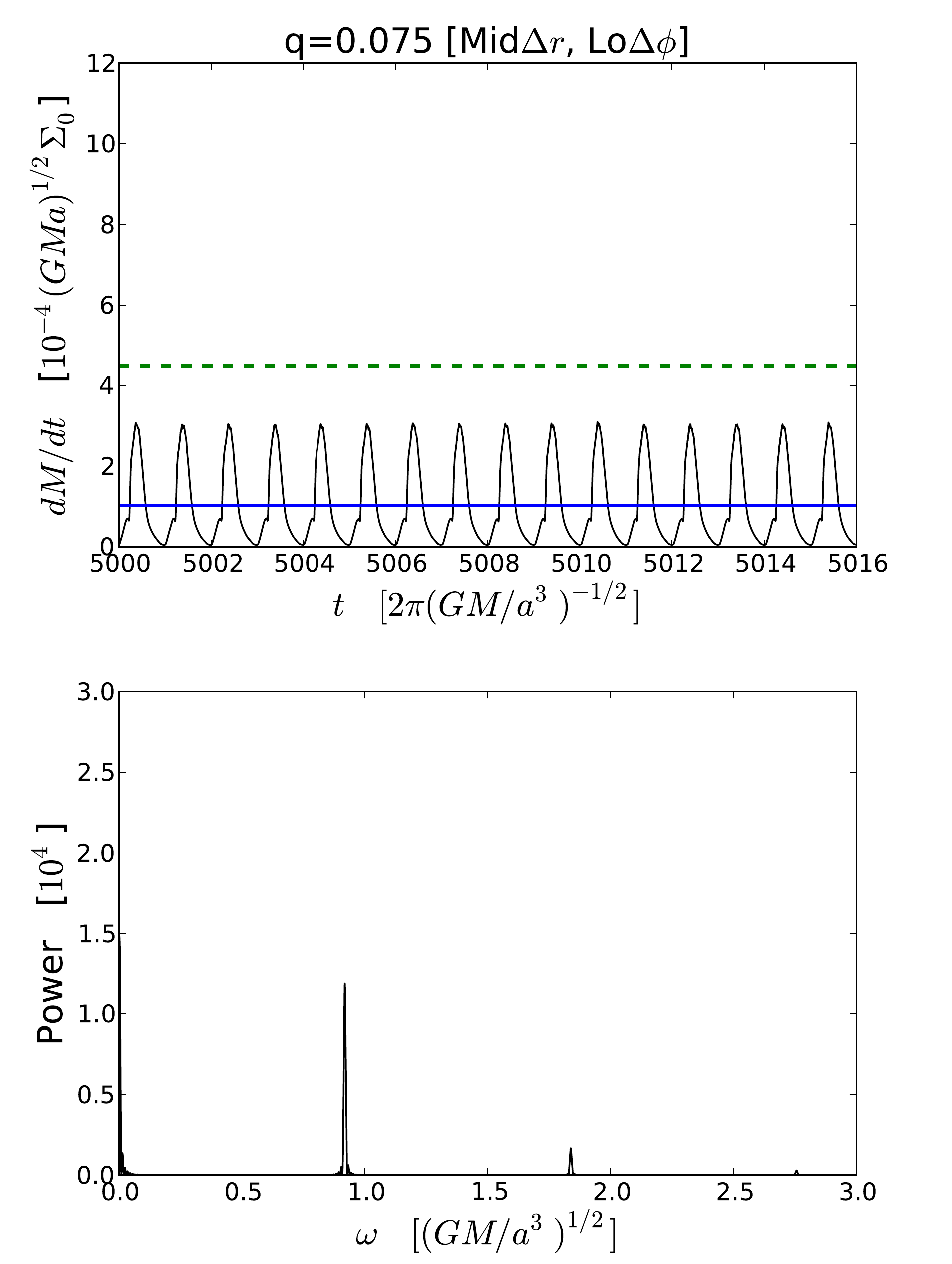} 
\end{array}$
\end{center}
\caption{ -- continued on next page}
\end{figure*}

\addtocounter{figure}{-1}

\begin{figure*}
\begin{center}$
\begin{array}{cc}
\includegraphics[scale=0.42]{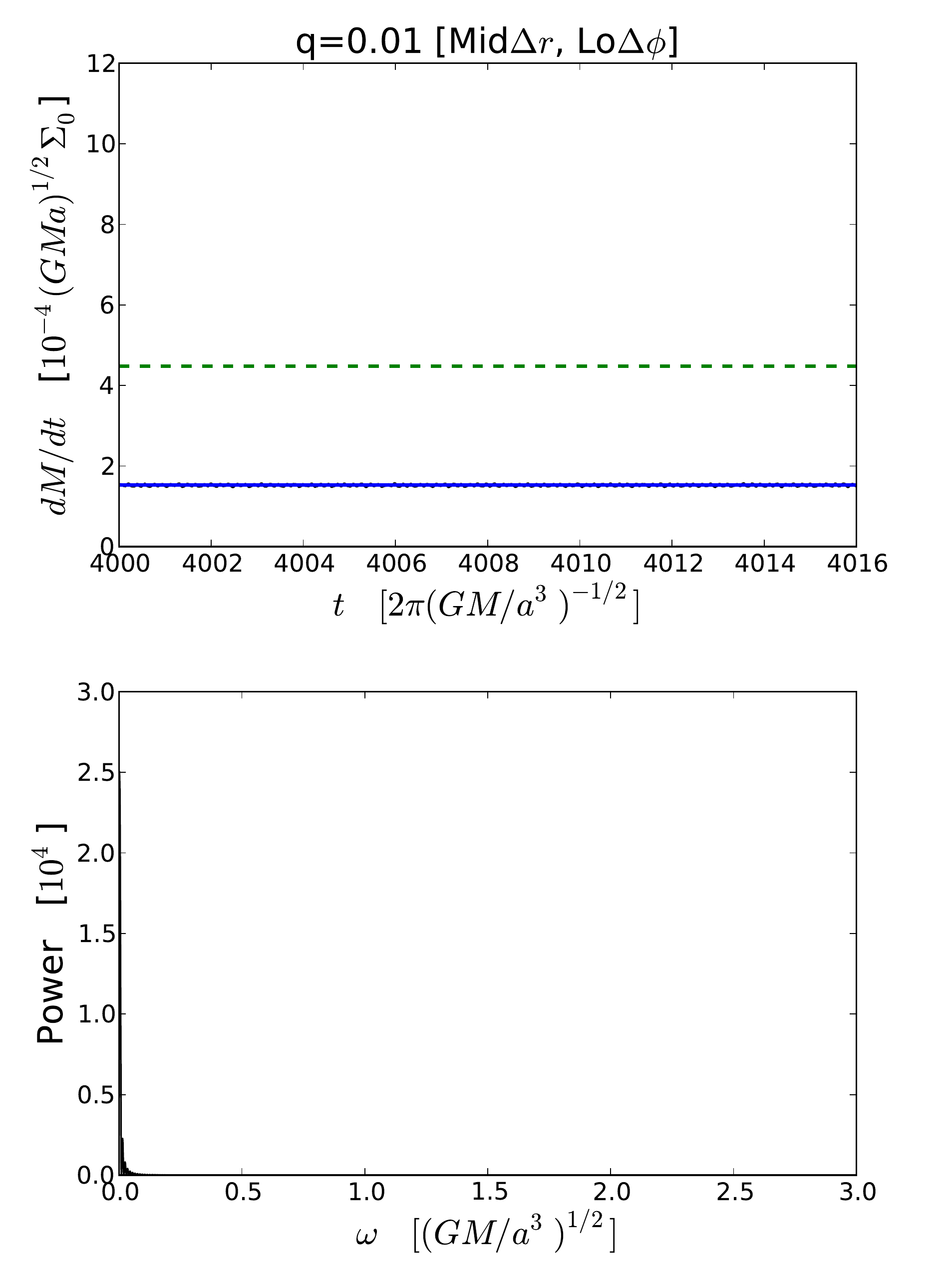} &
\includegraphics[scale=0.42]{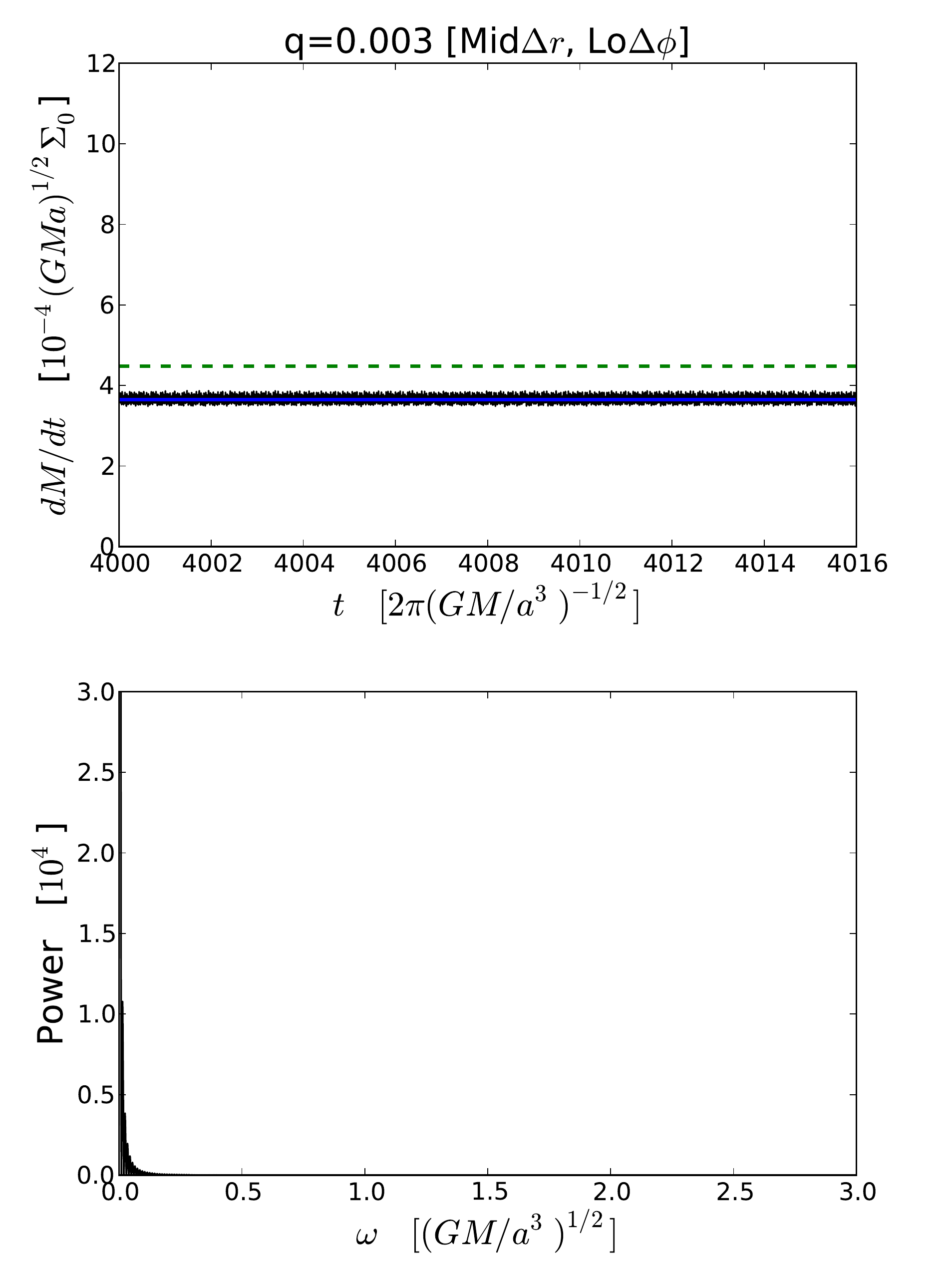} 
\end{array}$
\end{center}
\caption{The time variable accretion rate across the inner boundary of
  the simulation measured at $r=r_{\rm min}=a$ (top of each pair of
  panels) and the corresponding Lomb-Scargle periodogram (bottom of
  each pair) computed over 100 orbits.  Panels are displayed in order of decreasing binary mass
  ratio, starting from $q=1.0$ at top left (on the previous page) to
  $q=0.003$ at the bottom right (on this page). The average accretion
  rate in each panel is denoted by the solid [blue] horizontal line.
  The dashed [green] horizontal line in each plot shows the average
  accretion rate for the point mass ($q=0$) case for reference.  For
  $q>0.05$, the accretion rate is strongly modulated by the binary,
  with either one, two, or three distinct periods present
  simultaneously, depending on the value of $q$ (see text for detailed
  explanations). For $q \lsim 0.05$, the binary still reduces the mean
  accretion rate noticeably, but does not imprint strong periodicity;
  the $q=0.003$ binary is nearly indistinguishable from a single BH.}
\label{Mdott6}
\end{figure*}

The most interesting consequence of the lopsided cavity shape is on
the accretion rate.  In the top left pair of panels in
Figure~\ref{Mdott6}, we show the accretion rate, measured across the
inner boundary of the simulation ($r_{\rm min}=a$) during the
quasi-steady-state, after 4000 binary orbits.  The upper panel shows
the accretion rate as a function of time for $\sim$ 16 binary
orbits; the solid horizontal [blue] line shows the time-averaged
accretion rate during this time, and, for reference, the horizontal
dashed [green] line shows the accretion rate over the same orbits in
the $q=0$ reference simulation. The average accretion
rate onto the binary is approximately $2/3$ of the accretion measured 
for the $q=0$ run at the same resolution. Note that this ratio stays constant 
over the course of the quasi-steady-state.
The lower panel shows the corresponding Lomb-Scargle periodogram, measured over
100 binary orbits at a sampling rate of once per simulation time step
($\sim 500$ per orbit).

As mentioned above, once the quasi-steady-state is reached, the
accretion rate increases from that in the transient state by an order
of magnitude.  As Figure~\ref{Mdott6} shows, it also begins to exhibit
strong (factor of $\sim 3$ above average) variability.  Although this figure samples
the accretion only between 4000 and 4016 orbits, the pattern is
remarkably steady, and repeats itself until the end of the simulation (However, see bullet \ref{vi} in \S \ref{Viscosity Study}).

The accretion is clearly periodic, and displays two prominent
periods, $(1/2)t_{\rm bin}$ and $\sim5.7t_{\rm bin}$.  The stronger
variability at one half the orbital-time is due to the passage of each
black hole by the near side of the lopsided disc and the corresponding
stripping of gas streams from the cavity wall. These streams are then
driven into the opposite side of the cavity (as seen in the top right
panel in Figure~\ref{2DDensProf}; approximately 135 degrees from the
generation point).  The second, longer timescale corresponds to the
orbital period at the cavity wall.  As mentioned above, when the
non-accreted material from the streams hits the far-side of the
cavity, it creates an over-density which orbits at the disc's
orbital period there which ranges from $\sim 2 \pi (2.0a)^{3/2} (GM)^{-1/2} \sim
2.9\rm{t}_{\rm{bin}}$ out to $\sim2 \pi (3.3a)^{3/2} (GM)^{-1/2} \sim
6.0\rm{t}_{\rm{bin}}$. The larger streams pulled from the lump in turn create 
a new lump and the cycle repeats once every $\sim 5.7 \rm{t}_{\rm{bin}}$.
Similar over-dense lumps have also been found
and described in the 3D MHD simulations of \cite{ShiKrolik:2012}; more
recently, \cite{RoedigDotti:2011, Roedig:2012} have also mentioned the
contributions of such lumps to fluctuations in the accretion rate.

\subsection{Unequal-Mass Binaries}
\label{Unequal-Mass Binaries}

We next turn to the main new results of this paper, and examine the
disc behaviour as a function of the mass ratio.  We start with a
qualitative description of how the accretion pattern changes as we
decrease $q$.

\subsubsection{Three-Timescale Regime: $0.25 < q < 1$.}
\label{Three Timescale Regime}

\begin{table*}
\begin{center}
 \caption{The mean accretion rate $\dot{M}_{\rm bin}$, averaged over
    1000 orbits in the quasi-steady-state, for binaries with different
    mass ratios.  The rates are shown in units of the corresponding
    rate $\dot{M}_{q0}$ found in a single-BH ($q=0$) simulation.  
    This ratio is computed as an average from $3500$ to $4500$ orbits unless the quasi-steady 
    state isn't reached until after (or for large $\alpha$ much before) $3500$ orbits; in this case the value 
    in the table is denoted by $^*$.The first four rows show results for 
    different combinations of radial and azimuthal resolutions. 
    The first row is our fiducial resolution. The last three rows are for runs at the 
    fiducial resolution but different magnitudes of the viscosity parameter $\alpha$.}
\label{Mdot_ratios}
\begin{tabular}{l  |  c c c c c c c c c c}
                                     $q$                                   & 1.0   & 0.75 &  0.5 &  0.25   & 0.1   & 0.075  & 0.05 & 0.025 & 0.01 & 0.003 \\ \hline
    $\dot{M}_{\rm{bin}} / \dot{M}_{q0}  \left[\rm{Mid}\Delta r, \rm{Lo}\Delta \phi \right]$   & 0.671 &  0.700 & 0.655 & 0.382   & 0.355$^*$ & 0.228$^*$  & 0.025 & 0.100  & 0.341 &  0.814  \\                                   
  $\dot{M}_{\rm{bin}} / \dot{M}_{q0}  \left[\rm{Lo}\Delta r, \rm{Lo}\Delta \phi \right]$     & 0.544 & 0.518 & 0.499 & 0.406  & 0.021  & 0.027  & 0.055 & 0.147 & 0.381 &  \\  
  $\dot{M}_{\rm{bin}} / \dot{M}_{q0}  \left[\rm{Mid}\Delta r, \rm{Mid}\Delta \phi \right]$ & 0.821 &  &  &  &  0.426  &   & 0.028 & & &  \\
  $\dot{M}_{\rm{bin}} / \dot{M}_{q0}  \left[\rm{Hi}\Delta r, \rm{Hi}\Delta \phi \right]$      & 0.930 &  &  &  &  0.724 &   & 0.172$^*$  &   &  &  \\
  $(\alpha = 0.02$)                                            							        & 0.899 &    &    &     &   &    &  &   &  &    \\
  $(\alpha = 0.04$)                                           						    	        & 0.921 &    &    &     &   &    &  &   &  &    \\
  $(\alpha = 0.1$)                                    								        & 1.015$^*$ &    &    &     &   &    &  &   &  &    
  \end{tabular}
 \end{center}
\end{table*}

\begin{figure}
\begin{center}$
\begin{array}{cc}
\includegraphics[scale=0.3]{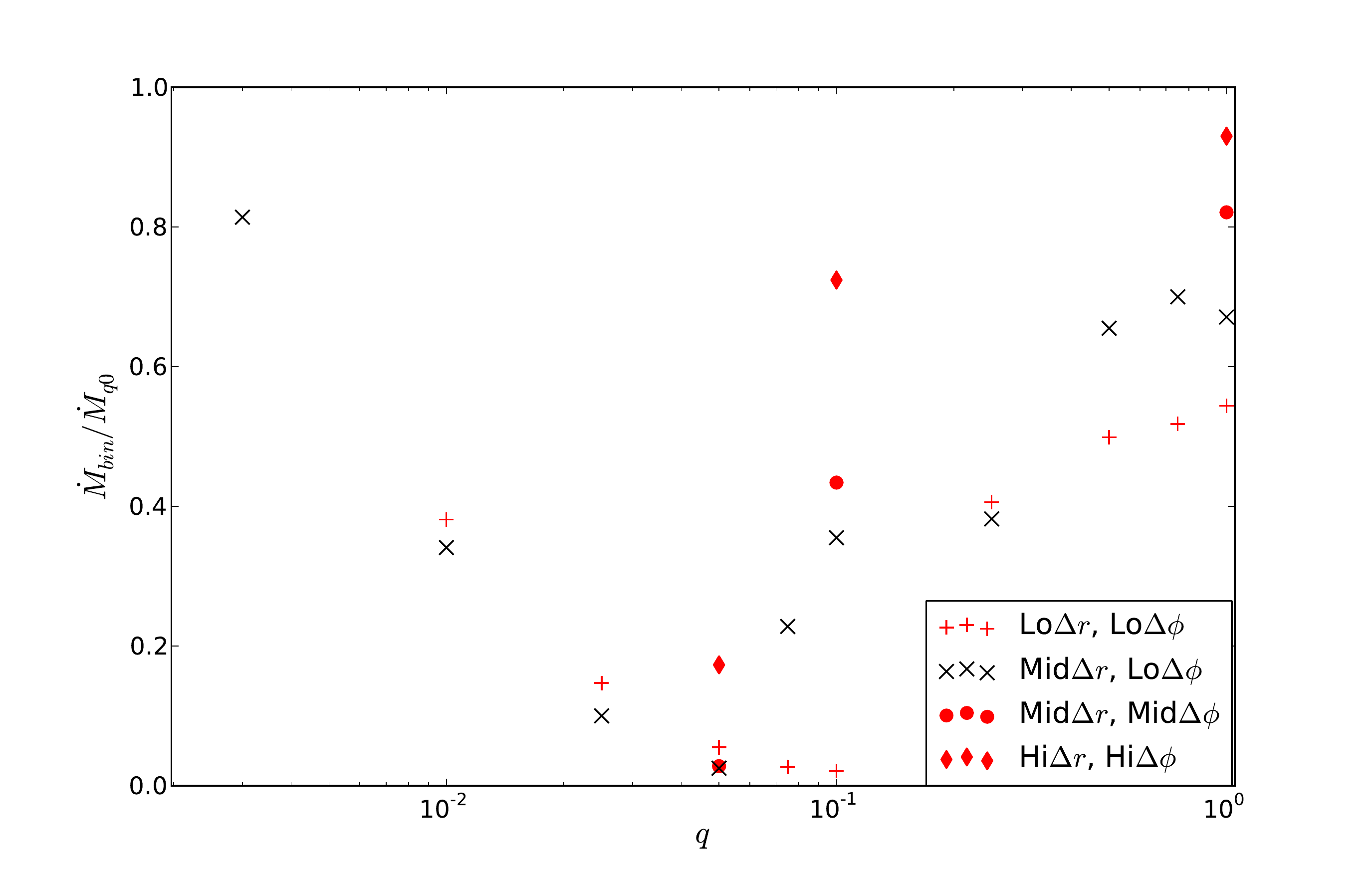} 
\end{array}$
\end{center}
\caption{The average mass accretion rate as a function of $q$, and for
  simulations with different spatial resolution.  This is the same
  information as in Table \ref{Mdot_ratios}, except here shown
  graphically and with only the simulations at the fiducial magnitude of viscosity.}
\label{Mdot_ratios_plot}
\end{figure}

The behaviour of systems with mass ratio in the range $0.25 \lsim q <
1$ are illustrated in Figures \ref{2DDensProf}, \ref{AzAvDens} and
\ref{Mdott6}.  These show snapshots of the 2D surface density in the
quasi-steady-state, the evolution of the azimuthally averaged density
profile, and the time-dependent accretion rates, respectively.
Additionally, in Table~\ref{Mdot_ratios}, and in the corresponding
Figure~\ref{Mdot_ratios_plot}, we show the time-averaged accretion
rate as a function~of~$q$.  In each case, the accretion rate is
averaged over 1000 binary orbits in the quasi-steady state (unless noted, from 3500-4500 binary orbits), and is quoted in
units of the corresponding rate for a single-BH $(q=0)$ disc. This ratio changes very little over the course of the quasi-steady-state regime.

These figures and table illustrate several trends as $q$ is lowered
from $q=1\rightarrow 0.75 \rightarrow 0.5 \rightarrow 0.25$:

\begin{enumerate}

\item The cavity becomes more compact, and less lopsided, as one
  naively expects when the binary torques are reduced. These effects
  are clearly visible in the middle row of Figure~\ref{2DDensProf},
  and also in the corresponding azimuthally averaged density profiles
  in Figure~\ref{AzAvDens}: the profiles look remarkably similar for
  $q=0.5$ and $0.25$, except the cavity for $q=0.25$ is smaller, and its
  wall is visibly sharper (as a result of azimuthally averaging over a
  less lopsided 2D distribution).

\item The secondary (primary) moves closer to (farther from) the cavity
  wall as $q$ is reduced. 
  Thus occurs for two reasons. First, the position of the secondary (primary) moves
  away from (towards) the binary's center of mass,
  \begin{align}
  r_s(q)= a(1+q)^{-1} \qquad \qquad  r_p(q) = a(1+1/q)^{-1}.
  \end{align}
  Second, as mentioned in (i), the size of the central cavity decreases. 
  (Fig.~\ref{AzAvDens}) shows the locations of
  the cavity edge $r_{\rm ce}$ expected from balancing the azimuthally
  averaged gravitational and viscous torques (Fig. \ref{TrqDq}), which
  agree well with the observed cavity sizes. In Figure
  \ref{RCE}, we explicitly show $r_{\rm ce}$ as a function of $q$.  The
  points in the figure have been obtained by balancing the azimuthally
  averaged viscous and gravitational torques measured in the simulation
  (equation \ref{TrqBalRCE}).  The black line is an empirical fit to the
  data points at fiducial resolution, given by
  \begin{align}
  r_{\rm ce}(q) &\simeq A + B \ \rm{ln} \left( q^{1/2} + 1\right) + C \ \mbox{ln} \left( q + 1\right)    \nonumber  \\ 
  A &=1.191 \qquad
  B =2.541 \qquad
  C =-1.350
  \label{EmpFitrce}
  \end{align}
  
\begin{figure}
\begin{center}$
\begin{array}{c}
\includegraphics[scale=0.4]{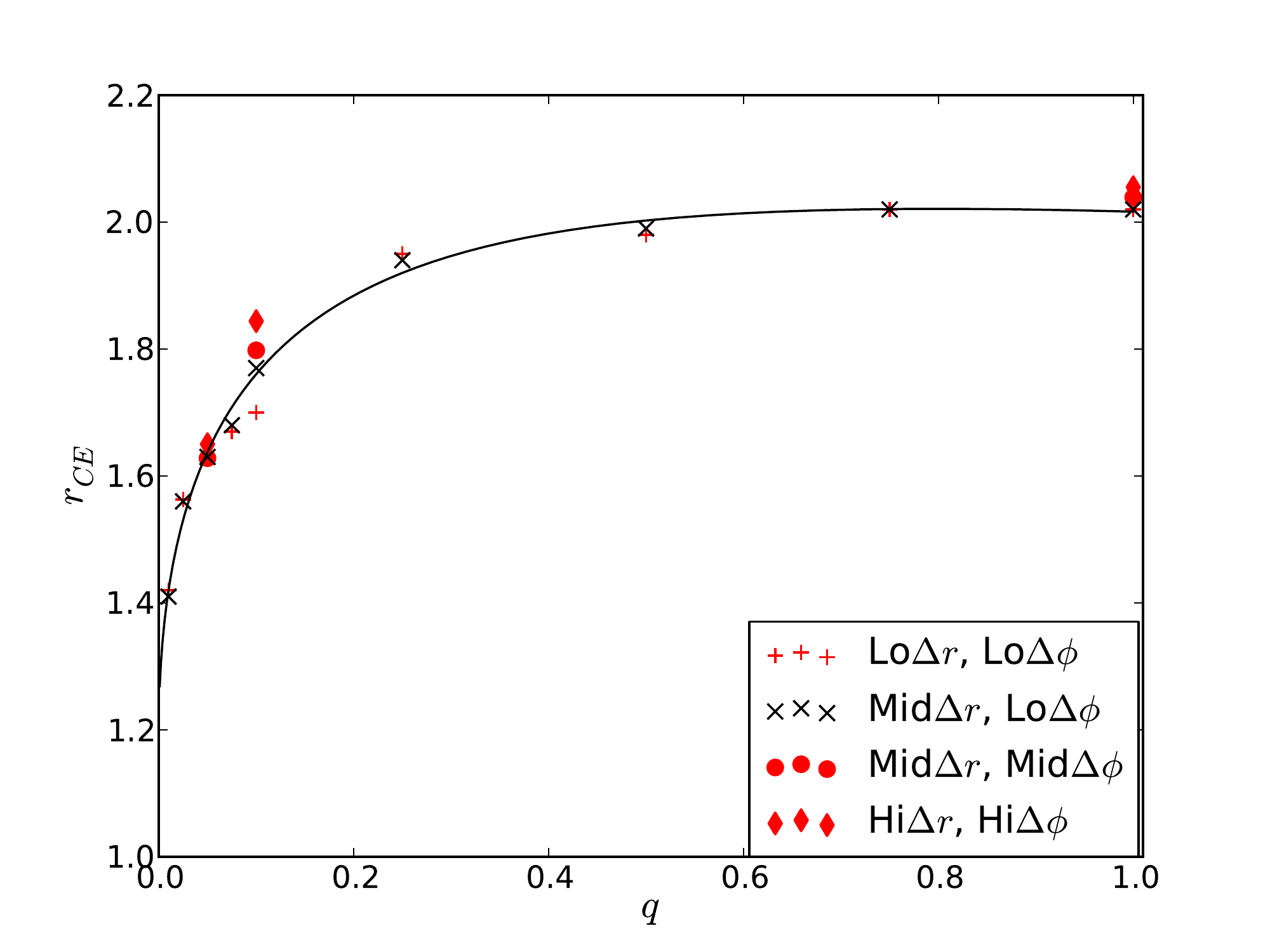}  
\end{array}$
\end{center}
\caption{The position of the cavity wall as a
  function of $q$, in runs with different radial and azimuthal 
  resolutions, as labeled.  The
  points mark the radii $r_{\rm ce}$ at which the azimuthally averaged
  viscous and gravitational torques balance (equation
  \ref{TrqBalRCE}). The black line is an empirical fit to the fiducial resolution data
  points (equation \ref{EmpFitrce}).  }
\label{RCE}
\end{figure}

\item The dense lump, created by the shocks due to the
  ``regurgitated'' stream-material thrown back out by the binary, is
  still present for $q=0.5$ (see the corresponding panel in
  Fig.~\ref{2DDensProf}, around 9 o'clock at $r\sim 2a$), but
  is much less discernible for $q=0.25$.  Again, this trend is
  unsurprising - as the torques diminish, one expects weaker shocks
  and smaller over-densities in any resulting lump.

\item The accretion streams in the $q=0.25$ panel of
  Figure~\ref{2DDensProf}, and the corresponding ripples in the azimuthally averaged density profiles of Figure~\ref{AzAvDens}, become noticeably weaker.  However, the
  average accretion rate, shown in Figures \ref{Mdott6}
  and~\ref{Mdot_ratios_plot} and in Table~\ref{Mdot_ratios}, stays at
  approximately $\simeq 0.7 \dot{M}_{q0}$ over the range $0.5 \lsim q \leq 1$. For
  $q \lsim 0.5$, the average accretion rate drops more rapidly,
  falling by nearly a factor of two to $\simeq 0.36 \dot{M}_{q0}$ by
  $q=0.1$. 
  
\item As the average accretion rate decreases, so does the maximum
  accretion rate (i.e., the amplitude of the spikes in
  Fig.~\ref{Mdott6}) keeping an approximately constant enhancement factor of $\sim3$
  as the mass ratio is decreased to
  $q\sim 0.1$.

\item The percentage of a stream which leaves the domain at $r_{\rm{min}} =a$ as opposed to being flung back out also decreases by a factor of $\sim2$ as $q$ decreases from 1.0 to 0.1. We measure this 
  percentage from the simulations by computing the ratio $\dot{M}(r_{95})/\dot{M}(r_{\rm min})$ averaged over 25 orbits in the quasi-steady-state. Here $r_{95} = 0.95 r_{\rm{ce}}$ (with $r_{ce}$ given by equation 
  (\ref{EmpFitrce})) is chosen to be just inside the cavity wall where the accretion rate is dominated by the streams. The percentage drops by approximately a factor of two, from $\sim 3.3 \%$ at $q=1$ to 
  $\sim1.8 \%$ at $q=0.1$ suggesting that the drop in average accretion rate in the three-timescale regime is due largely to the amount of stream material which can penetrate beyond the binary torque barrier at small $r$.

\item Perhaps the most interesting result is shown by the Lomb-Scargle
  periodograms in Figure \ref{Mdott6}. As $q$ decreases, power is
  traded from both the $(1/2)t_{\rm bin}$ and the $5.7t_{\rm
    bin}$ variability timescales into the $t_{\rm bin}$ timescale.
  This is because of the increased proximity between the secondary and
  the cavity wall, and a corresponding larger distance between the
  primary and the cavity wall noted above.  As a result, as $q$ is
  decreased, the secondary begins to dominate the variability, pulling 
  accretion streams off of the cavity wall once per binary orbit.

\end{enumerate}

Focusing on the last finding: in the $0.25 \lsim q < 1$ case, we find
that the time-dependent accretion rate displays {\em three distinct
  and sharply defined periods}, with well-defined ratios at $0.5, 1$,
and $5.7t_{\rm bin}$.  While the last of these reflects the orbit of the
dense lump at the elongated cavity wall and could depend on details of the disc
properties, the first two periods are fixed by the binary alone and
are independent of the disc.  {\em The 1:2 period ratio is therefore a
robust prediction; if observed, it could serve as a smoking gun
signature of a binary.}

\subsubsection{Single-Orbital-Timescale Regime: $0.05 \lsim q \lsim 0.25$.}
\label{Orbital Timescale Regime}

As $q$ is further decreased, the overall distortions to the disc
become less pronounced, and approach a nearly axisymmetric, tightly
wound spiral pattern (see the $q=0.075$ panel in
Figure~\ref{2DDensProf}).  The distance between the cavity wall and
the secondary further shortens, and the accretion variability becomes
dominated entirely by the streams created by the secondary's passage
once per orbit.  As Figure \ref{Mdott6} shows, for $q=0.075$ the accretion rate displays a nearly sinusoidal
variation, with the corresponding Lomb-Scargle periodogram showing a
single spike just offset from the orbital timescale of the binary.
For $q=0.075$,
the fluctuations are still large (factor of $\sim 3$), but by
$q=0.05$, the fluctuations disappear.  Interestingly,
we find that the average accretion rate is most strongly suppressed
among all of our runs in a narrow mass ratio range $q \sim0.05 \pm 0.0025$; by up to a factor of $\sim 40$ compared
to the single-BH case. However, for out highest resolution run at $q=0.05$, we no longer observer this extreme dip in average accretion rates (see \S \ref{Resolution Study}).

As mentioned above, the disappearance of the $(1/2) t_{\rm bin}$
variability timescale is easy to understand qualitatively: once the
primary BH remains very close to the center-of-mass, its compact
orbital motion no longer impacts the disc far away.  The disappearance
of the $5.7 t_{\rm{bin}}$ variability timescale is also clearly
attributable to the lack of any dense lump near the cavity wall for
$q\lsim 0.25$.  However, the reason this lump disappears is less
obvious, and warrants some discussion.

\begin{enumerate}
\item As $q$ decreases, the cavity becomes less lopsided, and the
  accretion rate spikes become weaker.  This suggests that when these
  weaker accretion streams are flung back to the cavity wall, they
  create less over-dense lumps.  For $q \leq 0.25$, the lump may not
  survive shear stresses and pressure forces, and may dissolve in less
  than an orbital time.
\item As can be eyeballed from Figure \ref{2DDensProf}, the stream
  impact zone (dense region outside the cavity) extends over an
  azimuth of $\Delta\phi \sim 100-120^{\circ}$. The orbital period at
  the cavity edge is $\leq 6 t_{\rm bin}$ (the orbital period at the furthest edge of the cavity), implying that multiple
  streams can hit parts of the same lump if streams are
  generated more than once per binary orbit.
\end{enumerate}

To test whether the strength or the frequency of the streams is more
important for lump generation, we repeat our simulation for an
equal-mass binary, but we placed one of the BHs artificially at what
would be the real binary's center of mass. The second hole still
orbits at $r=a/2$ as usual.  In this setup, the cavity wall is
perturbed by streams with a similar strength as in the real $q=1$
simulation (however much less of the streams reach the inner edge 
of the simulation domain in this one-armed perturber case), but now only once, rather than twice per orbit. We found
that, while this ``one-armed'' binary does generate significant stream impacts at the cavity edge, it does {\em not} create an orbiting
over-density at the cavity edge, nor does it excite a significant
elongation of the cavity.  We therefore conclude that multiple,
overlapping streams are required to generate a strong lump that
survives for an orbital time. However, as the $q=0.25$ accretion rate and periodogram show in Figure \ref{Mdott6}, simply generating two streams is not a sufficient condition for generating a cavity wall lump. Both streams must also be sufficiently large.
This explains the disappearance of the cavity wall frequency for $q\lsim0.25$; as the mass ratio decreases, the primary generates less significant streams and the overlap of two large streams crashing into the cavity wall can no longer occur to generate an over-dense lump there.
This result is also consistent with the cavity becoming less lopsided as $q$ decreases.

\subsubsection{Steady-Accretion Regime: $q \lsim 0.05$.}
\label{Non-Variable Regime }

As we continue to decrease $q$ from $0.075$ through $0.05$ to $0.01$, we find yet another
distinct regime.  The overall morphology of the snapshot of a $q=0.01$ and $q=0.05$ (not shown)
disc in Figure~\ref{2DDensProf} looks similar to the $q=0.075$ case,
except the nearly-concentric perturbations are even weaker, and the
cavity still smaller.  However, the similarity is quite deceptive,
with the movie versions of these figures showing a striking
difference.\footnote{Movie versions of the snapshots in
  Figure~\ref{2DDensProf} are available at http://www.astro.columbia.edu/$\sim$dorazio/moviespage}
In the $q=0.075$ case, accretion streams form and disappear
periodically, but in the $q=0.05$ case, the disc pattern becomes
constant and unchanging (in the frame co-rotating with the binary).
There is still a visible accretion stream, hitting the inner boundary
of the simulation just ahead of the secondary's orbit, but the stream
steadily co-rotates with the binary.

As Figure \ref{Mdott6} shows (see also Fig.~\ref{Mdot_ratios_plot} and
Table~\ref{Mdot_ratios}), the average accretion rate has reached its
minimum at $q=0.05$. For $q \leq 0.05$, the accretion rate
becomes steady, with no fluctuations, and its value {\em increases}
back towards the $q=0$ rate (dashed horizontal [green] line in Figure
\ref{Mdott6}).  For such a light secondary, the system
begins to resemble a disc with a single BH.  Although a cavity is
still clearly present (moving from $r\simeq1.6a$ at $q=0.05$ to
$r\simeq1.4a$ at $q=0.01$), it is being refilled, as for a single
BH. Indeed, after 4000 orbits, the $q=0.01$ azimuthally averaged density
profile is approaching the $q=0$ profile (see
Fig.~\ref{AzAvDens}). \\

Although the secondary still excites small but visible ripples in the
disc (Fig.~\ref{2DDensProf}), by $q\leq0.05$ it can no longer exert a
large enough torque to pull in large streams and drive them back out
to produce a lopsidedness in the circumbinary disc.  The ripples are 
in the linear regime, and they resemble the tightly-wound
spiral density waves launched in protoplanetary discs
(e.g. \citealt{GT80,DRS:2011:Linear,DM2012}), except here our
background discs have a pre-imposed central cavity.  In the 
linear regime, the waves are excited by resonant interactions with the
disc, and non-linear coupling to an $m=1$ mode is no longer
possible. Note the disappearance of any ripples in the azimuthally
averaged surface density for $q = 0.01$ (Fig.~\ref{AzAvDens}).

\section{Summary and Discussion}
\label{Summary and Discussion}
In summary, we find that the behavior of the accretion rate across the
circumbinary cavity as a function of $q$ can be categorized into four
distinct regimes:

\begin{enumerate}
\item {\em Two-timescale regime; $q=1$.}  Confirming previous
  results, an equal-mass binary maintains a central low-density cavity
  of size $r\sim 2a$ and the time-averaged accretion rate is $\sim2/3$ 
  of that for a point-mass case. There are up to factor of $\sim3$
  fluctuations around the average on two prominent time-scales,
  $(1/2)t_{\rm bin}$ and $\sim 5.7t_{\rm bin}$.
\item {\em Three-timescale regime; $ 0.25 < q < 1$.}  The
  time-averaged accretion rate drops by a factor of $\sim 1.8$ by
  $q=0.25$; however, the maximum fluctuations 
  continue to occur with amplitude $\sim3$ times the average rate . 
  There are three time-scales present, $(1/2)t_{\rm bin}$, $t_{\rm
    bin}$, and $\sim 5.7t_{\rm bin}$.
\item {\em Single-orbital-timescale regime; $0.05 \lsim q < 0.25$.} In
  this regime, the average accretion rate and its fluctuations continue to
  drop with decreasing $q$. Variability is dominated by the secondary, 
  and is nearly sinusoidal on the binary period $t_{\rm bin}$ though accompanied by small accretion spikes due to the primary with maxima below the average rate.
\item {\em Steady-Accretion regime; $q \lsim 0.05$.} The accretion 
  becomes steady while reaching its lowest rate at $q\sim0.05$. By $q=0.003$, the accretion rate rises 
  again to $\sim0.8$ of the $q=0$ case. The system overall resembles a 
  cavity-filling single-BH disc, with small perturbations due to 
  the secondary in the linear regime.
\end{enumerate}

\subsection{Comparison with MM08}
\label{Viscosity Dependence  and Comparison with MM08}
Our main qualitative conclusions for the equal-mass binary case,
including the morphology of the disc, and the accretion rate, are in
good agreement with MM08.  Nevertheless we do find a small 
discrepancy in the time averaged accretion rates.

\begin{figure}
\begin{center}
 \includegraphics[scale=0.45]{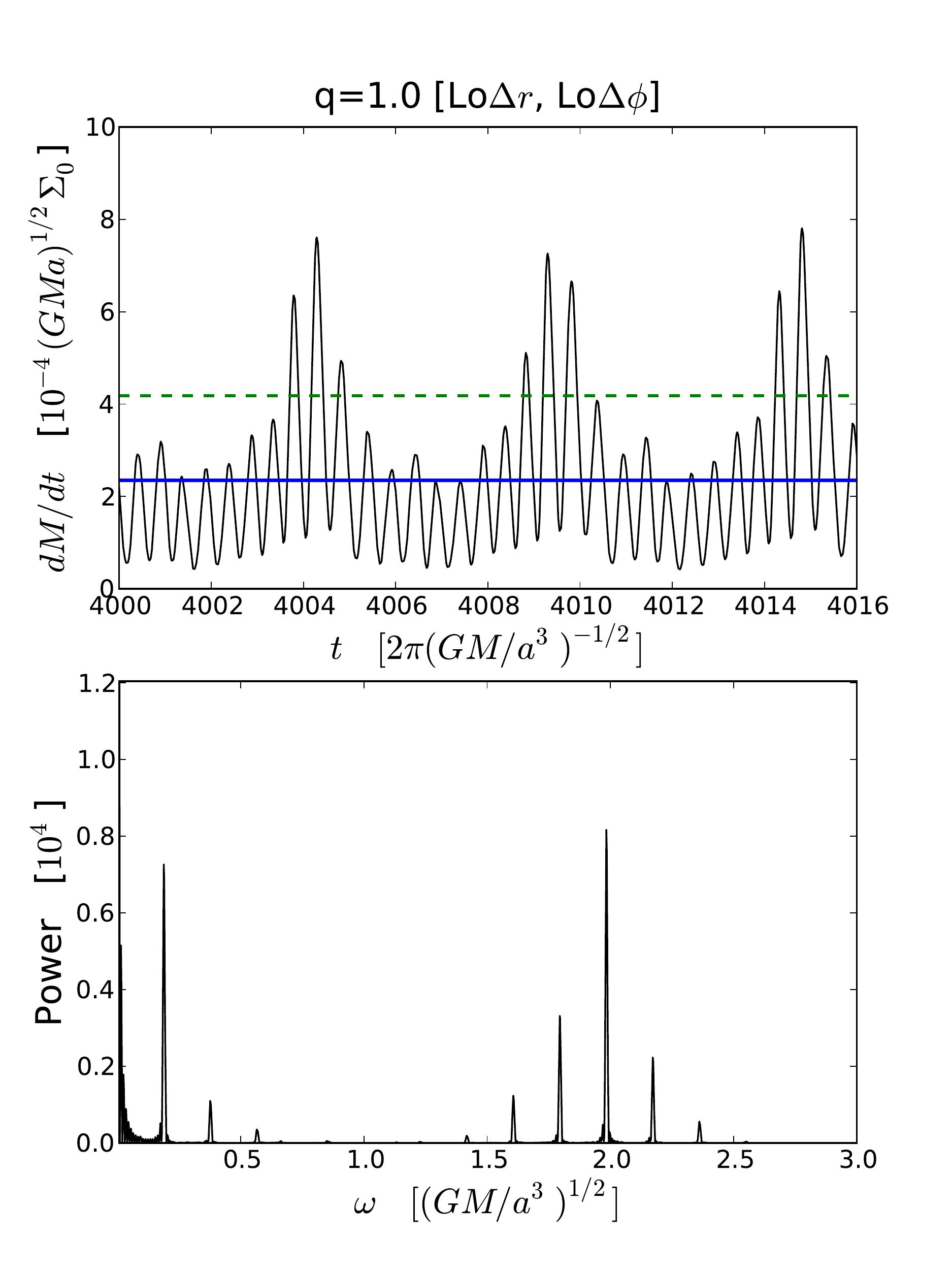}
\end{center}
\caption{Accretion rates at the inner boundary $r_{\rm min}=a$ for the
  equal-mass binary as in the top left panels of Figure \ref{Mdott6},
  except for the lowest resolution runs used in this study which 
  matches the highest resolution used in the disc simulated by MM08.  
  The solid horizontal [blue] line in the top panel is the average 
  accretion rate, and the bottom panel shows
  the Lomb-Scargle periodogram computed over 100 binary orbits.} 
\label{MM08_compare}
\end{figure}

Comparing MM08's Figures 7 and 8 with the top left panel of
our Figure \ref{MM08_compare}, which is run at the same 
resolution as the highest resolution used in the MM08 study,
\footnote{This refers to the highest resolution used by MM08 at the inner region of the disc. MM08 use a lower resolution far away from the binary where this study uses a uniform resolution throughout.}
we see that the magnitudes of the
time-averaged accretion rates, as well as the periodogram frequencies, agree. However, the detailed variability is not identical.

The primary difference is that there is more power at low frequencies in the MM08 periodograms. In-between the largest accretion spikes, the MM08 accretion rate drops closer to zero and becomes less uniform where as our accretion rate is a steady modulation of spikes occurring at the twice the binary orbital period. The accretion rate and periodogram observed in MM08 could be realized if viscous stresses are less efficient at breaking up the over-dense lump responsible for creating the $(5-6) t_{\rm bin}$ modulation. Then the lump will be more centralized and there will be a greater disparity between streams which are generated from the lump and streams which are not. Thus we conjecture that this small difference may be due to different treatments of viscosity and grid setup, or differences between FLASH2 (used by MM08) and FLASH3 (used here).

\subsection{Viscosity Study}
\label{Viscosity Study}
%
%
\begin{figure}
\begin{center}$
\begin{array}{cc}
\includegraphics[scale=0.29]{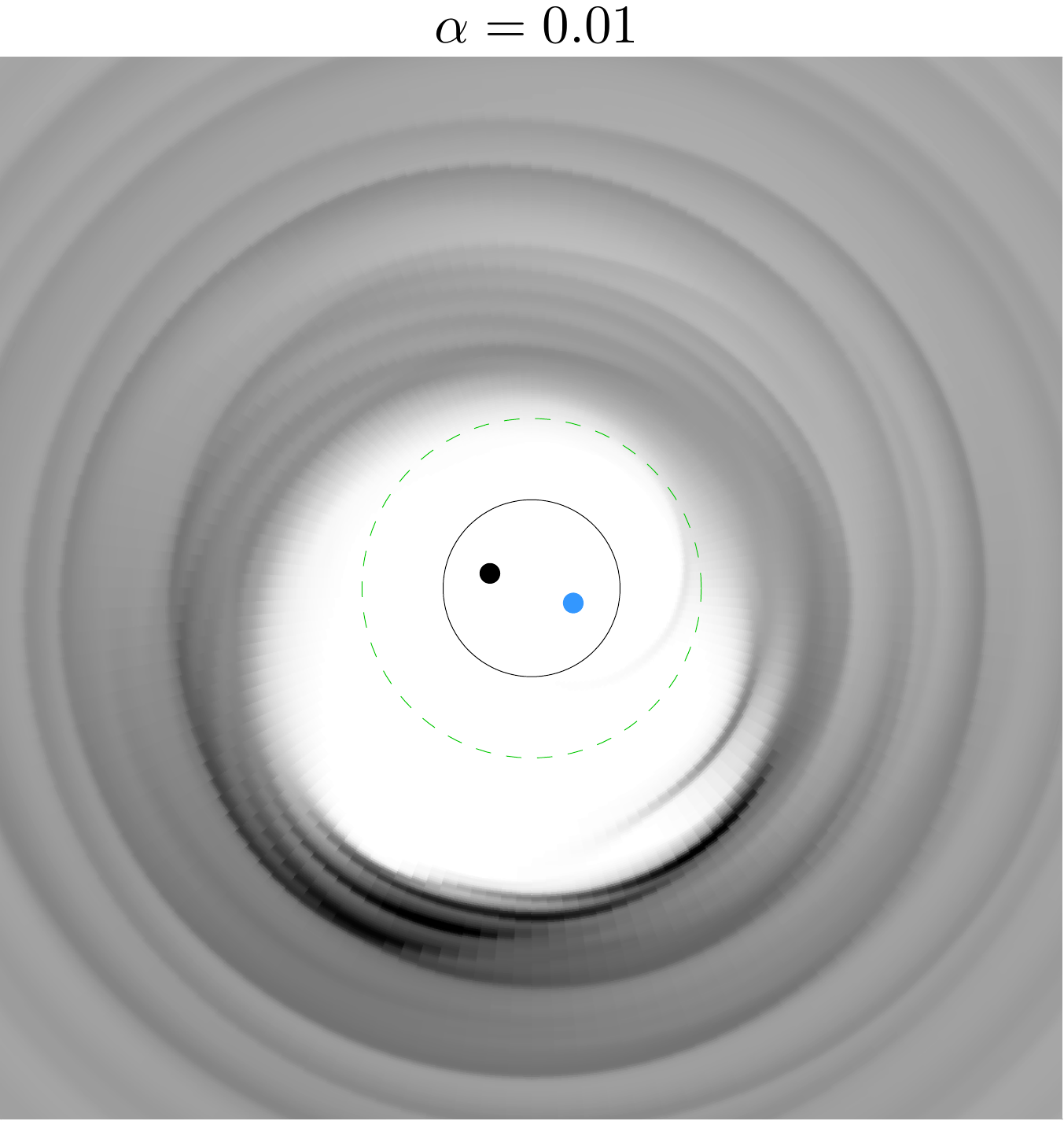} &
\includegraphics[scale=0.29]{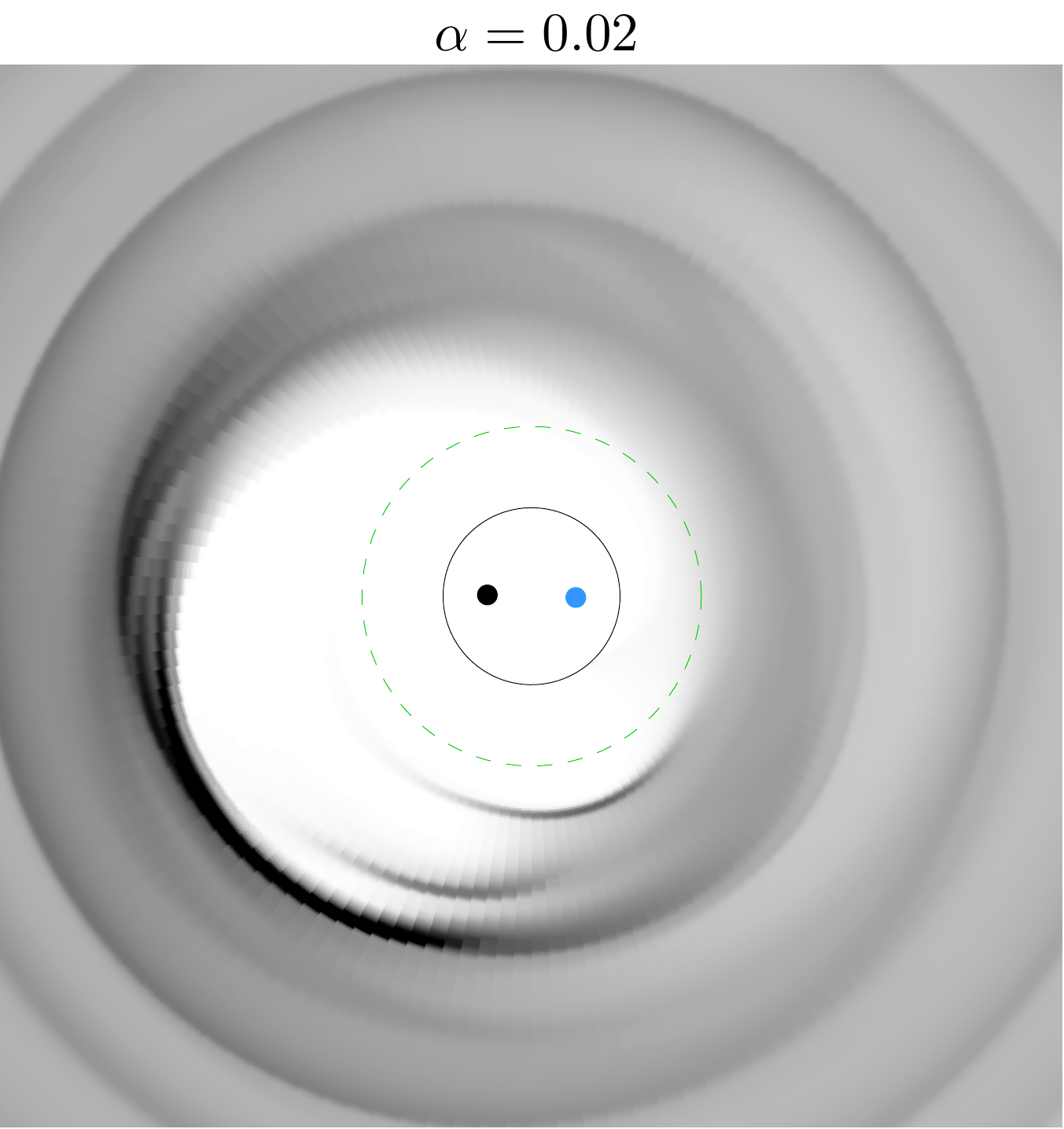} \\
\includegraphics[scale=0.29]{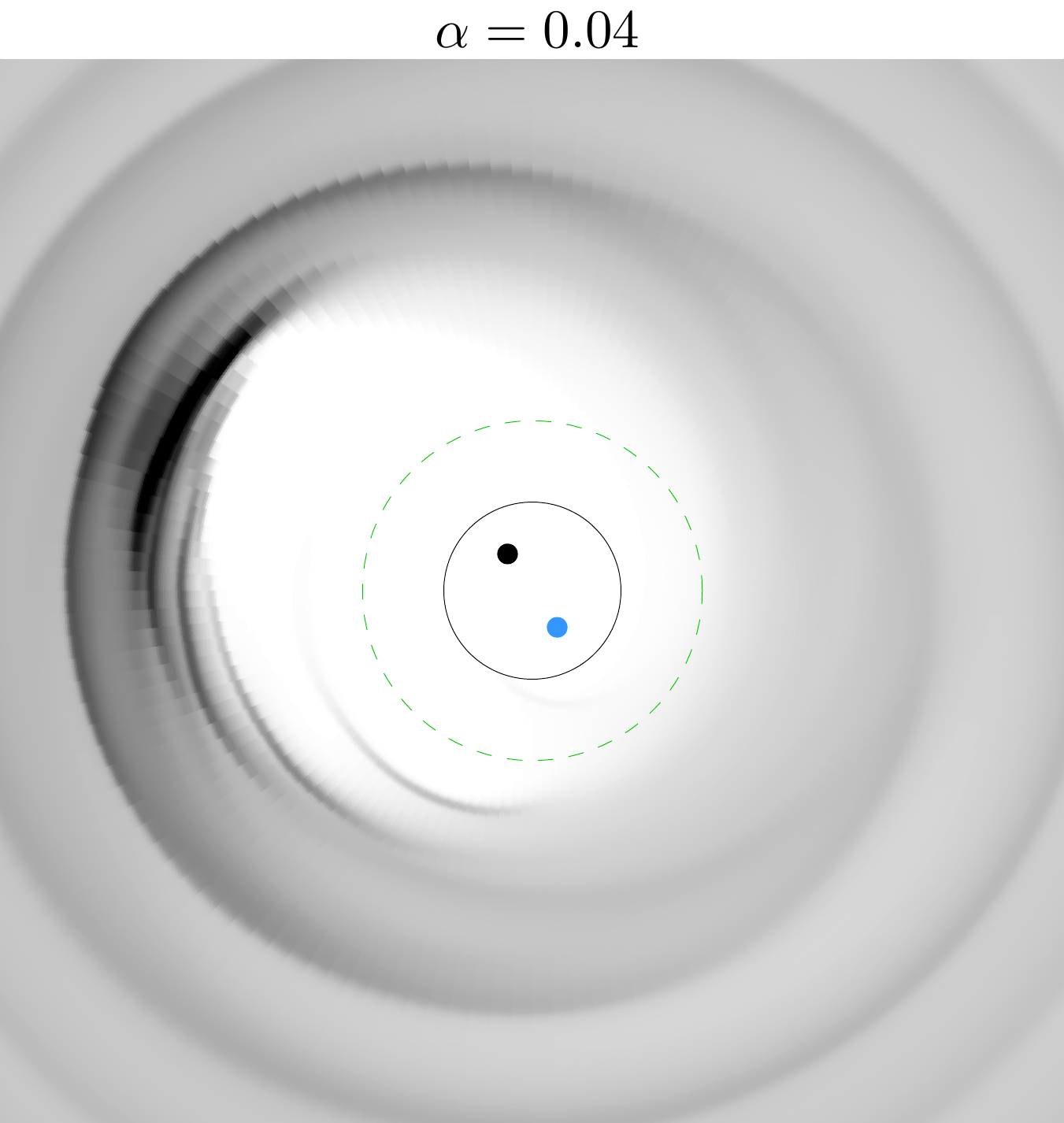} &
\includegraphics[scale=0.29]{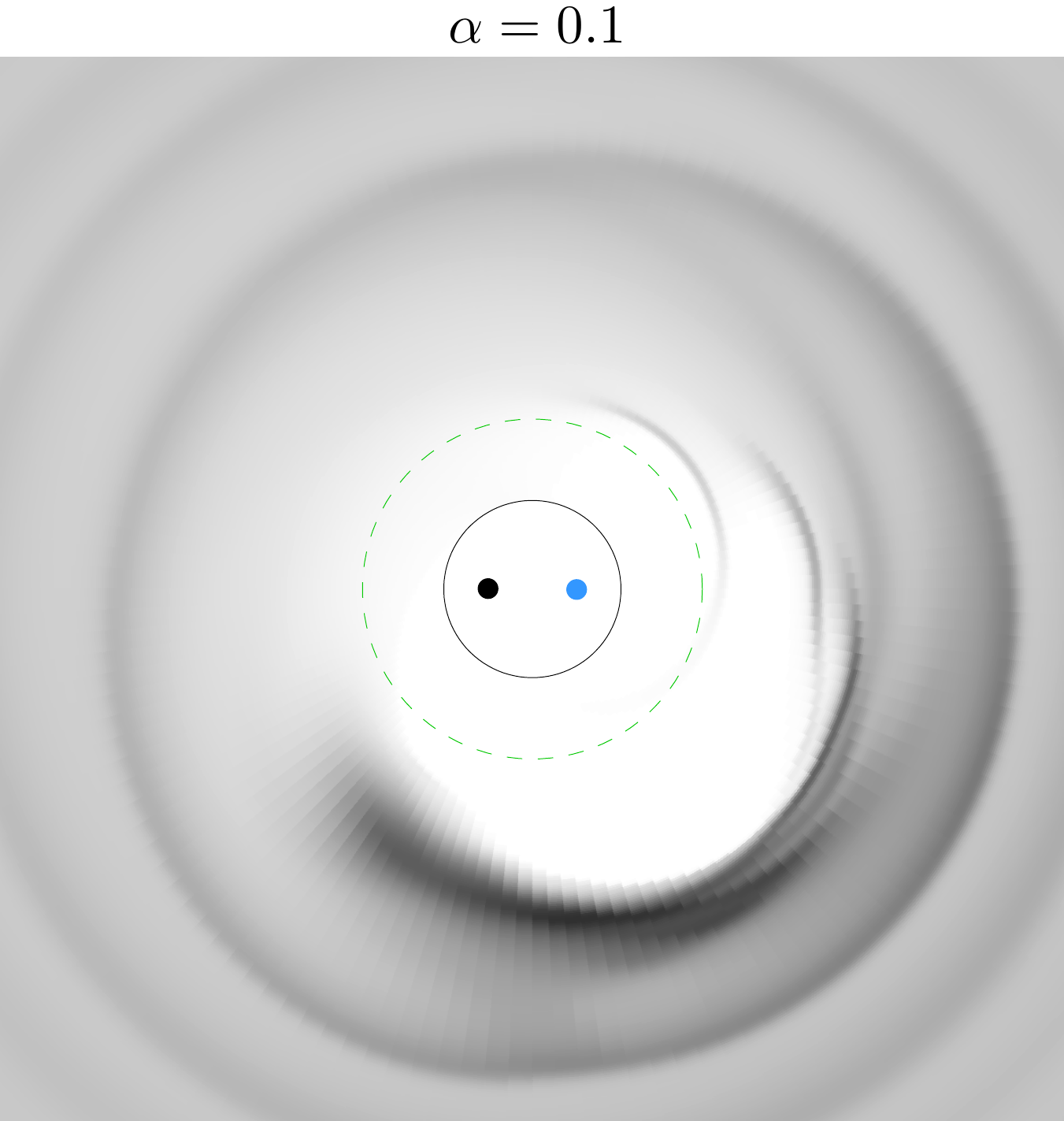} 
\end{array}$
\end{center}
\caption{Two-dimensional surface density distributions during the quasi steady-state phase,
  as in Figure~\ref{2DDensProf}, except for the single mass ratio
  $q=1.0$, and for four different values of the viscosity parameter $\alpha$, as labeled.  
  Increasing $\alpha$ causes ripples created by streams impacting the cavity wall to smear out more quickly causing the surface density snapshots to appear smoother. For all values of $\alpha$ shown here, the over dense lump still survives long enough to create the $\sim (5-6)t_{\rm bin}$ modulation of the accretion rate. Also for larger $\alpha$, the near side of the disc extends in closer to the binary causing a larger fraction of streams to exit the integration domain at $r=a$. This results in higher measured accretion rates relative to the point mass values for the same $\alpha$'s.}
 \label{Fig:2DDensVsc}
\end{figure}
%
%
%
%
\begin{figure*}
\begin{center}$
\begin{array}{cc}
\includegraphics[scale=0.45]{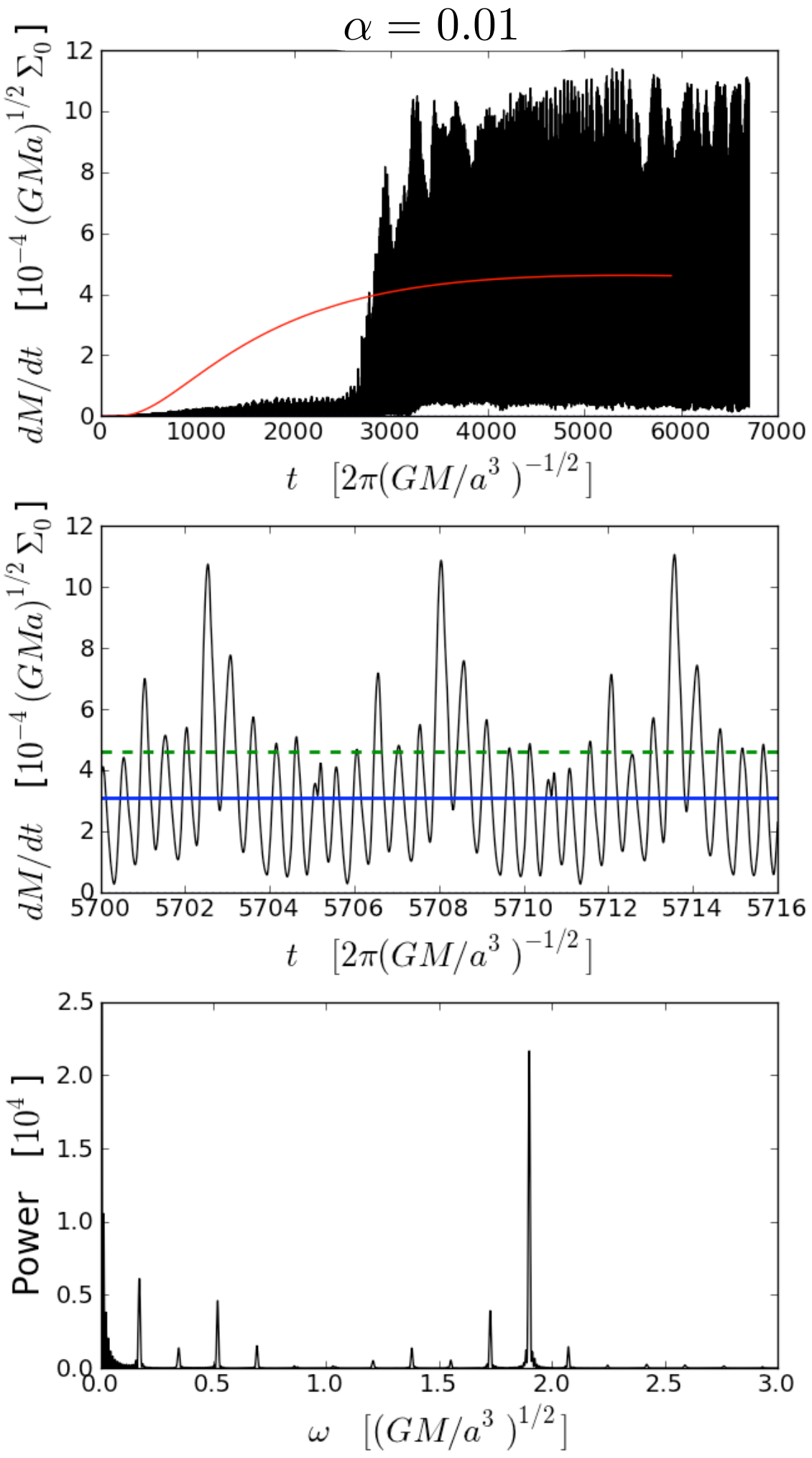} &
\includegraphics[scale=0.45]{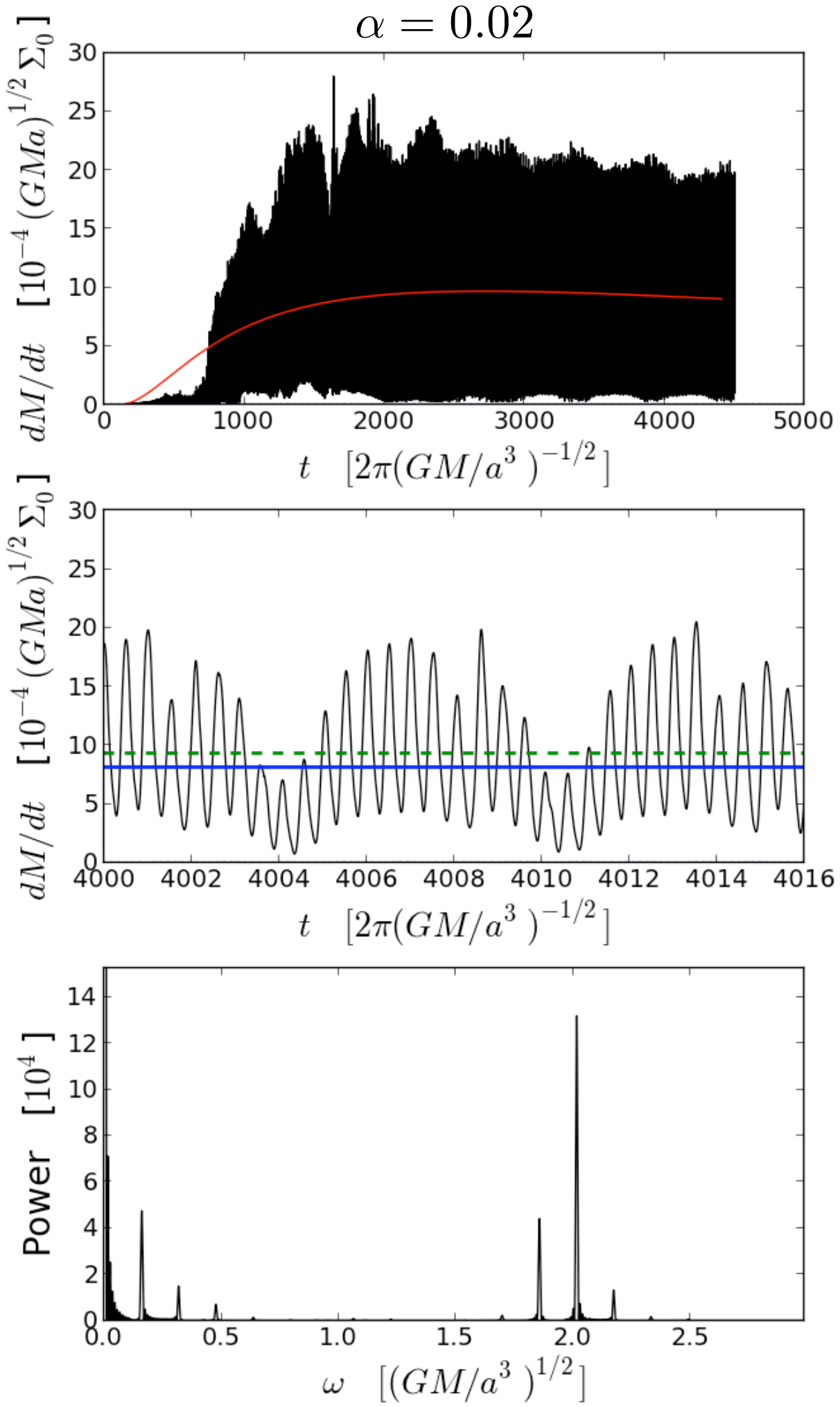} \\
\includegraphics[scale=0.45]{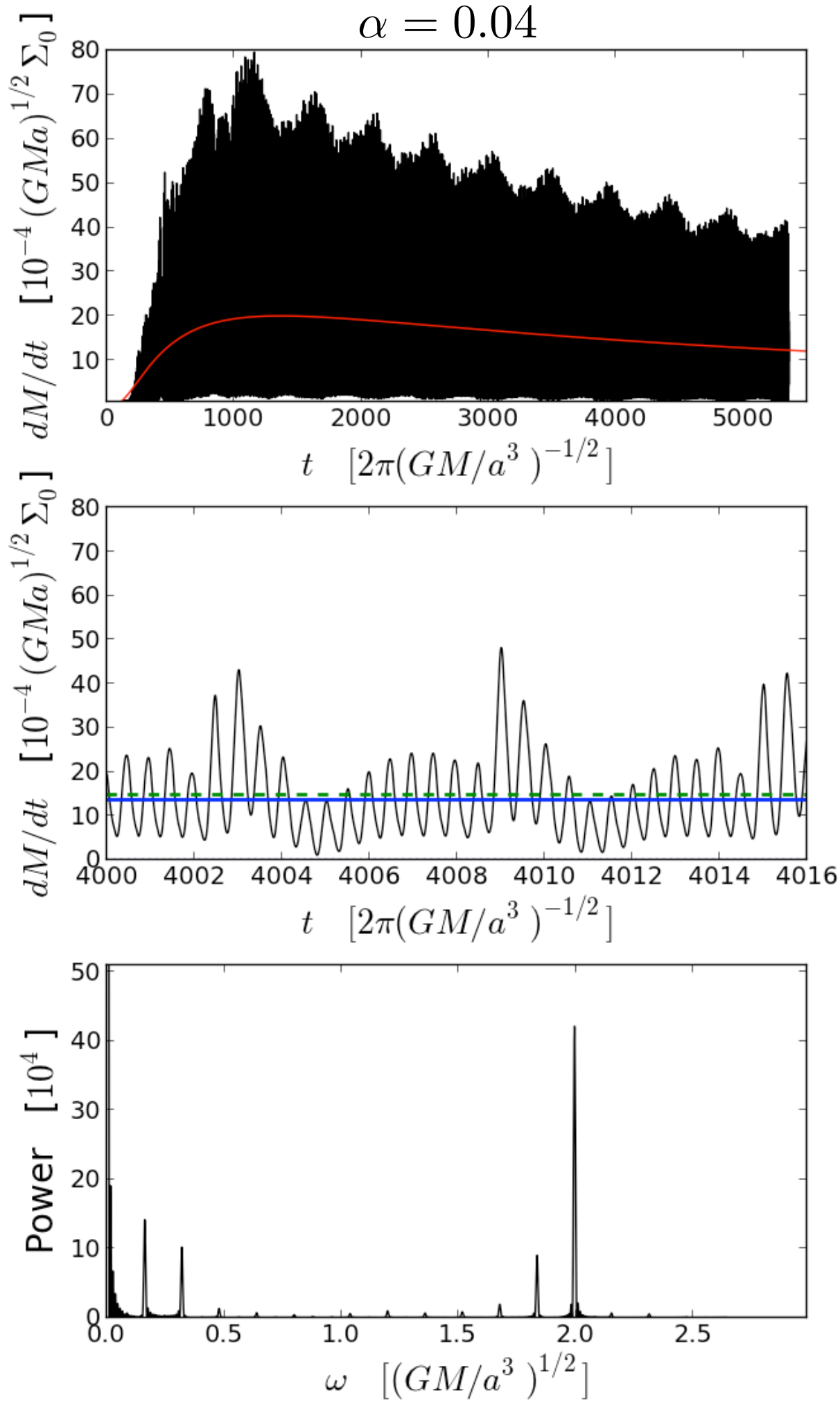} &
\includegraphics[scale=0.45]{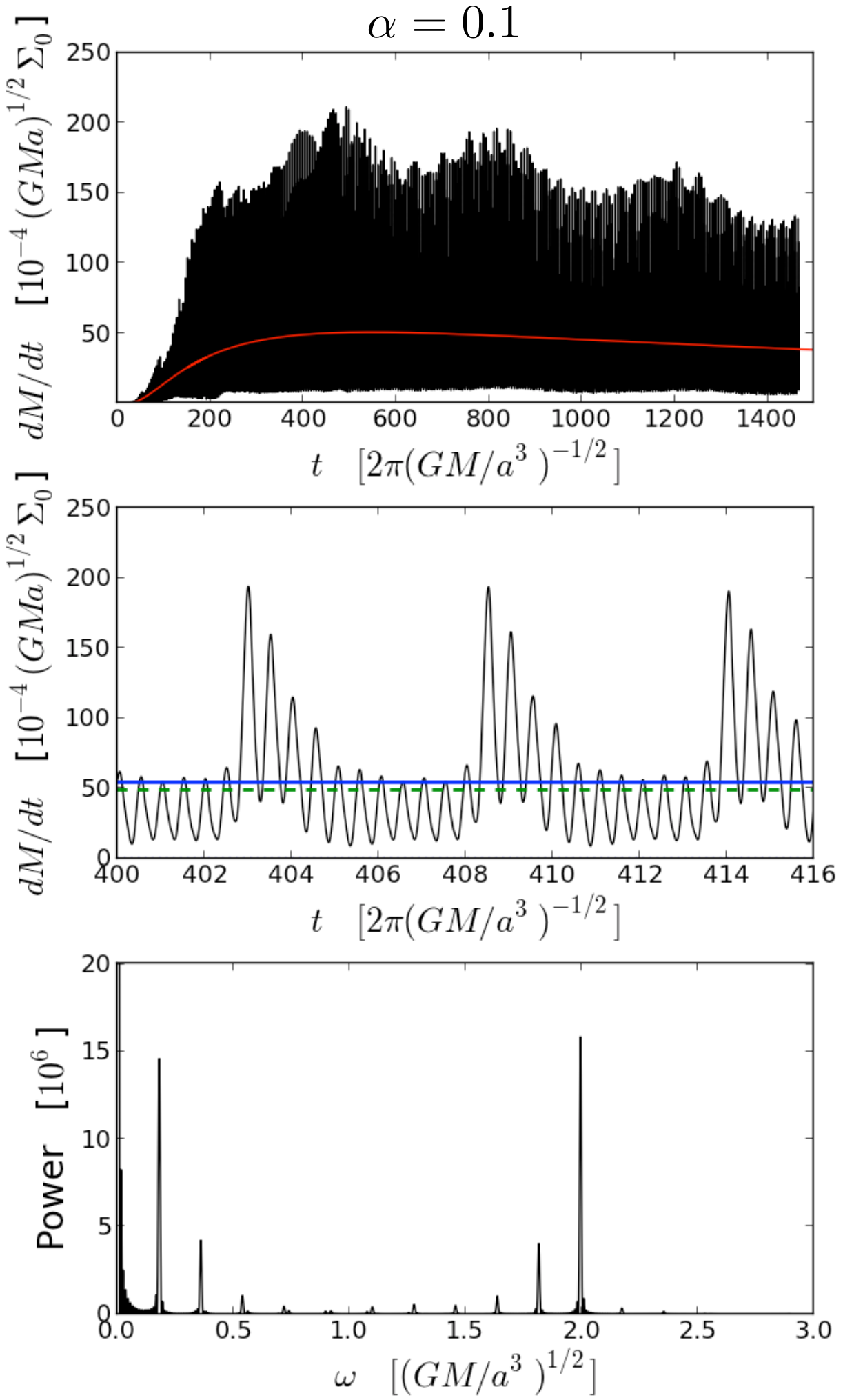}
\end{array}$
\end{center}
\caption{The time-dependent accretion rates as in Figure \ref{Mdott6},
  except for the single mass ratio $q=1.0$ and four different
 values of the viscosity parameter $\alpha$, as shown in Figure \ref{Fig:2DDensVsc}. Additionally the top panel of each figure shows the entire accretion rate history for the $q=1$ (black) and $q=0$ (red) cases. Notice that long-term (once per $\sim 400t_{\rm bin}$) variability appears in the larger $\alpha$ runs and is coincident with the period at which the elongated cavity precesses..}
  \label{Fig:Mdott_Vsc}
\end{figure*}
%
%
To evaluate the sensitivity of our results to the magnitude of viscosity, we run three additional equal-mass binary simulations for $\alpha=0.02$, $0.04$, and $0.1$. Figure \ref{Fig:2DDensVsc} plots snapshots of the 2D surface densities and Figure \ref{Fig:Mdott_Vsc} plots the accretion rates with corresponding periodograms for each of these runs once they have reached the quasi-steady-state regime. Table \ref{Mdot_ratios} records the average accretion rates as a fraction of the reference $q=0$ simulations (a new reference simulation is created for each $\alpha$). Our findings can be summarized as follows:
\begin{enumerate}
	\item{As $\alpha$ is increased, the near side of cavity spreads closer to the binary, as a result, larger portions of streams leave the simulation domain at $r=a$. This results in, not only a larger absolute 	accretion rate, but also an increased rate measured relative to the $q=0$ rate with the same $\alpha$ (See Table \ref{Mdot_ratios}).}
	\item{Despite the increase in average accretion rate, the ratio of maximum accretion rate spikes to the average accretion rate stays constant at $\sim 3$.}
	\item{A transition from a symmetric state to an elongated quasi-steady-state still occurs resulting in discs with similar  morphology including cavities which are of the same size and elongation. As can be seen from Figure \ref{Fig:2DDensVsc}, a difference is that density ripples observed in the fiducial $\alpha=0.01$ disc are progressively more smeared out for the higher $\alpha$ runs. This is due to larger viscous shearing forces more quickly diffusing over-densities due to stream impacts.}
	\item{Although there are larger shearing forces smearing out the small scale density structures seen in the fiducial case, the over-dense lump still survives for at least the necessary $\sim (1/2) t_{\rm bin}$ needed to modulate the accretion rate at the cavity wall orbital period. }
	\item{As a result of the above two points, the periodograms in Figure \ref{Fig:Mdott_Vsc} stay largely the same as $\alpha$ increases. If there is a trend, it is that the timescale associated with the orbital period of the over-dense lump at the cavity wall is more prominent for larger $\alpha$. However, it is possible that, for even larger $\alpha$, the over-dense lump could break up before it can seed another lump via stream generation.}	
	\item{For the larger $\alpha$ runs, the top panels of Figure \ref{Fig:Mdott_Vsc} show the appearance of a longer variability timescale with the same period as the lopsided cavity precession - once per $\sim400$ orbits. This variability manifests itself in a modulation of the maximum accretion rate achieved by the largest streams pulled from the cavity edge lump; every $\sim 400$ orbits the largest accretion rate spikes reach $30 \%$ higher above the average than they do $\sim 200$ orbits later. Note that in the fiducial $\alpha$ case, there is a similar long-term variation in the strength of the $5.7 t_{\rm bin}$ modulation, but it occurs more erratically and with approximately half of the total variation.}
	\label{vi}
	\item{Finally, the top panels of Figure \ref{Fig:Mdott_Vsc} also show that the quasi-steady, lop-sided mode occurs much earlier for larger $\alpha$. For the fiducial case the transition takes place after $\sim 2500  t_{\rm bin}$, for $\alpha=0.02$ after $	\sim 1000  t_{\rm bin}$ and for larger $\alpha=0.04, 0.1$ after less than a few 100 orbits which is set largely by the time for fluid to diffuse to the inner regions of the disc.}
\end{enumerate}
A more detailed investigation of the effects of viscosity should be carried out in future studies and understood self-consistently from simulations which generate turbulent viscosity via the magnetorotational instability (MRI) (See \citealt{ShiKrolik:2012,Noble+2012}).

\subsection{Resolution Study}
\label{Resolution Study}
Up to now we have discussed the results from our fiducial set of
medium-radial, low-azimuthal resolution runs ($\left[ \hbox{Mid}
  \Delta r, \hbox{Lo}\Delta \phi\right]$ in Table \ref{runs}).
Ideally, we would repeat these runs at increasingly high radial and
azimuthal resolutions, until the results converge.  Unfortunately,
this is computationally prohibitive, and we instead choose the
following approach.

\begin{enumerate}
\item For the $q=1.0$ case, we perform two higher-resolution
  runs ($\left[ \hbox{Mid} \Delta r, \hbox{Mid}\Delta \phi \right]$
  and $\left[ \hbox{Hi} \Delta r, \hbox{Hi} \Delta \phi \right]$ in
  Table \ref{runs}) and one lower resolution run ($\left[ \hbox{Lo}
    \Delta r, \hbox{Lo} \Delta \phi \right]$ in Table \ref{runs}) to
  look for signs of convergence.
\item We then explore the resolution sensitivity of the boundaries between each accretion regime: The $q=0.1$ case is at the cusp of
  the three--timescale and single--timescale regimes, where we expect the
  accretion behaviour to be particularly sensitive to $q$. The $q=0.05$ case is at the cusp of
  the single--timescale and steady accretion regimes, where the
  minimum accretion rates are achieved. Thus we also run the $q=0.1$ and $q=0.05$ cases for the same set of resolutions as the $q=1$ case. 
\item We repeated each of our runs at the lowest resolution.  Redoing
  the entire set of runs allows us to assess whether different massratios
  are affected by resolution differently.  
\end{enumerate}

\begin{figure}
\begin{center}$
\begin{array}{cc}
\includegraphics[scale=0.29]{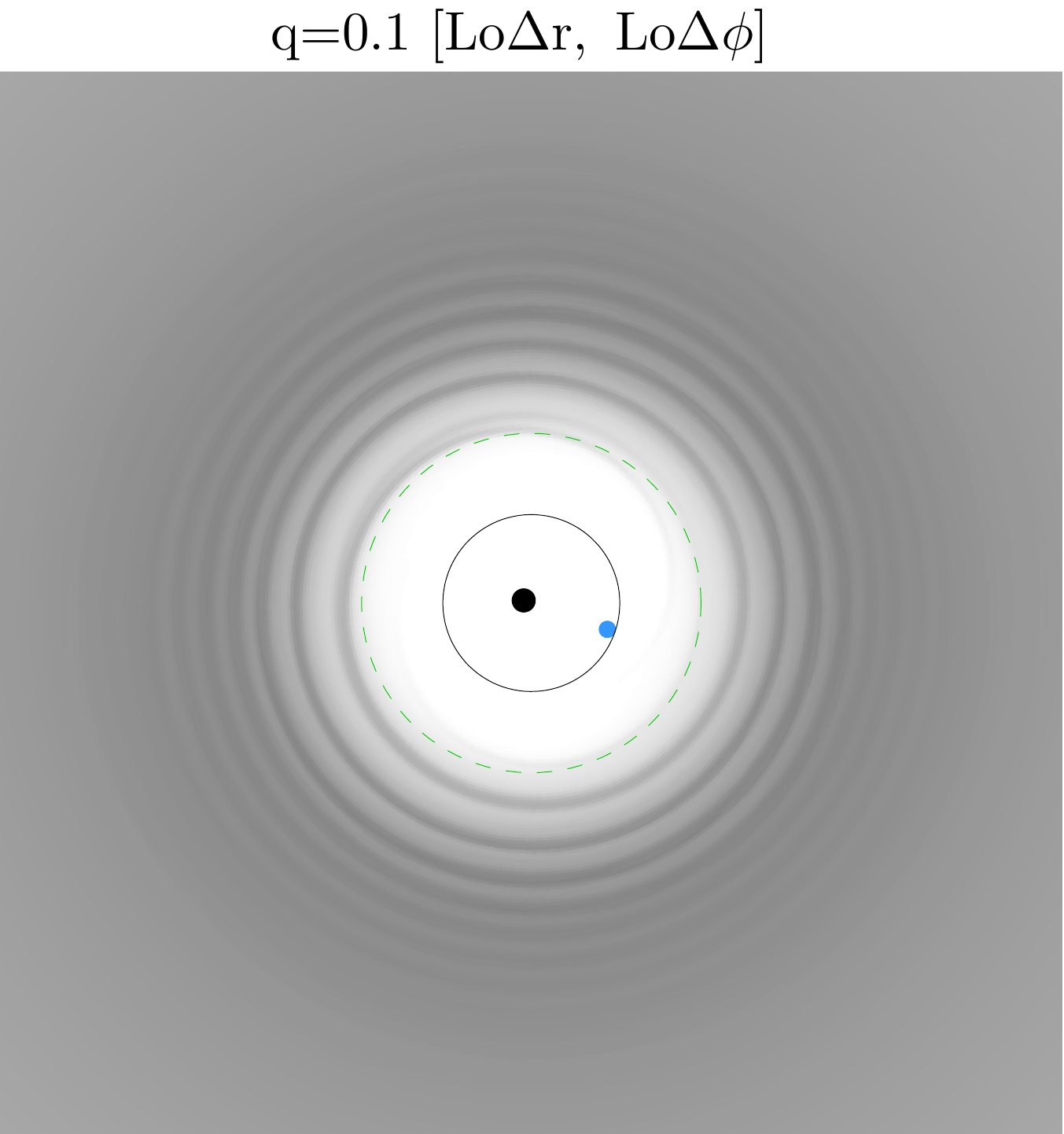} &
\includegraphics[scale=0.29]{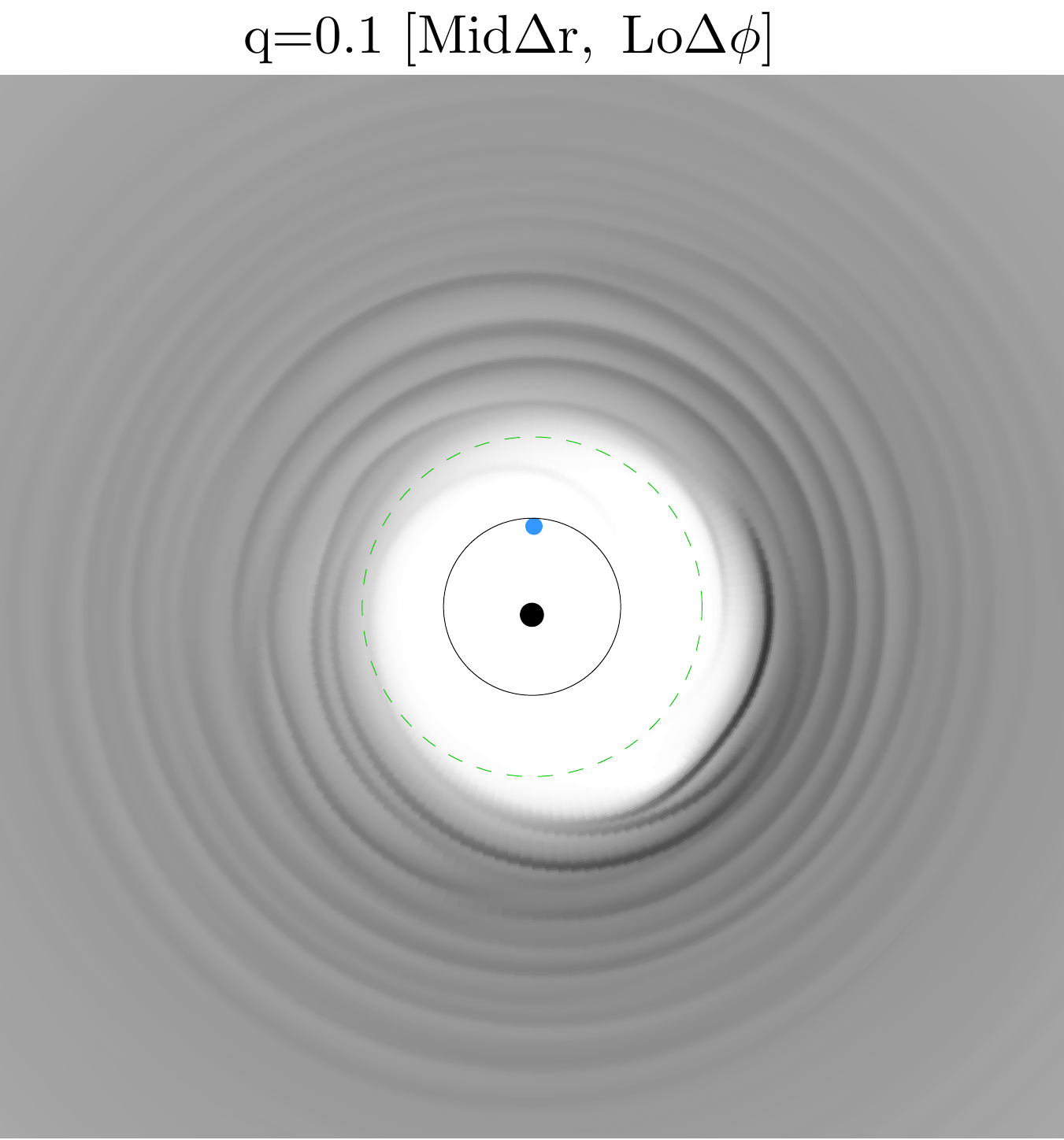} \\
\includegraphics[scale=0.29]{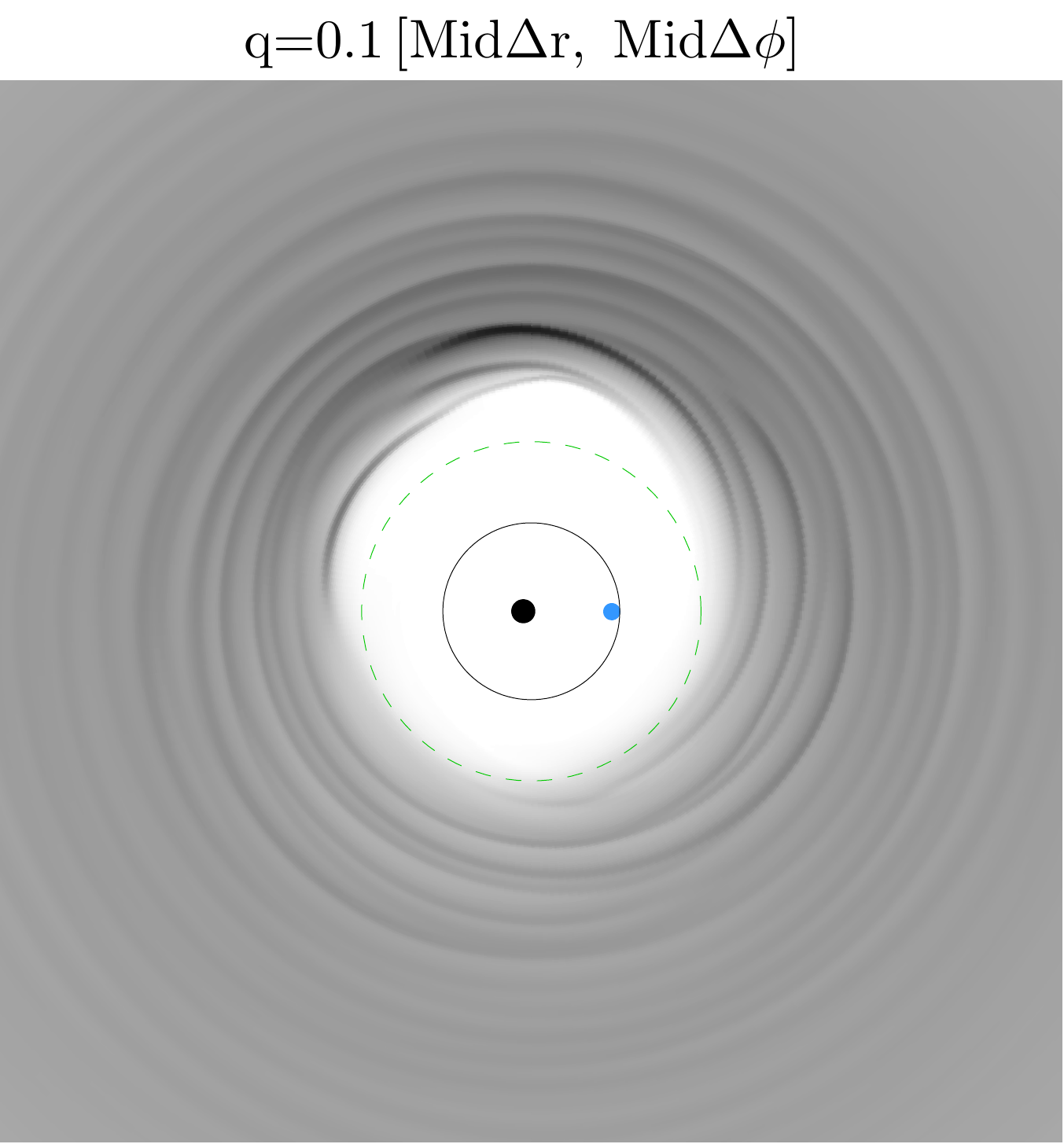} &
\includegraphics[scale=0.29]{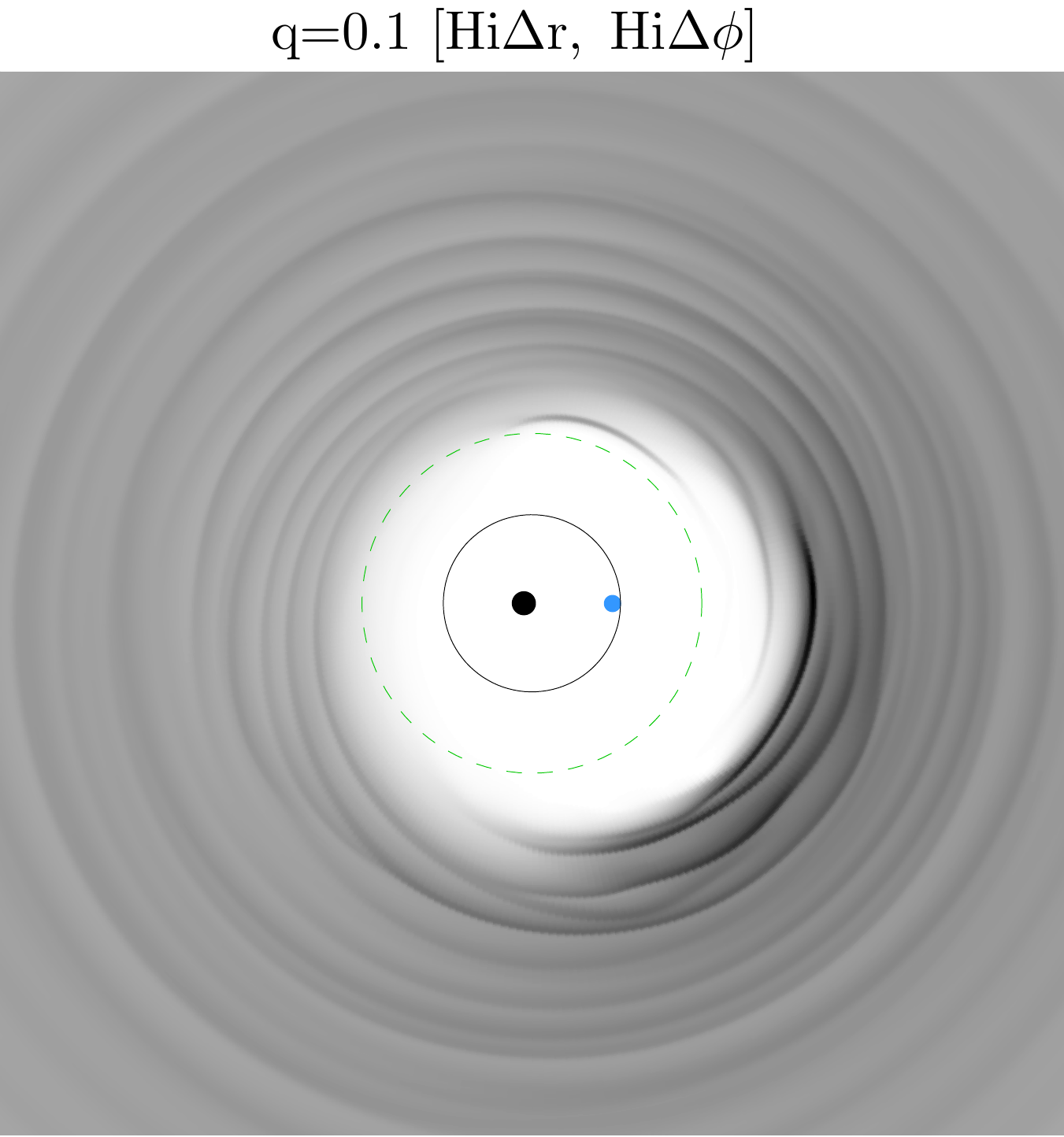} 
\end{array}$
\end{center}
\caption{Two-dimensional surface density distributions during the quasi steady-state phase,
  as in Figure~\ref{2DDensProf}, except for the single mass ratio
  $q=0.1$, and for four different combinations of high/low radial and
  azimuthal resolutions, as labeled.  Increasing the spatial
  resolution decreases numerical diffusion, leads to sharper features,
  and allows stronger accretion streams.  The stronger streams lead to
  more over-dense lumps where the regurgitated streams hit the cavity
  wall.  As a result, the cavity becomes larger and more lopsided as
  the resolution is increased. In the lowest resolution case, the cavity never becomes lopsided.}
\label{q012DDensRes}
\end{figure}

Figure~\ref{q012DDensRes} gives a visual impression of the surface
density distribution at the four different combinations of resolutions
for $q=0.1$.  They show a clear
trend: as the resolution is increased, the lumps near the cavity wall
become sharper and more over-dense, and the cavity becomes larger and
more lopsided.\footnote{Figure~\ref{RCE} shows that the {\em average}
  position of the cavity wall, found from the azimuthally averaged
  torques in equation (\ref{TrqBalRCE}), is much less affected over
  the range of resolutions studied here.}  This trend can be
attributed to numerical dissipation which is most prominent at shocks, \textit{i.e.} where regurgitated streams
impact the cavity wall.  Increasing the resolution implies weaker
numerical diffusion, stronger accretion streams, more momentum
carried by these streams into the disc, and an overall more efficient driving of the
$m=1$ mode. 
This is further evidenced by an earlier onset of the elongated mode for the higher resolution $q=0.1$ simulations. The $q=0.1$ simulations develop an elongated cavity after $\sim 1500$ (highest resolution), $\sim 2500$ (medium resolution), and $\sim 3500$ (fiducial resolution) binary orbits. The lowest resolution run never develops an elongated cavity even after $\sim 7000$ binary orbits. 

The 2D surface density profiles for the $q=1.0$ runs at different resolutions remain qualitatively the same as the fiducial resolution counterpart. We do find that as the azimuthal  resolution is increased the cavity becomes slightly more elongated likely due again to more efficient stream impacts.

The 2D surface density profiles for the $q=0.05$ runs at different resolutions remain qualitatively the same for all resolutions except the highest resolution. For the highest resolution $q=0.05$ run, the disc transitions in to the single timescale regime only after $\sim 4000$ binary orbits and resembles the $q=0.075$ disc at the fiducial resolution.

\begin{figure*}
\begin{center}$
\begin{array}{cc}
\includegraphics[scale=0.43]{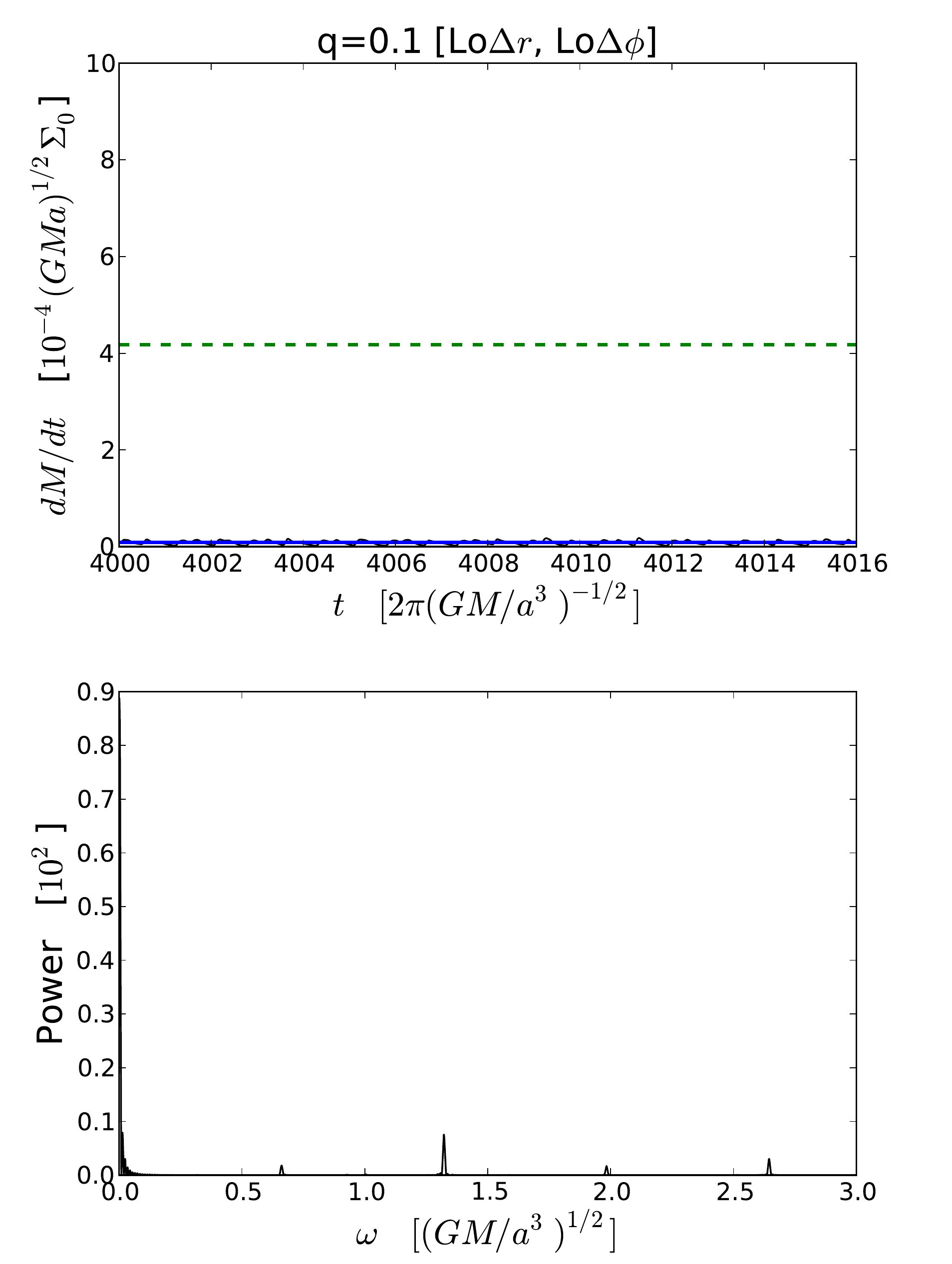}&
\includegraphics[scale=0.43]{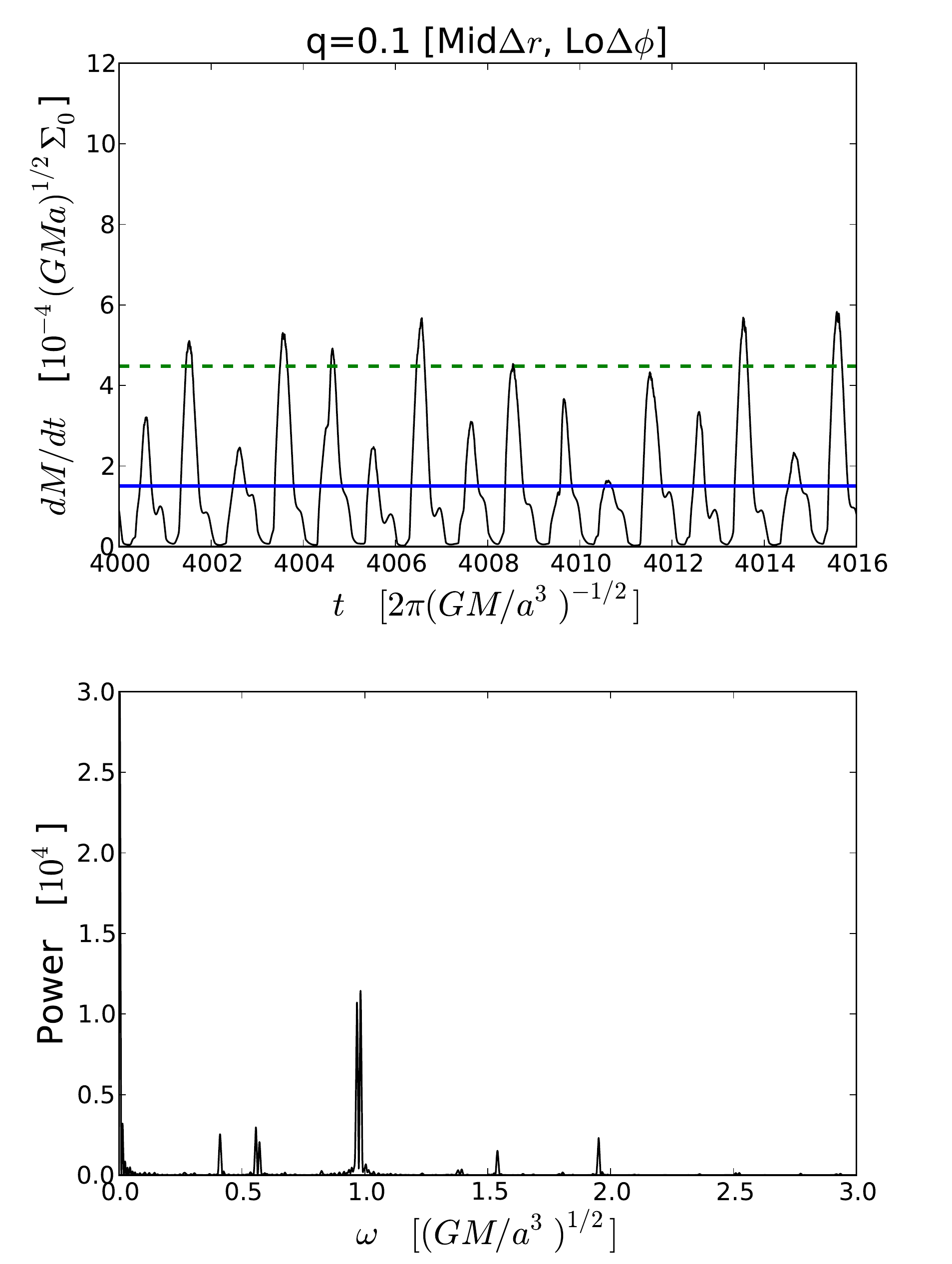} \\
\includegraphics[scale=0.43]{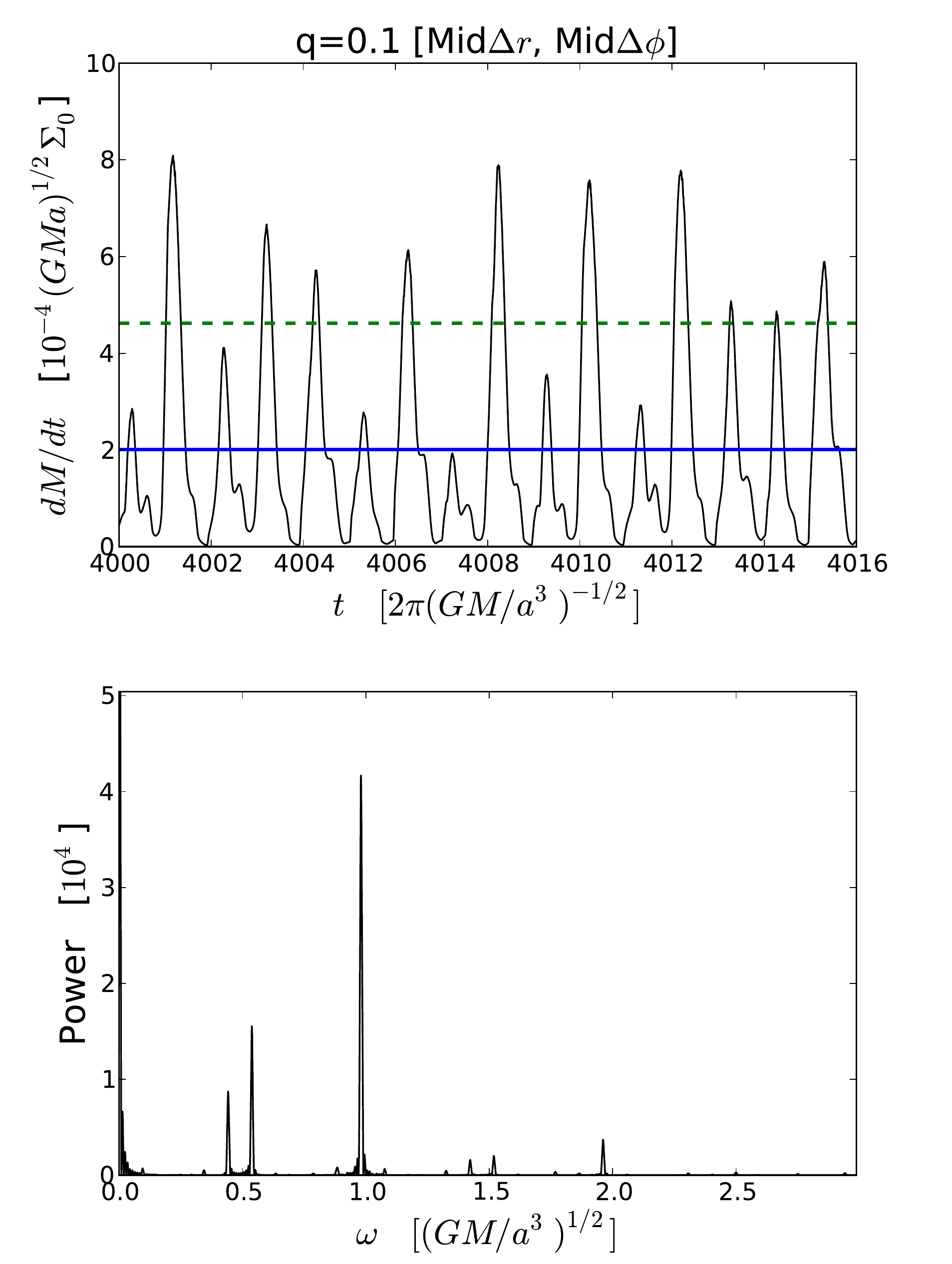} &
\includegraphics[scale=0.43]{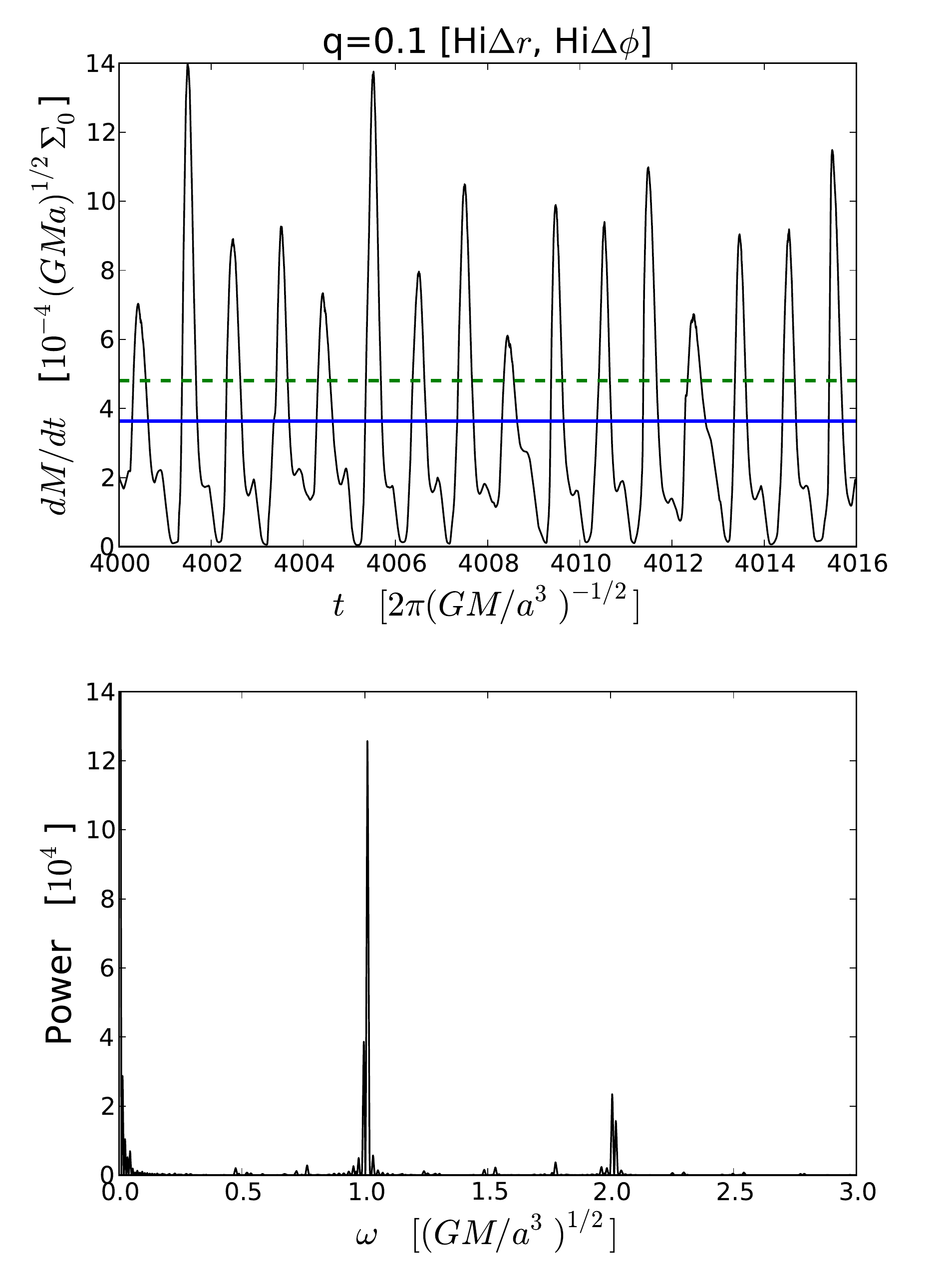}
\end{array}$
\end{center}
\caption{The time-dependent accretion rates as in Figure \ref{Mdott6},
  except for the single mass ratio $q=0.1$ and the four different
  resolutions also shown in Figure \ref{q012DDensRes}. The $q=0.1$
  binary is at the cusp of the transition from the three-timescale to
  the single-timescale regime, and is particularly sensitive to
  resolution, as seen especially in the maxima of the accretion
  spikes.}
\label{q01Res}
\end{figure*}

Figure~\ref{q01Res} shows the corresponding time-dependent accretion
rates at the inner boundary of the $q=0.1$ simulations at different resolutions. The accretion patterns
look visibly different for the lowest resolution run which, for reasons stated above never excites the lopsided cavity mode. 
However, encouragingly, the difference between our
fiducial [Mid$\Delta r$,Lo$\Delta \phi$] and the highest-resolution
[Hi$\Delta r$, Hi$\Delta\phi$] cases is modest with a primary trend of increasing accretion rate relative to the $q=0$ rate, with increasing resolution (see Fig.~\ref{Mdot_ratios_plot} and
Table~\ref{Mdot_ratios}).

The accretion rates for the $q=1.0$ runs at different resolutions remain qualitatively the same as their fiducial resolution counterparts. One difference is that the higher and lower (Figure \ref{MM08_compare}) resolution $q=1.0$ runs have more power at the cavity wall periodicity than the fiducial run; the highest resolution run having more power at the cavity wall frequency than at the $2 t_{\rm bin}$ frequency. 
Also, the cavity wall period becomes slightly longer as the resolution increases, from $5.3 t_{\rm bin}$ to $5.7 t_{\rm bin}$ to $5.9 t_{\rm bin}$, to $6.4 t_{\rm bin}$ for the  [Lo$\Delta r$,Lo$\Delta \phi$],  [Mid$\Delta r$,Lo$\Delta \phi$], [Mid$\Delta r$,Mid$\Delta \phi$], and [Hi$\Delta r$,Hi$\Delta \phi$]  runs respectively. A larger cavity again indicates that higher resolution allows more efficient elongation of the cavity. 
This is further evidenced by an earlier onset of the elongated mode for the higher resolution $q=1$ simulations. The two highest resolution $q=1$ simulations develop an elongated cavity at $\sim 1500$ binary orbits where as the two lower resolution runs develop the elongated cavity only after $\sim 2500$ binary orbits.

The accretion rates for the $q=0.05$ runs at different resolutions again remain nearly identical for all resolutions except the highest resolution. For the highest resolution $q=0.05$ run, the disc transitions in to the single, orbital-timescale regime after $\sim 4000$ binary orbits and exhibits modulation of the accretion rate at the orbital frequency, mimicking the $q=0.075$ accretion rates at the fiducial resolution.

Table~\ref{Mdot_ratios} and the corresponding
Figure~\ref{Mdot_ratios_plot} shows the time-averaged accretion rates
as a function of $q$ at different resolutions.  In all cases, at the
same fixed $q$, we find that increasing resolution produces a higher
accretion rate.  This is consistent with the interpretation above that
higher resolution allows stronger accretion streams.  Interestingly,
we find a strong correlation between the values of $\dot{M}_{\rm
  bin}/\dot{M}_{q0}$ listed in Table~\ref{Mdot_ratios} and the
accretion patterns seen in Figure~\ref{q01Res}: runs at different
resolutions but with similar values of $\dot{M}_{\rm bin}/\dot{M}_{q0}$ 
have very similar accretion patterns (including
the variability and the values of the maxima).  The result of
increasing [decreasing] the resolution can therefore be interpreted as
a shift of the accretion behavior to lower [higher] mass ratios.

Comparing the full set of mass ratio runs for the lowest resolution to the fiducial resolution runs, we observe the same progression of the accretion rate through each of the accretion variability regimes discussed above; a difference being that, as discussed in the previous paragraph, the boundaries between each regime are delineated at larger mass ratios in the lowest resolution runs. We also notice that in the three-timescale regime, there is more power in the periodogram peak associated with the cavity wall orbital period (\textit{e.g.} compare the top left of Figure \ref{Mdott6} with Figure \ref{MM08_compare}). The cavity wall peak still disappears for $q=0.25$ when the stream due to the primary becomes much smaller than the stream due to the secondary and the overlapping of large streams at the cavity wall no longer generates an over-dense lump.

Encouragingly, the two higher resolution runs at $q=1.0$,
also plotted in Figure~\ref{Mdot_ratios_plot}, lie closer to each
other than the two lower resolution runs. Since the resolution steps
are evenly spaced, we consider this evidence that the simulations are
converging monotonically with resolution. Since the $q=0.1$ and 
$q=0.05$ discs are positioned at the boundary of different accretion 
regimes we find a large dependence of disc response and accretion rate on resolution, but with a clear 
trend of increasing resolution moving the boundaries between the 
accretion regimes described in this study to slightly lower values of the binary mass ratio.

\subsection{Physical Regime: Black Hole Binary Parameters}
\label{Physical Regime}

\begin{figure*}
\begin{center}$
\begin{array}{cc}
\includegraphics[scale=0.45]{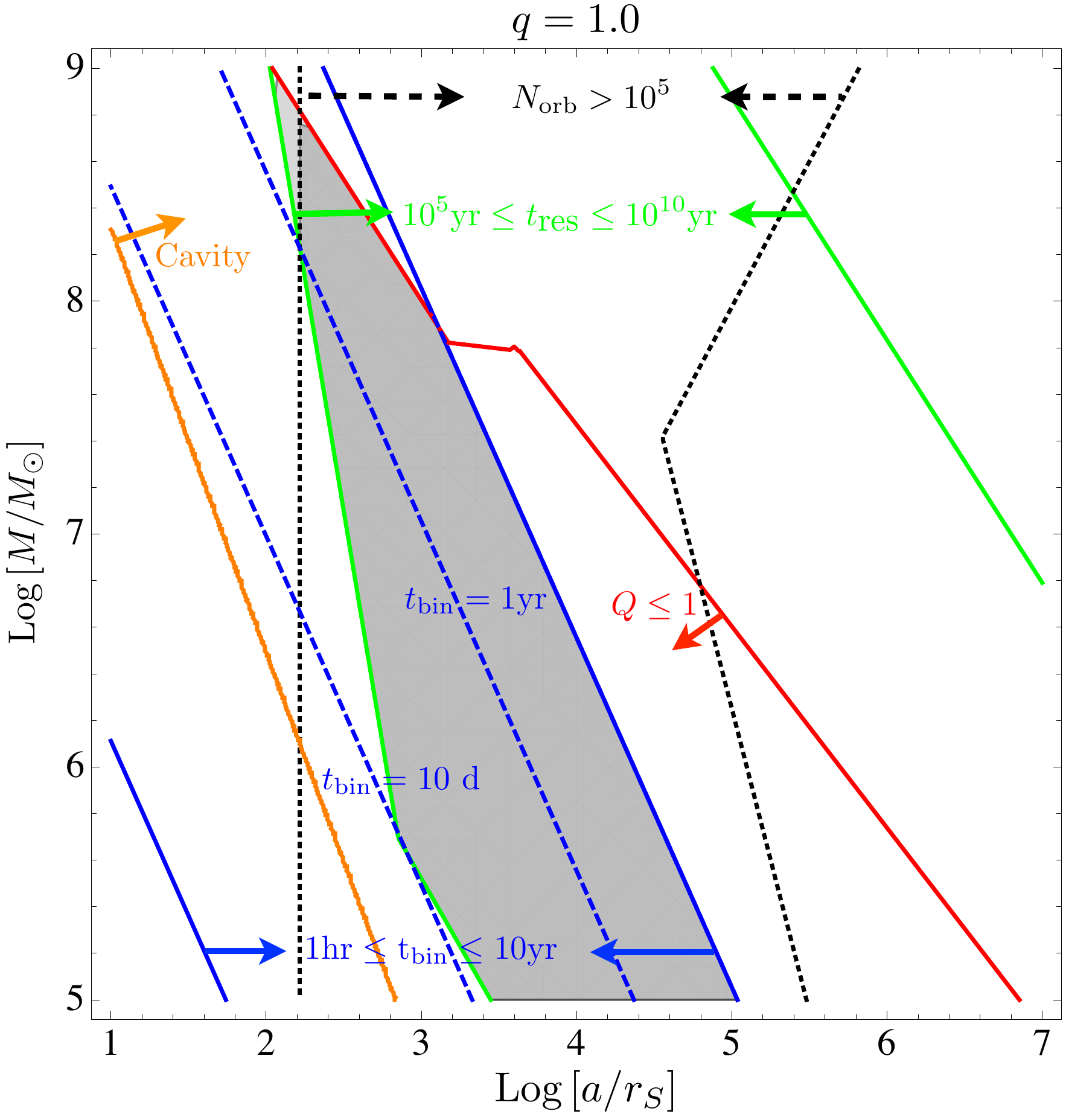}&
\includegraphics[scale=0.45]{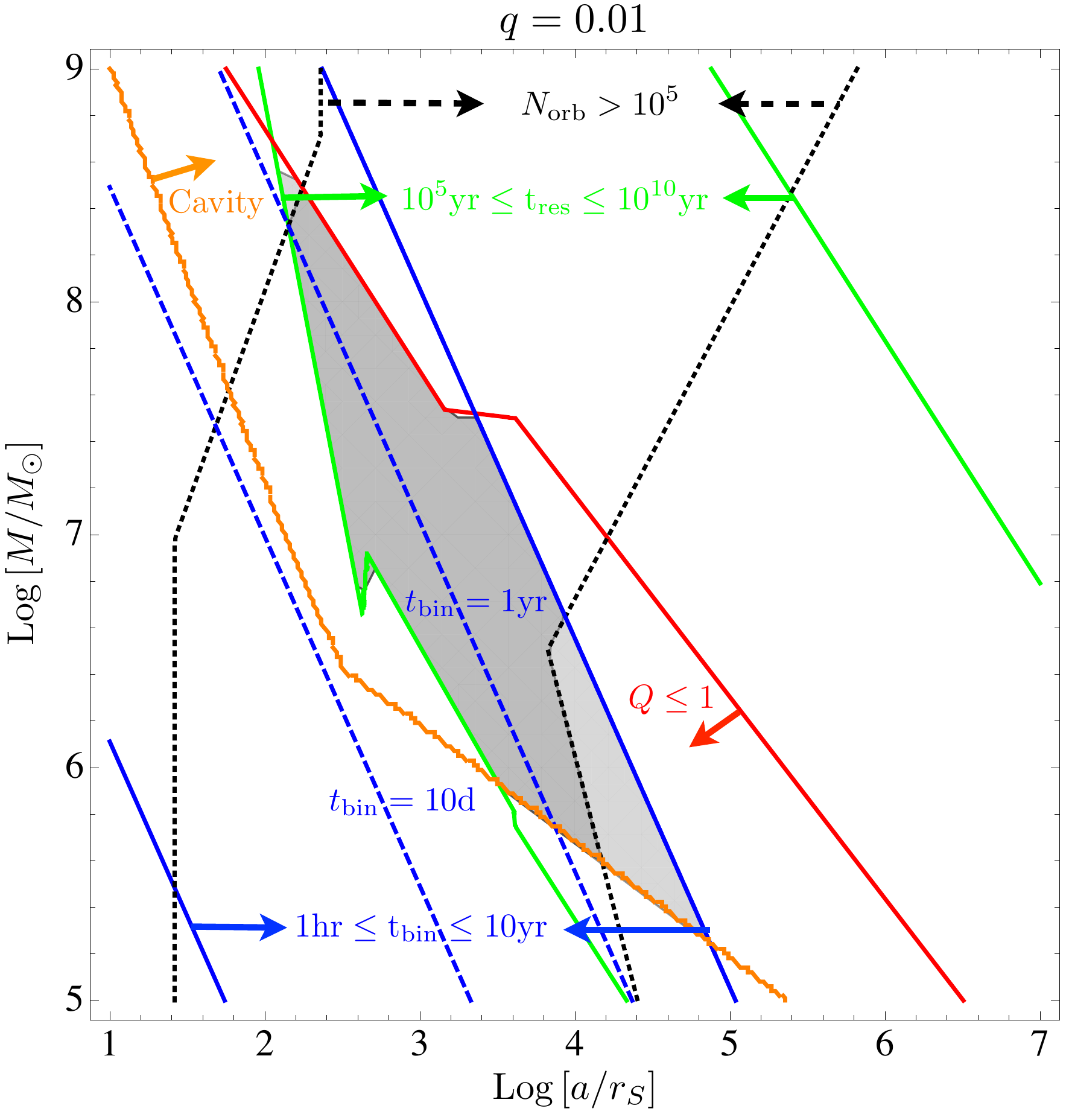}  
\end{array}$
\end{center}
\caption{The shaded regions in each panel denote the values of binary total mass $M$ and separation $a$ for which a binary + disc system would be physically viable and observationally interesting as well as meet simulation specific constraints. Blue lines denote contours of binary orbital time (a characteristic variability timescale). Green lines denote contours of binary residence time $t_{\rm{res}}\equiv -a (da/dt)^{-1}$ computed from \citealt{HKM09} for migration of the secondary through a gaseous disc as well as gravitational radiation. Dashed black lines denote the simulation specific constraint  that the binary separation not change appreciably over the course of a simulation time. Red lines denote the boundary between gravitationally stable and unstable disc regions via the Toomre Q parameter. Orange lines denote the boundary between binaries which can maintain a cavity and those which will not (computed via the steady-state solutions given by \citealt{Kocsis+2012b}). The left panel is for an equal-mass ratio binary and the right panel is for a binary with mass ratio $q=0.01$. For both plots we use $\alpha=0.1$. Note that for all mass ratios, the most massive binaries do not fit into a gravitationally stable disc. However, this is determined for an undisturbed $\alpha$-disc surrounding the primary; perturbations due to a large secondary would increase the stability of the disc out to larger radii \citep{HKM09}.
In the $q=0.01$ case, less massive, close binaries do not maintain cavities and do not represent systems which are consistent with the initial conditions adopted in this study.}
\label{ObsCBD}
\end{figure*}

The simulations presented above can be scaled, in principle, to any black hole mass
and orbital separation.  In this section, we discuss the physical
scales for which our simulations could be relevant (i.e. physically
viable and observationally interesting). The shaded region in 
Figure \ref{ObsCBD} plots this relevant portion of parameter 
space by imposing the following restrictions.

\begin{enumerate}
	\item{$10^5~{\rm M_\odot} \lsim M_p \lsim 10^9~{\rm M_\odot}$. It is not clear whether smaller BHs exist in galactic nuclei, and, in any case,
	the radiation from such a low--mass BHs would likely be too faint to
	detect.  Likewise, much more massive BHs are known to be rare.} 
	\item{$10^{-2}\lsim q \leq 1$. As we have shown, for the set-up we study (i.e. with a cavity inserted
	by hand into the disc), the accretion pattern converges as we decrease
	the mass ratio to $q=0.01$ and below. In practice, a physical lower limit of $q\sim
	0.01$ may arise from the fact that bound binary BHs can be
	created only in relatively major mergers. 
	In a minor merger, the smaller satellite galaxy may be tidally stripped
	during the early stages of the merger, decreasing the efficiency of dynamical friction, 
	and aborting subsequent binary formation \citep{Callegari+:2009}.
 	Coupled with the well--established correlations
	between the mass of a MBH and its host galaxy, this suggests that the
	$q$--distribution may not extend to values significantly below $q\sim
	0.01$.  \footnote{In principle, less massive BHs may grow from the
 	 accretion discs around the primary \citep{McKernan+2012}; the
	  long-term evolution of such systems would be worthy of further  study.}
	 Figure \ref{ObsCBD} plots the restriction of $a, M$ parameter space for 
	 these limiting mass ratios $q=1$ (left) and $q=0.01$ (right).}
	\item{\textit{The binary is embedded in a thin gaseous disc.} Following a merger of two MBH--harboring galaxies, the MBHs sink to
	the bottom of the new galactic potential via dynamical friction in
	approximately a galactic dynamical timescale
	\citep{Begel:Blan:Rees:1980}. In addition to stellar interactions
	(e.g. \citealt{Preto:2011}), many studies have shown that gas in the
	vicinity of the binary could aid in hardening the binary down to $\ll$
	pc separations \citep[\textit{e.g.}][]
	{Escala:2005,Dotti:2007,Mayer+2007,Lodato:2009,Cuadra:2009,Nixon:2011:LongSim,Chapon+2011}.
	We have assumed that such gas in the vicinity of the binary cools
	efficiently and forms a rotationally supported thin disc.  For a given
	$M_p$ and $M_s$, we are still free to choose a physical distance for
	the orbital radius $a$, which could correspond to a snapshot of the
	binary anywhere along its orbital decay.  The assumption that the binary is embedded in a thin disc allows us to make the following additional constraints on $a$ given $M$:
	\begin{enumerate}
		\item{\textit{The accretion disc is gravitationally stable.} Accretion discs become
		self-gravitating, and unstable to fragmentation, beyond a radius of
		order $ \gsim 10^4 (M/10^7 {\rm M_\odot})^{-1} r_S$ (where $r_S$ is
		the Schwarzschild radius; see, e.g., \citealt{Goodman:2003,HKM09} for the formulae used to generate the $Q\leq1$ criteria in Figure \ref{ObsCBD}).
		Since the binary has to fit inside a gravitationally stable disc, this
		puts an upper limit on the orbital separation denoted by the red lines in Figure \ref{ObsCBD}.}
		\item{\textit{Variability occurs on an observable timescale.} The binary orbital time is given by
		\begin{equation}
		 t_{\rm bin} = \frac{2\pi}{\Omega} = 0.88\, 
		\left(\frac{M}{10^7{\rm M_\odot}}\right)
		\left(\frac{a}{10^3 r_S}\right)^{3/2}\, {\rm yr}.
		\label{eq:torb}
		\end{equation}
		As we have shown, the accretion rate shows periodicity on a timescale
		of $\sim t_{\rm bin}$.  In a realistic survey, it will be feasible to
		look for periodic variations between $0.1\,{\rm hr} \lsim
		t_{\rm bin} \lsim {\rm few}\, {\rm yr}$ denoted by the solid blue lines in Figure \ref{ObsCBD}.  
		Here the lower limit comes from the integration time required to measure the flux variations for
		MBHBs in the above mass range (for a survey instrument with a
		sensitivity similar to LSST; \citealt{HKM09}), and the upper limit
		comes from the duration of proposed time-domain surveys.  As a guide, 
		the dashed blue lines in Figure \ref{ObsCBD} are contours of constant orbital times drawn at 10 days and 1 year.}
		\item{\textit{The binary spends a long time at a given separation.} 
		Assuming that the binary is embedded in a thin disc,
		\citealt{HKM09} compute residence times, $t_{\rm res}\equiv -a
		(da/dt)^{-1}$, as a function of the binary separation $a$, due to 
		migration of the secondary through the disc and due to gravitational wave 
		decay at small enough binary separations. In Figure \ref{ObsCBD} 
		the green lines denote the requirement that $10^5 \leq t_{\rm res} \leq 10^{10}$ years. 
		A residence time of greater than $10^{10}$ years does not on its own exclude a 
		binary system from observation. Nevertheless, we include this limit in order to show 
		which binaries will not merge (due to migration through a gaseous disc) in a 
		Hubble time. Note also that there is a trade-off: a longer residence time is 
		desirable since it increases the probability of finding such a system; however, 
		longer residence times occur at larger separations and longer orbital times, 
		which will make it more difficult to verify any periodic behavior.}
		\item {\textit{A cavity is maintained.} For consistency with the initial conditions 
		adopted here, we require that the binary + disc systems
		will indeed form a cavity during earlier stages of
		their evolution. The region of parameter space for which a cavity 
		may be maintained is denoted by the orange lines in Figure \ref{ObsCBD} 
		and is calculated using the {\em steady-state} disc solutions detailed 
		in \cite{Kocsis+2012b}. }
		\item{\textit{The orbital separation is fixed.} Throughout our simulations, we fix the binary separation; we
		therefore require that the orbital decay should be slow enough for the
		binary's orbit not to change significantly over a few thousand
		orbits. This is denoted by the dashed black lines in Figure \ref{ObsCBD} 
		which are drawn where $N_{\rm orb} = t_{\rm res}/t_{\rm bin} =10^5$.} 
	\end{enumerate}  }
	\item{Though not expressed in Figure \ref{ObsCBD}, for our simulations to be self-consistent, we also require $a\gsim 100
	r_S$, since our Newtonian treatment ignores general relativity.
	Furthermore, at approximately the same binary separation, the orbital
	decay of the binary due to gravitational wave emission becomes more
	rapid than the viscous time at the edge of the cavity. As a result,
	the disc decouples from the binary and is `left behind', rendering our
	initial conditions inconsistent in this regime
	(e.g. \citealt{Milos:Phinney:2005}; although see
	\citealt{FarrisGold:2012} and \citealt{Noble+2012} whose MHD
	simulations suggest that the gas can follow the binary down to smaller
	separations).}
\end{enumerate}

\subsection{Caveats}
\label{Caveats}

However instructive, the simulations presented here are still of
course simplified models of a real binary disc system, and it is worth
listing some major caveats.

\begin{enumerate}
\item Our simulations are two dimensional -- we expect that the 3D
  vertical structure could modify the structure of the accretion
  streams, including their $q$-dependence. However, note that \citealt{Roedig:2012} 
  find similar features in the accretion rate periodograms measured from 3D simulations.
  
\item Our discs are assumed to have angular momentum co-aligned with that of the binary (prograde discs). In principle, a random accretion event onto the binary could result in a misaligned disc which will eventually be torqued into co- or counter-alignment with the binary angular momentum \citep{Nixon:2011:Lett}. 

\item Our simple $\alpha$-viscosity prescription may be inaccurate,
  especially in the nearly radial accretion streams. Recent MHD
  simulations in the $q=1$ case find a larger effective $\alpha$
  than the fiducial value adopted here \citep{ShiKrolik:2012, Noble+2012}.  
  It is interesting, however, that our highest $\alpha=0.1$ simulation for 
  an equal mass binary exhibits the same variability as our fiducial case but a larger accretion rate, 
  in agreement with the above mentioned 3D MHD simulations.

\item We assumed a locally isothermal equation of state; more
  realistic equations of state could have an especially large impact
  on the strength and dissipation of shocks at the cavity wall.
  
\item Our initial conditions correspond to an unperturbed,
  near-Keplerian, circular disc, and circular binary orbit, with a
  significant pile-up of gas.  In reality, accretion onto the binary
  could produce significant binary (as well as disc) eccentricity
  \citep[\textit{e.g.}][]{Cuadra:2009, Lodato:2009, RoedigDotti:2011}.
  In this case, the variability we find would most likely have a more
  complex structure \citep[\textit{e.g.}][]{Hayasaki:2007} due to the
  plethora of resonances available in an eccentric binary potential
  \citep[\textit{e.g.}][]{Artymowicz:1994}. 
In a study of circumbinary discs around eccentric binaries (with mass ratio $1/3$), \citealt{RoedigDotti:2011} find similar accretion rate periodograms as in this study; periodogram peaks exist at the orbital frequency, twice the orbital frequency and the cavity wall orbital frequency. They find that an increasingly eccentric binary: (a) increases the size of the cavity and thus decreases the overall magnitude of the accretion rate, (b) enhances power at harmonics of the orbital frequency, and (c) increases power in a peak located at the beat frequency of the orbital frequency and the cavity wall frequency.
\item As previously stressed, these simulations do not allow the binary orbit to evolve in response to forces exerted by the gas disc. When this assumption of a massless disc is lifted, in addition to changes in binary eccentricity and semi-major axis, the binary center of mass could oscillate around the disc center of mass due to the orbiting eccentric disc. For massive discs, the above effects could alter the description of disc evolution, and hence accretion, presented in this zero disc-mass study.
\item Although we begin with an ``empty'' cavity, this cavity may 
  overflow already at a large radius \citep{Kocsis+2012b}.
  Future studies should construct a self-consistent initial density
  profile, by evolving the binary's orbit from large radius through
  gap clearing, and thus determining whether a true pile-up occurs.
\item We have ignored the radiation from the gas accreted onto the
  BHs.  Given that we find high accretion rates -- comparable to those
  for a single BH -- the secondary BH can be fed at super-Eddington
  rates, and the flow dynamics can then be strongly affected by the
  radiation.
\item We have not allowed accretion onto the BHs, and have excised the
  inner region $r<a$ from the simulation domain.  This could have an
  impact on the dynamics of the streams that are flung back towards
  the cavity wall, and therefore on the formation of the dense lump,
  the lopsidedness of the cavity, and the variable accretion patterns.
\end{enumerate}

These caveats should all be pursued in future work, to assess the
robustness of our results.  We expect that our main conclusions,
namely that the accretion rate is strongly modulated by the binary,
and that the power-spectrum of the accretion shows distinct periods,
corresponding to the orbital periods of the binary and the gas near
the location of the cavity wall, will be robust to all of the above
caveats.  However, the numerical values, such as the mean accretion
rate, and the critical value of $q$ for the transition between
variable and steady accretion, will likely be affected.

\section{Conclusions}
\label{Conclusions}

We have investigated the response of an accretion disc to an enclosed
binary via two-dimensional, Newtonian, hydrodynamical simulations. As
previous work has shown (\citealt{Artymowicz:1994, Hayasaki:2007};
MM08; \citealt{Cuadra:2009, ShiKrolik:2012, Roedig:2012}), for
non-extreme mass ratios, the binary carves out a cavity in the disc,
but gas still penetrates the cavity in streams which possibly accrete
onto the binary components. Here we have followed up on the work of
MM08 by investigating the nature of this inflow across the
circumbinary cavity, as a function of binary mass ratio $q$.  We have
simulated 10 different mass ratios in the range $0.003 \leq q \leq
1$. This corresponds to the expected range of $q$-values for massive
BH binaries produced in galaxy-galaxy mergers.

We find that while the binary `propellers' are effective at
maintaining a low-density cavity at the center of the disc, they can
not efficiently suppress accretion across the cavity.  
For $q=1$, the average accretion rate is on order of $2/3$ that of a singe BH 
with accretion spikes of $\sim3$ times larger.  As long
as the circumbinary disc is fueled at a near-Eddington rate from large
radius, these binaries could therefore have quasar-like luminosities.
This should facilitate finding counterparts to GW events
\citep{Kocsis+2006}, and should also allow their detection in
electromagnetic surveys \citep{HKM09}.

We have found that the accretion is not only strong, but can be
strongly variable (by a factor of $\sim3$), with a characteristic $q$-dependent frequency
pattern.  While the accretion for $q<0.05$ is steady, for $q\gsim
0.05$ there is a strong modulation by the binary, and a clear
dependence on $q$ of both the variability pattern, and the magnitude
of the time-averaged accretion rate.  For an equal-mass binary, the
accretion rate is modulated at twice the orbital frequency and $\sim
1/6$ the orbital frequency.  As the mass ratio is lowered, the power
in the $1/2 t_{\rm bin}$ and $(5-6)t_{\rm bin}$ variability timescales is
reduced, and traded for a third variability timescale at $t_{\rm
  bin}$.  In the range $0.05 \lsim q \lsim 0.25$, the single
$t_{\rm bin}$ timescale is dominant.

Increasing the magnitude of viscous forces has little effect on the above findings except to increase the magnitude of the accretion rate (both absolute and relative to $q=0$) and to bring out a long-term accretion variability timescale with a periodicity of $400 t_{\rm bin}$. However, accretion discs with even larger viscous forces could quench the $(5-6)t_{\rm bin}$ variability timescale if the over-dense lump responsible for its generation can be broken up before it repeats $\sim$ an orbit at the cavity wall. Hence, further investigation into the effects of viscosity are warranted.

Strong and highly variable accretion, with characteristic
frequencies, should aid in identifying massive BH binaries in galactic
nuclei.  The presence of two frequencies, in the ratio 1:2 for unequal-mass binaries ($0.1 \leq q < 1$), is an
especially robust prediction that is {\em independent of disc
  properties}, and could serve as a `smoking gun' evidence for the
presence of a binary.  Our results suggest that the ratio of the power
at these two frequencies could probe the mass ratio $q$, while other
features of the periodogram could probe properties of the disc, such
as its viscosity.

The variability time-scales are on the order of the orbital period,
and we have argued that the most promising candidates in a blind
electromagnetic search would be those with total mass and separation contained in the shaded regions of Figure \ref{ObsCBD}; $10^{6-7} {\rm M_\odot}$ binaries,
preferably with orbital periods of days to weeks.  The time-variable accretion to
the central regions could produce corresponding variability in
broad-band luminosities, allowing a search in a large time-domain
survey, such as LSST, without spectroscopy.  Additionally, the
emission lines could exhibit periodic shifts in both amplitude and
frequency; kinematic effects from the binary's orbit could be
distinguished from those due to the fluctuating accretion rate,
whenever the latter contains multiples of the binary period.

A few percent of the accretion streams generated periodically fuel the BHs,
but the majority of the stream material is flung back and hits the accretion
disc farther out. The shocks produced at these impact sites are
prominent for $q\gsim 0.1$, and can provide additional observable
signatures.  In particular, radiation from these shocks should be
temporally correlated with the luminosity modulations arising near the
secondary and/or primary BH, with a delay time on the order of a
binary orbit.

GW observatories, such as eLISA, and Pulsar Timing Arrays will be able to constrain the mass
ratios of in-spiraling MBHB's at the centers of galactic nuclei
ab-initio, providing a template for the expected variability pattern.
This should be helpful in identifying the unique EM counterpart among
the many candidates (with luminosity variations) in the eLISA/PTA error
box, as the source with a matching period.

In summary, our results imply that massive BH binaries can be both
bright and exhibit strongly luminosity variations, at the factor of
several level. This raises the hopes that they can be identified in a
future, suitably designed electromagnetic survey, based on their
periodic variability.  Although encouraging, these conclusions are
drawn from simplified 2D hydrodynamical models of a real binary disc
system, and should be confirmed in future work.

\section*{Acknowledgments}

We thank Brian Farris and Bence Kocsis for useful discussions.  We thank Paul
Duffell for useful discussions, for verifying some of our results
in independent runs with the code DISCO, and also for suggesting 
the Cartesian shear flow test of our viscosity implementation. We also thank 
the anonymous referee for constructive comments which have improved this manuscript. 
We acknowledge support from NASA grant NNX11AE05G (to ZH and AM). 
DJD acknowledges support by a National Science Foundation Graduate 
Research Fellowship under Grant No. DGE1144155.

\bibliography{Dan}

\appendix
\section{Viscous implementation in polar coordinates}
\subsection{Viscosity and the Momentum Equation}
We may write the momentum equation in component form and with respect to an arbitrary basis as
\begin{equation}
\partial_t(\rho v_i) = -\nabla_j \Pi_{ i}^{\ j}
\label{MvEqn} 
\end{equation}
where $\nabla$ represents the covariant derivative,
\begin{equation}
\Pi_{ij}  = P g_{ij} + \rho v_i v_j - \sigma_{ij}
\label{MvTensor}
\end{equation}
are the components of the momentum flux density vector, P is the mechanical pressure, and $g$ is the metric tensor. The first two terms represent a reversible momentum flux due to pressure forces and mechanical transport of the fluid. The last term expresses a non-reversible momentum flux due to viscous forces via the \textit{viscous stress tensor} $\sigma$. Writing out (\ref{MvEqn}) with (\ref{MvTensor}),
\begin{equation}
\partial_t(\rho v_i) = \nabla_j (P g_{ i}^{\ j} ) +  \nabla_j (\rho v_i v^j - \sigma_{ i}^{\ j} ) \nonumber
\end{equation} 
or in vector notation
\begin{equation}
\partial_t(\rho \mathbf{v}) = -\nabla P -  \nabla \cdot (\rho \mathbf{v v} - \sigma) 
\label{MvVector}
\end{equation} 
Thus we see that the effects of viscosity can be incorporated by computing viscous momentum fluxes from $\sigma$ and subtracting them from the mechanical transport term $\rho \mathbf{v v}$. This is what FLASH does currently to incorporate the effects of viscosity in Cartesian coordinates. However, in non-Cartesian coordinates, there will also be geometric source terms from taking the divergence of the rank-two-tensors $\mathbf{v v}$ and $\sigma$. Thus we must compute not only the components of these terms, but also the divergence in order to identify geometric source terms.

\subsection{The Form of the Viscous Stress Tensor}
The viscous stress tensor has components \citep{LandauLifschitz:Fluids}
\begin{equation}
\sigma_{ij} = \rho \left[ \nu\left(\nabla_i v_j + \nabla_j v_i \right) + \left(\zeta - \frac{2}{D} \nu \right) g_{ij} \nabla_l  v^l   \right]
\label{ViscTensor}
\end{equation}
where $\nu$ is the kinematic coefficient of viscosity, $\zeta$ is the bulk coefficient of viscosity, and D is the number of spatial dimensions. The $2/D$ factor is chosen so that only the bulk viscosity term survives upon taking the trace of $\sigma$.

\subsection{Components of $\sigma$ in Polar Coordinates}
To compute the viscous stress tensor components we work in a coordinate basis to evaluate the covariant derivatives in terms of Christoffel symbols,
\begin{equation}
\nabla_j T_i = \partial_j T_i - \Gamma^k_{\ ij}T_k \nonumber
\end{equation} 
Working in 2D polar coordinates, $(r, \phi)$, the non-zero Christoffel symbols are
\begin{equation}
 \Gamma^r_{\ \phi \phi} = -r \qquad  \Gamma^{\phi}_{\ r \phi}=\Gamma^{\phi}_{\  \phi r} = \frac{1}{r} \nonumber
\end{equation} 
In the polar coordinate basis Eqs.\ (\ref{ViscTensor}) become
\begin{align}
\sigma_{rr} &= \Sigma \left[ 2 \nu \partial_r v_r + \left( \zeta -1 \right) \nabla \cdot v \right]   \nonumber \\
\sigma_{r \phi} &= \Sigma \left[  \nu \left( \partial_r v_{\phi} + \partial_{\phi}v_r  - \frac{2}{r} v_{\phi} \right) \right] \nonumber \\
\sigma_{\phi \phi} &= \Sigma \left[ 2 \nu \left(\partial_{\phi} v_{\phi} + r v_r \right) + \left( \zeta - 1\right) r^2 \nabla \cdot v \right]  \nonumber 
\end{align}
where $\Sigma$ is the height integrated 2D surface density, $v^{\phi}=\Omega$ and $v_{\phi}= r^2 \Omega$, $\Omega$ being the angular frequency.
Transforming to an orthonormal basis (used in FLASH)
\footnote{For this conversion one needs to contract the coordinate tensor components with the orthonormal components of the coordinate basis vectors. This simply amounts to multiplying each $\phi$-up component by $r$ and each $\phi$-down component by $1/r$ (\textit{e.g.} $v^{\hat{\phi}} = r v^{\phi} = v_{\hat{\phi}} = v_{\phi}/r = r \Omega$).}
these components become,
\begin{align}
\sigma_{\hr \hr} &= \Sigma \left[ 2 \nu \partial_r v_{\hr} + \left( \zeta -1 \right) \nabla \cdot v \right]  \nonumber \\
r \sigma_{\hr \hp} &= \Sigma \left[  \nu \left( \partial_r \left(r v_{\hp} \right)+ \partial_{\phi}\left(r v_{\hr} \right) - \frac{2}{r}\left( r v_{\hp} \right) \right) \right]  \nonumber \\
r^2 \sigma_{\hp \hp} &=  \Sigma \left[ 2 \nu \left(\partial_{\phi}(r v_{\hp}) + r v_{\hr} \right) + \left( \zeta - 1 \right) r^2 \nabla \cdot v \right]  \nonumber 
\end{align}
where $v^{\hat{\phi}}=\vp= r \Omega$. Since the value of the bulk viscosity coefficient $\zeta$ is somewhat arbitrary, 
we set it to 0. Then simplifying the above using $\nabla \cdot v = \partial_r v_r + \frac{1}{r} \partial_{\phi} v_{\phi} + \frac{v_r}{r}$,
\begin{center}
\begin{align}
\sigma_{\hr \hr} = \Sigma \nu \left[  \partial_r \vr -  \frac{1}{r}\partial_{\phi} \vp -  \frac{\vr}{r} \right]  \nonumber \\
\sigma_{\hr \hp} =  \Sigma \nu \left[  \partial_r \vp +  \frac{1}{r}\partial_{\phi} \vr -  \frac{\vp}{r} \right]  \nonumber \\
\sigma_{\hp \hp} = \Sigma \nu \left[   \frac{1}{r}\partial_{\phi} \vp -  \partial_r \vr + \frac{\vr}{r} \right] 
\label{OrthComps}
\end{align}
\end{center}

\subsection{Divergence of $\sigma$ in Polar Coordinates}
To compute the geometric source terms which will modify the 2D polar momentum equation we compute the divergence of the second rank tensor $\sigma$. Starting again in a coordinate bases we may write,
\begin{equation}
\left( \nabla \cdot \sigma \right)^i = \partial_j \sigma^{ji} + \Gamma^j_{\ kj} \sigma^{ki} + \Gamma^i_{\ kj} \sigma^{kj}
\end{equation}
giving us the components of the viscous force
\begin{align}
\left( \nabla \cdot \sigma \right)^r &=  \partial_r \sigma^{rr} + \partial_{\phi} \sigma^{\phi r} + \frac{1}{r}\sigma^{rr} - r\sigma^{\phi \phi} \nonumber \\
\left( \nabla \cdot \sigma \right)^{\phi} &=  \partial_r \sigma^{r\phi} + \partial_{\phi} \sigma^{\phi \phi} + \frac{3}{r}\sigma^{r \phi} \nonumber
\end{align}
Transforming again to an orthonormal basis for implementation in FLASH,
\begin{align}
\left( \nabla \cdot \sigma \right)^{\hr} &=  \partial_r \sigma^{\hr \hr} + \partial_{\phi}  \frac{1}{r}\sigma^{\hp \hr} + \frac{1}{r}\sigma^{\hr \hr} - r \left( \frac{1}{r^2} \sigma^{\hp \hp} \right) \nonumber \\
\frac{1}{r}\left( \nabla \cdot \sigma \right)^{\hp} &=  \partial_r  \left( \frac{1}{r}\sigma^{\hr \hp} \right) + \partial_{\phi}  \frac{1}{r^2} \sigma^{\hp \hp} + \frac{3}{r}  \left( \frac{1}{r} \sigma^{\hr \hp}\right) \nonumber.
\end{align}
Simplifying,
\begin{center}
\begin{align}
\left( \nabla \cdot \sigma \right)^{\hr} =  \frac{1}{r} \partial_r \left( r \sigma^{\hr \hr} \right) + \frac{1}{r}\partial_{\phi} \left( \sigma^{\hp \hr} \right) - \frac{1}{r} \sigma^{\hp \hp}  \nonumber \\
\left( \nabla \cdot \sigma \right)^{\hp} = \frac{1}{r} \partial_r \left( r \sigma^{\hr \hp} \right) + \frac{1}{r}\partial_{\phi} \left( \sigma^{\hp \hp} \right)+ \frac{1}{r} \sigma^{\hr \hp}    .
\label{NonAngDiv}
\end{align}
\end{center}
Plugging in the values of the components from (\ref{OrthComps}) we have,
\begin{center}
\begin{align}
(\nabla \cdot \sigma)_{\hat{r}} = \frac{1}{r} \partial_{r} \left[ r \Sigma \nu \left(  \partial_r \vr - \frac{1}{r}\partial_{\phi} \vp  - \frac{\vr}{r} \right)   \right]  \nonumber \\
+ \frac{1}{r}\partial_{\phi} \left[  \Sigma \nu \left(  \partial_r \vp + \frac{1}{r}\partial_{\phi}\vr  - \frac{\vp}{r}  \right)   \right]  \nonumber \\
- \frac{\Sigma \nu}{r} \left[  \frac{1}{r} \partial_{\phi} \vp  -  \partial_r \vr +  \frac{\vr}{r} \right]  \nonumber \\
(\nabla \cdot \sigma)_{\hat{\phi}} = \frac{1}{r} \partial_{r} \left[ r \Sigma \nu \left( \partial_r \vp + \frac{1}{r}\partial_{\phi}\vr  - \frac{\vp}{r}  \right)   \right]  \nonumber \\
+ \frac{1}{r}\partial_{\phi} \left[ \Sigma \nu \left( \frac{1}{r} \partial_{\phi} \vp  -  \partial_r \vr +  \frac{\vr}{r} \right) \right]  \nonumber \\
+ \frac{\Sigma \nu}{r} \left[  \left( \partial_r \vp + \frac{1}{r}\partial_{\phi}\vr  - \frac{\vp}{r}   \right)   \right]  
\label{Div_sig_Full}
\end{align}
\end{center}
The first two terms in each of the above look like a normal divergence of a rank-one-tensor and the third terms are the geometric source terms that we must add to the momentum equation. They are akin to the centrifugal and Coriolis terms which arise form a similar exercise performed on the $\rho \mathbf{v v}$ term of (\ref{MvVector}). Note that each term has the units of force per volume while the components of the stress tensor have units of density times velocity squared which matches the $\rho \mathbf{v v}$ term in the hydro equations.

\subsection{Implementation in FLASH}
In FLASH 3.2, the components (with respect to the orthonormal basis) of the viscous stress tensor are computed in the routine $\rm{Diffuse\_visc.F90}$. This routine then subtracts the flux indicated by the viscous stress tensor from the $\Sigma \mathbf{v v}$ flux. Note that since FLASH computes fluxes on cell boundaries, the stress tensor components in $\rm{Diffuse\_visc.F90}$ must be computed on the lower face of the current sweep direction (See \textit{e.g.} \citep{Edgar:2006}). We compute the viscous source terms along with the centrifugal and Coriolis source terms in the routine $\rm{hy\_ppm\_force.F90}$.

We test the above implementation in FLASH with two tests, the viscously spreading ring of \cite{Pringle:1981} and a Cartesian shear flow. 

\subsection{Viscously Spreading Ring}
The viscously spreading ring test begins with a delta function initial density distribution which spreads solely due to viscous forces. This test assumes axis-symmetry $\partial_{\phi} \equiv 0$ and exercises only the terms with $\sigma^{\hr \hp}$ in (\ref{Div_sig_Full}). The analytic solution assuming a constant coefficient of kinematic viscosity is well known and given by \cite{Pringle:1981}.

\begin{figure}
\begin{center}
\includegraphics[scale=0.5]{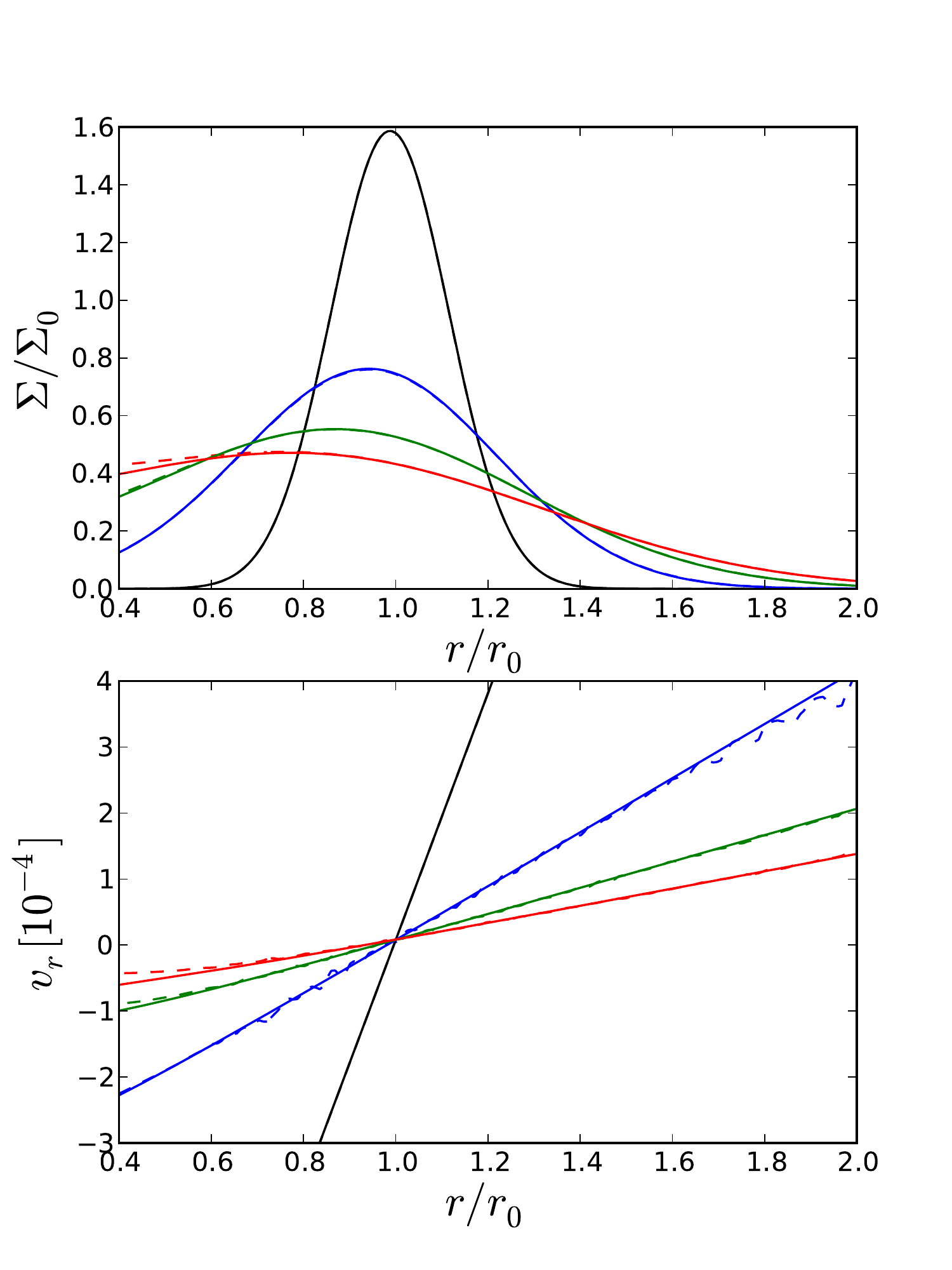} 
\end{center}
\caption{Viscously spreading ring test with constant coefficient of kinematic viscosity. See text for details.}
\label{ViscRing}
\end{figure}

To implement the viscously spreading ring test we choose $\nu = \rm{cst.} =\alpha/\mathcal{M}^2$ with dimensionless parameter $\alpha = 0.1$, and mach number of the disc $\mathcal{M}=100$. This choice of Mach number mitigates pressure effects and also allows time for a small initial transient to pass through the simulation domain without greatly affecting the evolution of the solution (wiggles in the earliest blue-dashed radial velocity curve of Figure \ref{ViscRing}). As required by the analytic solution, we also turn off all terms in (\ref{Div_sig_Full}) which do not include $\sigma^{\hr \hp}$. The outer and inner boundaries are at $r_{\rm{min}}=0.2$ and $r_{\rm{max}}=2.2$ where boundary values are set by the time-dependent analytic values. We use a spatial resolution of 128 radial cells by 64 azimuthal cells and we start with initial conditions corresponding to the dimensionless initial time parameter $\tau = 12 \nu t /r^2_0 = 0.032$, where $r_0$ is the initial position of the delta function ring.

Figure \ref{ViscRing} shows the result of the test. Besides some expected deviation at the inner boundary, the numerical solution (dashed lines) agree well with the analytic solution (solid lines).

\subsection{Cartesian Shear Flow}
The viscously spreading ring test confirms that the most important terms for thin disc accretion (those containing $\sigma^{\hr \hp}$) are implemented properly. The Cartesian shear flow tests all of the terms in (\ref{Div_sig_Full}). The idea is to choose a problem which is analytic in Cartesian coordinates and use the computer to solve it in polar coordinates thereby exercising all of the polar derivatives to make up the simple Cartesian derivatives. 

To set-up the Cartesian shear flow problem, we start with the Navier-Stokes equation for an incompressible fluid with constant coefficient of viscosity $\nu$,
\begin{align}
\partial_t \mathbf{v} + \left(\mathbf{v} \cdot \nabla \right) \mathbf{v} = -  \frac{1}{\Sigma} \nabla P + \nu \nabla^2 \mathbf{v}
\label{NavStoke}
\end{align}
Choosing constant pressure and $\mathbf{v} = v^x(y,t) \mathbf{e_x}$ reduces (\ref{NavStoke}) to a simple 1D diffusion equation in $v_x(y,t)$
\begin{align}
\partial_t v_x(y,t)  =  \nu \partial^2_y v_x(y,t)
\end{align}
A solution is
\begin{align}
v_x(y,t) =\frac{v_0}{\sqrt{2 \pi \nu t}} \rm{exp}\left[{\frac{-(y-y_0)^2}{4 \nu t}}\right]
\label{CShrSoln}
\end{align}
while $v_y$ is 0 for all time.

\begin{figure}
\begin{center}$
\begin{array}{cc}
\includegraphics[scale=0.4]{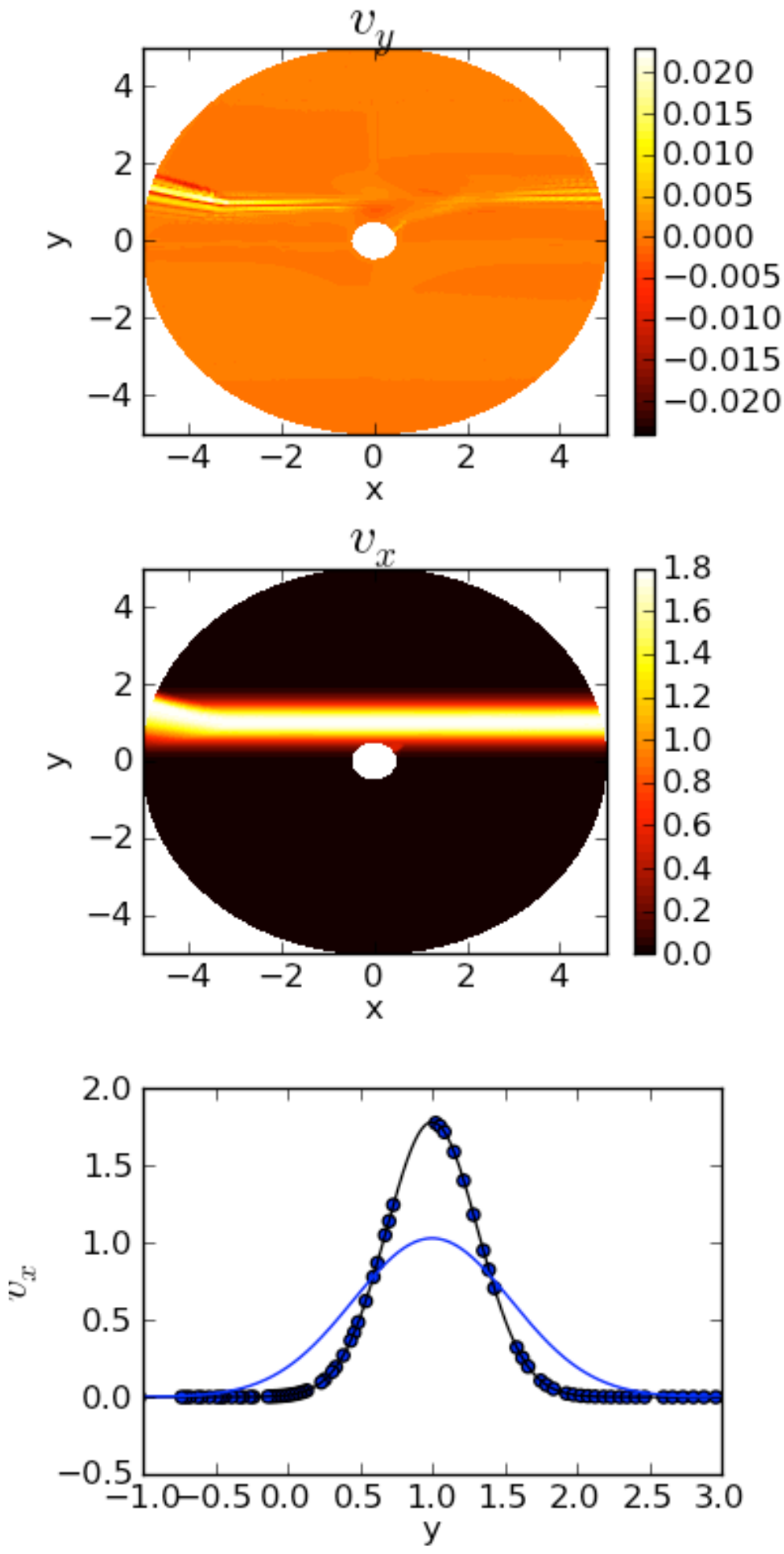} &
\includegraphics[scale=0.4]{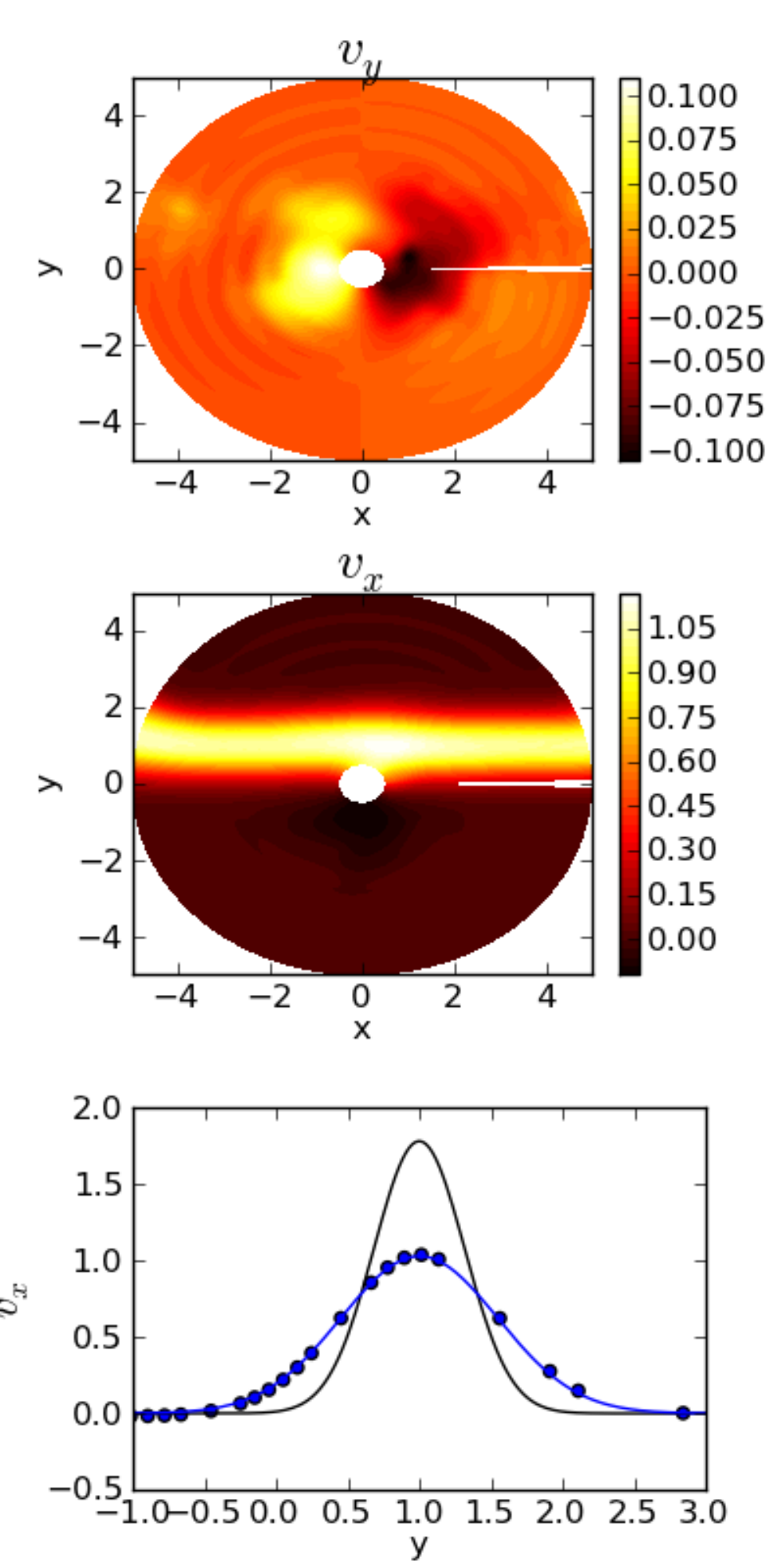} \\
\includegraphics[scale=0.4]{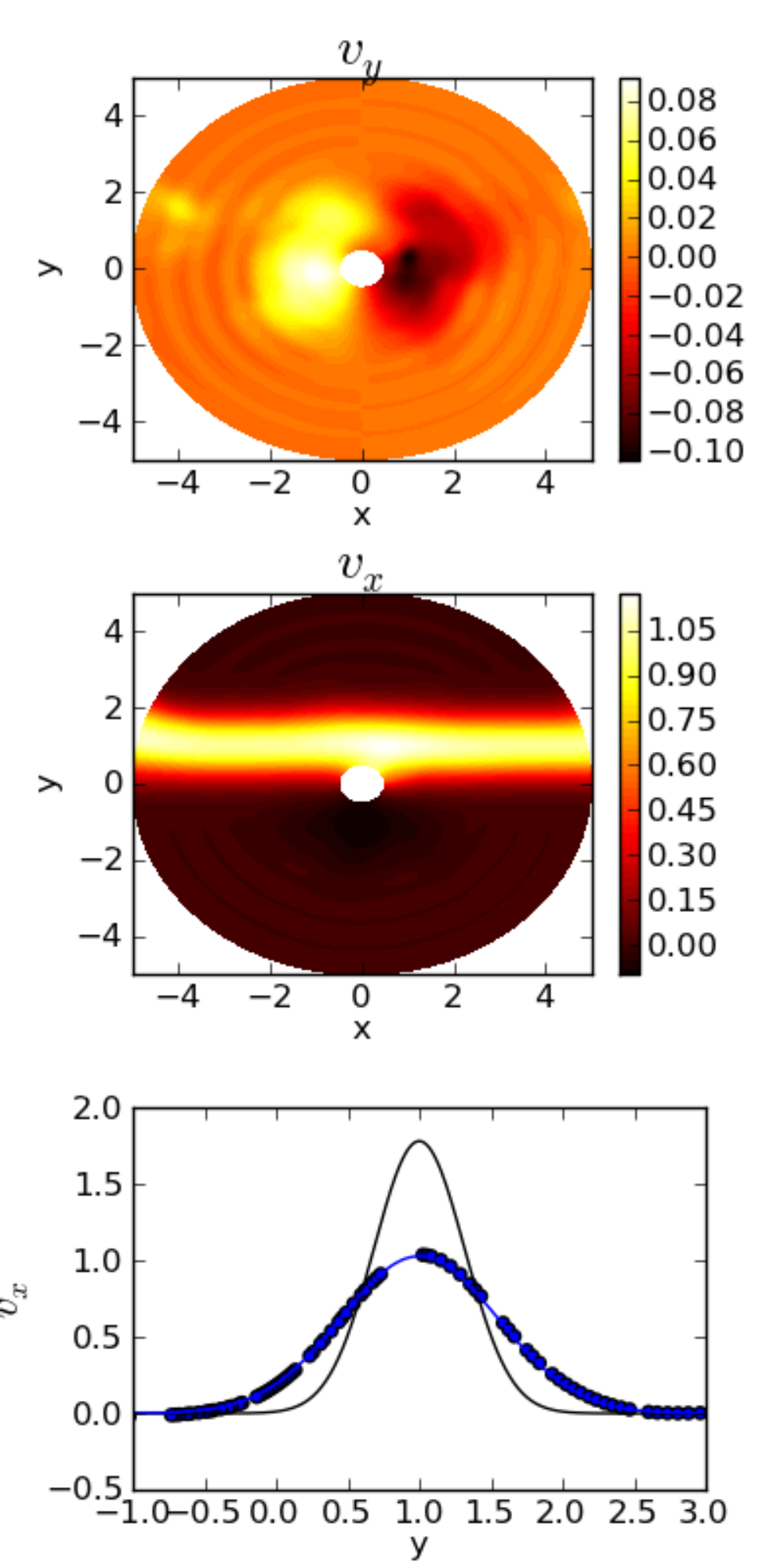} &
\includegraphics[scale=0.4]{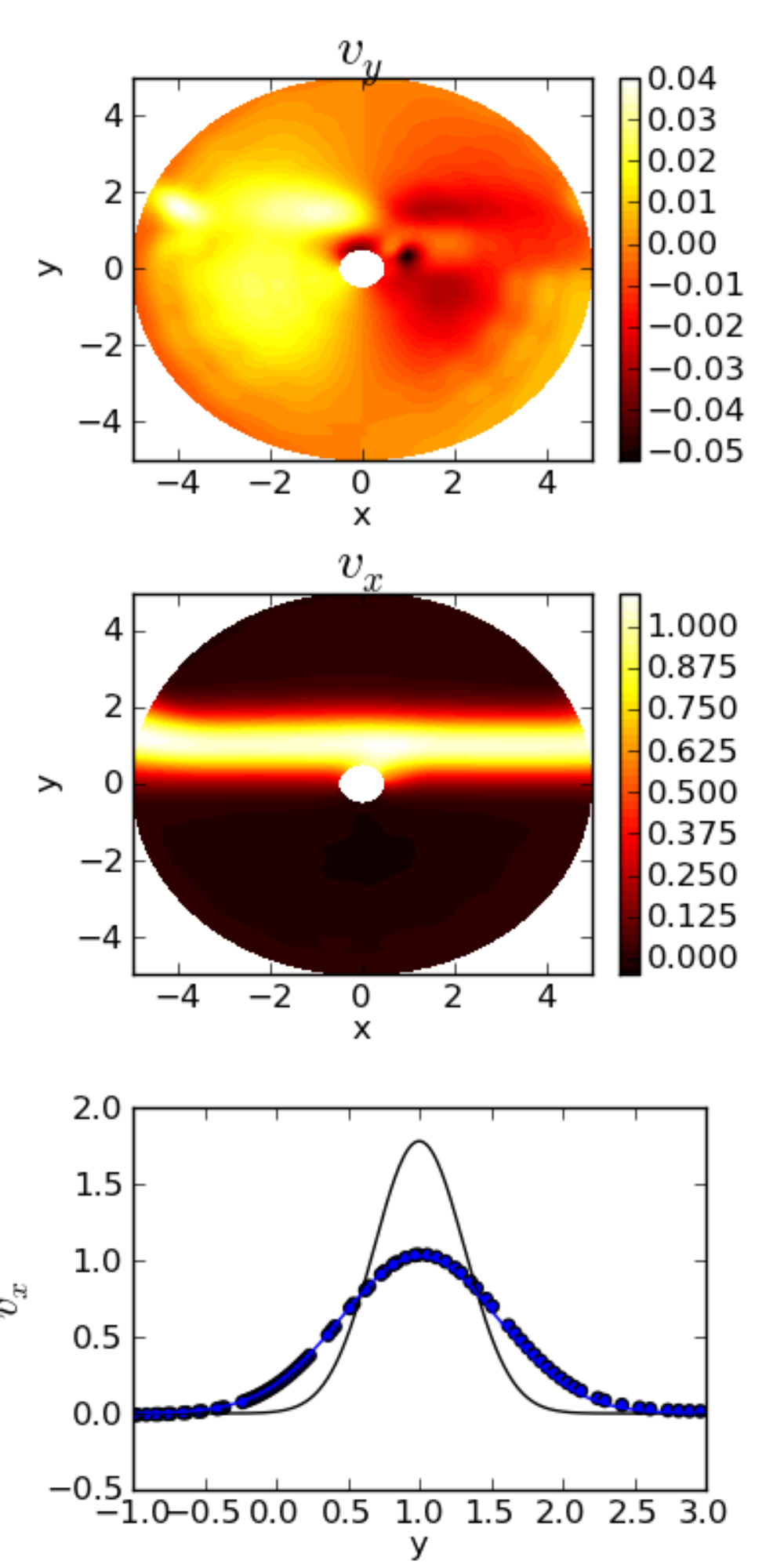} 
\end{array}$
\end{center}
\caption{Results of the Cartesian shear flow test. Each panel is a snapshot taken at $t=1.0$ ($t_0 =  0.5$) of components of velocity in the y-direction (top image in the panel) and components of velocity in the x-direction (middle image in the panel). The bottom image in each panel shows the initial x-velocity (black solid line) and the x-velocity at time t=1.0 (blue solid line) given by (\ref{CShrSoln}). The blue dots plotted in the bottom panels are the simulation values of the x-velocity sampled along the line $x=-2$. The top left panel is run with viscosity turned off at a spatial resolution of 64 radial by 512 azimuthal cells (making cells square at r=1.0).  The other three panels have viscosity turned on at different spatial resolutions. The top right panel has 16 radial by 128 azimuthal cells (making cells square at r=1.0), the bottom left panel has 64 radial by 512 azimuthal cells (making cells square at r=1.0), and the bottom right panel has 24 radial by 512 azimuthal cells (making cells square at r=2.5). We see that the numerical solution follows well the analytic solution (\ref{CShrSoln}) for the evolution of the velocity. Also, the non-zero components of the y-velocity (which should stay zero), decreases with higher resolution and better chosen cell aspect ratio.}
\label{CartShear}
\end{figure}

In practice we implement the Cartesian shear flow test with initial conditions,
\begin{align}
 \Sigma (r,\phi,t_0) &= 1.0  \nonumber \\
 v_r(r, \phi, t_0) &=   v_x(y, t_0) \cos{\phi} \nonumber \\
 v_\phi(r, \phi, t_0)& =   -v_x(y, t_0) \sin{\phi} 
 \end{align}
with $y_0=1.0$, $v_0=1.0$, $\nu =0.1$, $t_0 = 0.5$. Choice of outer and inner boundaries of $r_{\rm{min}}=0.5$, $r_{\rm{min}}=5.0$ allow the solution to not be greatly affected by the boundaries while supplying reasonable resolution requirements. Figure \ref{CartShear} plots the results of this test set up in FLASH for a number of different resolutions and cell aspect ratios.

\label{lastpage}

\end{document}